%
%
%
%
%
%
\documentstyle{amsppt}
\loadbold
\newcount\mgnf\newcount\tipi\newcount\tipoformule\newcount\greco 
\tipi=2          
\tipoformule=0   

\global\newcount\numsec\global\newcount\numfor
\global\newcount\numapp\global\newcount\numcap
\global\newcount\numfig\global\newcount\numpag
\global\newcount\numnf
\global\newcount\numtheo

\def\SIA #1,#2,#3 {\senondefinito{#1#2}%
\expandafter\xdef\csname #1#2\endcsname{#3}\else
\write16{???? ma #1,#2 e' gia' stato definito !!!!} \fi}

\def \FU(#1)#2{\SIA fu,#1,#2 }

\def\etichetta(#1){(\veroparagrafo.\veraformula)%
\SIA e,#1,(\veroparagrafo.\veraformula) %
\global\advance\numfor by 1%
\write15{\string\FU (#1){\equ(#1)}}%
\write16{ EQ #1 ==> \equ(#1)  }}
\def\etichettaa(#1){(A\veraappendice.\veraformula)
 \SIA e,#1,(A\veraappendice.\veraformula)
 \global\advance\numfor by 1
 \write15{\string\FU (#1){\equ(#1)}}
 \write16{ EQ #1 ==> \equ(#1) }}
\def\getichetta(#1){Fig. \verafigura
 \SIA g,#1,{\verafigura}
 \global\advance\numfig by 1
 \write15{\string\FU (#1){\graf(#1)}}
 \write16{ Fig. #1 ==> \graf(#1) }}
\def\retichetta(#1){\numpag=\pgn\SIA r,#1,{\verapagina}
 \write15{\string\FU (#1){\rif(#1)}}
 \write16{\rif(#1) ha simbolo  #1  }}
\def\etichettan(#1){(n\verocapitolo.\veranformula)
 \SIA e,#1,(n\verocapitolo.\veranformula)
 \global\advance\numnf by 1
\write16{\equ(#1) <= #1  }}
%
%
\def\etichettat(#1){\veroparagrafo.\veratheorema%
\SIA e,#1,{\veroparagrafo.\veratheorema} %
\global\advance\numtheo by 1%
\write15{\string\FU (#1){\thu(#1)}}%
\write16{ TH #1 ==> \thu(#1)  }}
%
%
\def\etichettapg(#1){
\SIA e,#1,{\number\pageno} 
\write15{\string\FU (#1){\pgu(#1)}}%
\write16{ PG #1 ==> \pgu(#1)  }}

\newdimen\gwidth
\gdef\profonditastruttura{\dp\strutbox}
\def\senondefinito#1{\expandafter\ifx\csname#1\endcsname\relax}
\def\BOZZA{
\def\alato(##1){
 {\vtop to \profonditastruttura{\baselineskip
 \profonditastruttura\vss
 \rlap{\kern-\hsize\kern-1.2truecm{$\scriptstyle##1$}}}}}
\def\galato(##1){ \gwidth=\hsize \divide\gwidth by 2
 {\vtop to \profonditastruttura{\baselineskip
 \profonditastruttura\vss
 \rlap{\kern-\gwidth\kern-1.2truecm{$\scriptstyle##1$}}}}}
\def\verapagina{
{\romannumeral\number\numcap}.\number\numsec.\number\numpag}}

\def\alato(#1){}
\def\galato(#1){}
\def\veroparagrafo{\number\numsec}\def\veraformula{\number\numfor}
\def\veraappendice{\number\numapp}
\def\verapagina{\number\pageno}\def\veranformula{\number\numnf}
\def\verafigura{{\romannumeral\number\numcap}.\number\numfig}
\def\verocapitolo{\number\numcap}\def\veranformula{\number\numnf}
\def\veratheorema{\number\numtheo}
\def\Eqn(#1){\eqno{\etichettan(#1)\alato(#1)}}
\def\eqn(#1){\etichettan(#1)\alato(#1)}
\def\ver{\veroparagrafo}
\def\Eq(#1){\eqno{\etichetta(#1)\alato(#1)}}
\def\eq(#1){\etichetta(#1)\alato(#1)}
\def\Eqa(#1){\eqno{\etichettaa(#1)\alato(#1)}}
\def\eqa(#1){\etichettaa(#1)\alato(#1)}
\def\dgraf(#1){\getichetta(#1)\galato(#1)}
\def\drif(#1){\retichetta(#1)}
\def\TH(#1){{\etichettat(#1)\alato(#1)}}
\def\thv(#1){\senondefinito{fu#1}$\clubsuit$#1\else\csname fu#1\endcsname\fi} 
\def\thu(#1){\senondefinito{e#1}\thv(#1)\else\csname e#1\endcsname\fi}

\def\PG(#1){{\etichettapg(#1)\alato(#1)}}
\def\pgv(#1){\senondefinito{fu#1}$\clubsuit$#1\else\csname fu#1\endcsname\fi} 

\def\pgu(#1){\senondefinito{e#1}\pgv(#1)\else\csname e#1\endcsname\fi}
\def\eqv(#1){\senondefinito{fu#1}$\clubsuit$#1\else\csname fu#1\endcsname\fi}
\def\equ(#1){\senondefinito{e#1}\eqv(#1)\else\csname e#1\endcsname\fi}
\def\graf(#1){\senondefinito{g#1}\eqv(#1)\else\csname g#1\endcsname\fi}
\def\rif(#1){\senondefinito{r#1}\eqv(#1)\else\csname r#1\endcsname\fi}
\def\bib[#1]{[#1]\numpag=\pgn
\write13{\string[#1],\verapagina}}

\def\include#1{
\openin13=#1.aux \ifeof13 \relax \else
\input #1.aux \closein13 \fi}

\openin14=\jobname.aux \ifeof14 \relax \else
\input \jobname.aux \closein14 \fi
\openout15=\jobname.aux
\openout13=\jobname.bib

\let\EQ=\Eq

\ifnum\tipoformule=1\let\Eq=\eqno\def\eq{}\let\Eqa=\eqno\def\eqa{}
\def\equ{}\fi


{\count255=\time\divide\count255 by 60 \xdef\hourmin{\number\count255}
        \multiply\count255 by-60\advance\count255 by\time
   \xdef\hourmin{\hourmin:\ifnum\count255<10 0\fi\the\count255}}

\def\oramin{\hourmin }

\def\data{\number\day/\ifcase\month\or january \or february \or march \or
april \or may \or june \or july \or august \or september
\or october \or november \or december \fi/\number\year;\ \oramin}

\setbox200\hbox{$\scriptscriptstyle \data $}

\newcount\pgn \pgn=1
\def\foglio{\number\numsec:\number\pgn
\global\advance\pgn by 1}
\def\foglioa{A\number\numsec:\number\pgn
\global\advance\pgn by 1}

\footline={\rlap{\hbox{\copy200}}\hss\tenrm\folio\hss}


\global\newcount\numpunt

\magnification=\magstephalf
\baselineskip=16pt

\def\righthoffset=1truemm
\def\lefthoffset=-1truepc
\voffset=2.5truepc
\hoffset={\ifodd\pageno\righthoffset \else \lefthoffset \fi}
\hsize=6.1truein 
\vsize=8.5truein 
%
%
%


%
%
%
%
\newif\iftitle
\def\titlepage{\global\titletrue}   




\def\a{\alpha}
\def\b{\beta}
\def\d{\delta}
\def\e{\epsilon}

\def\g{\gamma}
\def\k{\kappa}
\def\l{\lambda}

\def\s{\sigma}
\def\t{\tau}

\def\o{\omega}

\def\G{\Gamma}
\def\O{\Omega}

\def\del #1{\frac{\partial^{#1}}{\partial\l^{#1}}}

\def\1{{1\kern-.25em\hbox{\rm I}}}
\def\eu{{1\kern-.25em\hbox{\sm I}}}

\def\R{{\Bbb R}}  
\def\N{{\Bbb N}}  
\def\P{{\Bbb P}}  
\def\C{{\Bbb C}}  
\def\E{{\Bbb E}}  

\def\del{\partial}


\let\cal=\Cal
\def\AA{{\cal A}}
\def\BB{{\cal B}}
\def\CC{{\cal C}}
\def\DD{{\cal D}}

\def\FF{{\cal F}}
\def\GG{{\cal G}}
\def\II{{\cal I}}
\def\JJ{{\cal J}}

\def\SS{{\cal S}}

\def\NN{{\cal N}}
\def\MM{{\cal M}}
\def\OO{{\cal O}}
\def\PP{{\cal P}}

\def\UU{{\cal U}}
\def\LL{{\cal L}}

\def\QQ{{\cal Q}}
%
%
%
\def\specskip{\vskip8pt} 

\def\chap #1#2{\line{\ch #1\hfill}\numsec=#2\numfor=1\numtheo=1}
\def\subchap #1{\line{\bf #1\hfill}\vskip8pt}

\def\un #1{\underline{#1}}

\def\ba{{\backslash}}
\def\intl{\int\limits}

\def\wt{\widetilde}


\def\note#1{\footnote{#1}}

\def\frac#1#2{{#1\over #2}}

\def\text#1{\quad{\hbox{#1}}\quad}
\def\newpage{\vfill\eject}
\def\proposition #1{\specskip\noindent{\thbf Proposition #1:}}

\def\theo #1{\specskip\noindent{\thbf Theorem #1: }}
\def\lemma #1{\specskip\noindent{\thbf Lemma #1: }}

\def\corollary #1{\specskip\noindent{\thbf Corollary #1: }}
\def\proof{{\noindent\pr Proof: }}
\def\proofof #1{{\noindent\pr Proof of #1: }}
\def\endproof{$\square$}
\def\remark{\specskip\noindent{\bf Remark: }}
\def\thanks{\noindent{\bf Acknowledgements: }}
\def\inrm#1{\hbox{\sevenrm{#1}}}
\font\pr=cmbx10  
\font\thbf=cmbx10 scaled\magstephalf
\font\tit=cmbx12
\font\aut=cmbx12
\font\aff=cmsl12

\font\ch=cmbx12

\font\it=cmti10
\font\bf=cmbx10
\font\sm=cmr7

\overfullrule=0pt

\hyphenation{Hei-del-berg Kirck-pat-rick Sherring-ton}
%
%
%
%
%
\titlepage
\centerline{\tit THE SPIN-GLASS PHASE-TRANSITION IN THE HOPFIELD MODEL}
\vskip.2truecm
\centerline{\tit WITH $p$-SPIN INTERACTIONS}
\vskip.2truecm 

\centerline{\aut Anton Bovier 
\note{ e-mail:
bovier\@wias-berlin.de} \note{work supported in part by DFG 
Schwerpunktprogramm ``Interacting stochastic systems of
high complexity''.}
}
\vskip.1truecm
\centerline{\aff Weierstra\ss {}--Institut}
\centerline{\aff f\"ur Angewandte Analysis und Stochastik}
\centerline{\aff Mohrenstrasse 39, D-10117 Berlin, Germany}
\vskip.4truecm
\centerline{\aut  Beat Niederhauser\note{
e-mail: beat\@ime.usp.br}${}^,$\note{
supported by DFG in the  Graduiertenkolleg ``Stochastische Prozesse und 
probabilistische Analysis''and FASPES under grant No. 00/05134-5.}}
\vskip.1truecm
\centerline{\aff IME-USP}
\centerline{\aff Caixa Postal 66.281}
\centerline{\aff  05315-970 S\~ao Paulo - SP, }
\centerline{\aff Brasil}
\vskip1cm
\def\s{\sigma}
\noindent {\bf Abstract:} We study the Hopfield model with pure $p$-spin
interactions with even $p\geq 4$,
  and a number of patterns, $M(N)$  growing with the system size, 
$N$, 
as $M(N) = \a N^{p-1}$. 
We prove the existence of a critical temperature $\b_p$ characterized as the 
first time quenched and annealed free energy differ. We prove that as 
$p\uparrow\infty$, $\b_p\rightarrow\sqrt {\a 2\ln 2}$. Moreover, we show that 
for any $\a>0$ and for all inverse temperatures $\b$,
the free energy converges to that of the REM at inverse temperature 
$\b/\sqrt\a$.  Moreover, above the critical temperature the
distribution of the  {\it replica
overlap} is concentrated at zero. We show that for
large enough $\a$, there exists a non-empty interval
of in the low temperature regime where   the distribution has mass 
both near zero and near $\pm 1$. 
As was first shown by M. Talagrand in the case
of the $p$-spin SK model, this
 implies the the Gibbs measure at
low temperatures is concentrated, asymptotically for large $N$,
 on a countable union of disjoint sets, no finite subset of which has full 
mass. 
 Finally, we show that there is $\a_p\sim 1/p!$ such that for  $\a>\a_p$
the set carrying almost all mass does not contain the original patterns. 
In this sense we describe a genuine spin glass transition.

Our approach follows that of Talagrand's  analysis of the $p$-spin 
SK-model. The more complex structure of the random interactions necessitates, 
however, considerable technical modifications. 
In particular, various results that follow easily in the Gaussian case from 
integration by parts fromulas have to be derived by expansion techniques.



\newpage

\chap{1.~Introduction and Results}1 
 
\bigskip 
 
\noindent 
In a recent paper [T4] (see also [T6] for a more pedagogical 
 exposition) Talagrand has presented for the first time a rigorous 
 analysis  
of a phase transition from a high temperature phase to what could be called 
a "spin glass phase". This was done in the context of the so called $p$-spin  
Sherrington-Kirkpatrick (SK) model [SK] for $p\geq 3$.  
From the heuristic analysis on the basis of the  
replica method (see [MPV]), it is known that this model should have a 
 spin glass phase that   
is much simpler than in the case $p=2$, the standard SK model, and this fact is 
to be expected to be related to the success of Talagrand's approach.  
In any event, this important new result has highlighted the $p$-spin 
interaction model as an important playground to develop new techniques and 
to gain more insight into the fascinating world of spin glasses.

The Hamiltonian of the  
$p$-spin SK model can most simply be described as a Gaussian process $X_\s$ 
 on the  
hypercube $\SS_N\equiv \{-1,1\}^N$ with mean zero and covariance function 
$$ 
\E X_\s X_{\s'}=NR_N(\s,\s')^p 
\Eq(0.1) 
$$ 
where $R_N(\s,\s')\equiv \frac{1}N\sum_{i=1}^N=1-dist_{Ham}(\s,\s')$ 
where $d_{Ham}$ denotes the Hamming distance. Seen from this point of view, 
the distinction between different values of $p$ is in the speed of decrease 
of the correlation of the process $X_\s$ with distance.  
 
Talagrand's methods use heavily the Gaussian nature of the SK model, and  
in particular the fact the $X_\s$ can be represented in the form  
$$ 
X_\s=\sum_{1\leq i_1<i_2<\dots<i_p\leq N}J_{1_i,\dots,i_p}\s_{i_1},\dots 
\s_{i_p} 
\Eq(0.2) 
$$ 
where $J_{1_i,\dots, i_p}$ is a family of i.i.d. standard Gaussian random  
variables.  
It is therefore natural to ask whether and to what extent his approach can be  
generalized to other models that have similar correlation  
decay properties as processes on $\SS_N$, but that are not Gaussian and do not  
have the simple structure as \eqv(0.2). A natural candidate to test  
this question on and   whose investigation 
has considerable interest in its own right, is the so-called $p$-spin  
Hopfield model which we shall describe below. These models have been introduced 
in the context of neural networks by Peretto and  Niez [PN] and Lee et 
al. [Lee] as generalizations   
of the standard Hopfield model [Ho] which corresponds to the case 
$p=2$. This latter case has been studied heavily and since its first 
introduction by Figotin and Pastur [FP1,FP2] has become, on the 
rigorous level, one of the best 
understood mean field spin glass models 
[N1,ST,Ko,BGP1,BGP2,BG1,BG2,BG3,BG4,T3,T7]. It should be noted, however, 
that all the results obtained for this model so far concern the 
high-temperatures phase and the so-called retrieval phase, while next 
to nothing is known about the supposedly existing {\it spin glass 
phase}.  The investigation of this phase in the $p\geq 4$ version of the 
model is the main concern of the present paper.

We now give a precise definition of the models we will study.  
Let $(\O,\FF,\P)$ be an abstract probability  
space and $\{\xi_i^{\mu}\}_{i,\mu \in \N}$ a family of i.i.d.\ Bernoulli variables, taking  
values $1$ and $-1$ with equal probability. 
 
Define for each $N \in \N$ a (finite) random {\it Hamiltonian}, that is, a function 
$H_N: \O \times \SS_N \rightarrow \R$ by 
$$ 
H_N[\o](\s) \equiv - \left(\frac{p!}{N^{2p-2}}\right)^{\frac{1}{2}}\, 
  \sum_{\mu=1}^{M(N)} \sum_{i_1 < \ldots < i_p} \prod_{l=1}^p \xi_{i_l}^{\mu} \s_{i_l}. 
\Eq(PR.1) 
$$ 
The value of $p$ is considered a fixed parameter of the model, and will in the following 
be even and at least be 4.  
While this model can be analyzed rather easily along the lines of the
standard   
Hopfield model if $M\sim N$ (see [BG1]), the results of Newman [N1] on the  
storage capacity suggest that the model should have a good behavior even if 
$M(N)$ scales as $N^{p-1}$, i.e.\note{ In the sequel, we will write
with slight abuse of notation      
$M(N)= \a N^{p-1}$ even for finite $N$. } 
$$ 
\lim_{N \uparrow \infty} \frac{M(N)}{N^{p-1}} = \alpha < \infty. 
\Eq(PR.2) 
$$ 
In this paper we will always be concerned with this case. 
The limit $\alpha$ will also turn out to be a crucial parameter  
for the behavior of 
the system. In the standard Hopfield model, it has been proven that for 
small values of $\a$, the model at low temperatures is in a retrieval phase, 
where there are Gibbs measures that are concentrated on small neighborhoods 
of the stored patterns. It is believed that for large values of $\a$ (or smaller 
 values of $\b$) this  
property fails and that in fact the model should then be very similar to the 
Sherrington-Kirkpatrick  model; however, there exist no rigorous results to  
that effect. While in the present paper we do not present results concerning  
the retrieval phase in the $p\geq 4 $ case, the results we shall present   
show that for reasonably large values of $\a$  a phase transition occurs from the 
 high-temperature phase to a "spin glass phase" that is 
strikingly similar to those of the corresponding SK models.

We will use the following multi-index notation. For finite 
subsets $\II$ of the natural numbers, and real numbers $(x_n)_{n \in \N}$, 
let by $x_{\II} = \prod_{l \in \II} x_{l}$. Let furthermore $\PP_N$ be the set 
of subsets of $\NN = \{1,\ldots,N\}$ of cardinality $p$. The Hamiltonian \eqv(PR.1) 
can then be written as  
$$ 
H_N[\o](\s) = - \left(\frac{p!}{N^{2p-2}}\right)^{\frac{1}{2}}\, 
        \sum_{\mu=1}^{M(N)} \sum_{\II \in \PP} 
        \xi^{\mu}_{\II}\s^{\vphantom{\mu}}_{\II}. 
\Eq(PR.2bis) 
$$ 
\noindent 
These Hamiltonians define random, finite volume Gibbs measures $\GG_{N,\b}[\o]$  
by assigning each configuration $\s \in \SS_N$ a weight proportional to its Boltzmann factor,  
that is 
$$ 
\GG_{N, \b}[\o](\s) = 2^{-N}\frac{e^{- \b H_N[\o](\s)}}{Z_{N,\b}[\o]}. 
\Eq(PR.3) 
$$ 
Consider now the Hamiltonian as a random process indexed by $\s \in \SS_N$. 
Simple calculations allow to verify that the  mean of $H_N$ with respect  
to $\P$ vanishes for all $\s$, that is 
$
\E\,H_N(\s) = 0,\quad \forall \s \in \SS_N, 
$ 
whereas the variance satisfies (for some number $C$ depending on $p$ only) 
$$ 
\a N (1 - CN^{-1}) \leq \E\, H_N(\s)^2 = \frac{p!}{N^{2p-2}} \sum_{\mu=1}^{M(N)}  
        \sum_{\II \in \PP_N} 
\leq \a N, 
\Eq(PR.5) 
$$ 
which motivates our choice of normalization in the definition of $H_N$. The  
covariance is given as 
$$ 
\E\,H_N(\s) H_N(\s') = \frac{p!}{N^{2p-2}} \sum_{\mu=1}^{M(N)} \sum_{\II \in \PP_N} 
    \s^{\vphantom{'}}_{\II}\s'_{\II}  
= \a N R^{p}(\s,\s')(1+ \cal{O}(N^{-1})), 
\Eq(PR.6) 
$$ 
where $R_N(\s,\s')\equiv \frac{1}{N} \sum_{i=1}^N \s_i \s'_i$ is the (normalized)  
{\it replica overlap}. Note that this covariance is in leading order and up to 
the factor $\a$ the same as the covariance for the $p$-spin SK-model ([T4]). 
 
The normalizing factor $Z_{N,\b}$ in \eqv(PR.3) is called {\it partition function} and it is  
given by 
$$ 
Z_{N,\b}[\o] = \E_\s e^{- \b H_N[\o](s)}, 
\Eq(PR.7) 
$$ 
where $\E_{\s}$ is the expectation with respect to the uniform  distribution  
on $\SS_N$. We will call the mean of $Z_{N,\b}$ under $\P$ the  
{\it annealed partition function}.  
 
We define the {\it free energy} $F_{N,\b}[\o]$ by  $F_{N,\b}[\o] \equiv \frac{1}{N}\ln Z_{N,\b}[\o]$.  
\note{Note that physicists often use a different  
normalization, $F_{N,\b} = - \frac{1}{\b N} \ln Z_{N,\b}$.  
We use Talagrand's choice  
convention to facilitate comparison with [T4].} Customarily one calls the mean  
of the free energy, $\E F_{N,\b}$,  the {\it quenched free energy}, while the  
normalized logarithm of the annealed partition function is called the  
 {\it annealed free energy} $F^{\hbox{\sevenrm an}}_{N, \b}\equiv 
\frac{1}{N} \ln \E\,Z_{N,\b}$.  
Observe that by H\"older's  inequality, both  
the quenched free energy and the annealed free energy are  
convex functions of $\b$.

Let us briefly mention a variant of the above model. On the same configuration space and 
with the same random variables $\xi$, we define macroscopic random order parameters  
$$ 
m^{\mu}[\o](\s) \equiv \frac{1}{N} \sum_{i = 1}^N \xi^{\mu}_i \s^{\vphantom{\mu}}_i. 
\Eq(RO.1001) 
$$ 
These parameters are considered as components of a vector in $\R^{M(N)}$ with $M(N)$ as in 
\eqv(PR.2). New Hamiltonians are now defined through 
$$ 
\bar{H}_N[\o](\s) - \frac{N}{s_p}\left(\|m[\o](\s)\|_p^p - \E\,\|m[\o](\s)\|_p^p\right), 
\Eq(PR.1002) 
$$ 
where $s = s_p > 0 $ is defined such that the covariance of $\bar{H}$ is in leading order 
in $N$ equal to $\a N$. The interaction $\bar{H}$ is a straightforward  
generalization 
of the usual $p=2$ case.  
However, computing the 
resulting covariance function one sees that it decreases  
only quadratically with  
the Hamming distance.  Therefore it will not share the special features  
of the $p$-spin SK model.  
An analysis of the high-temperature phase for  $\bar{H}$ has been  presented  
in  [Ni1].  
 
We will now state our results. They will always concern the model 
with Hamiltonian  \eqv(PR.1) and $p\geq 4$.  
 
The first result we prove for both choices of the Hamiltonian is that 
for high enough  
temperatures (that is, low values of $\b$), the limit of the annealed 
free energy exists.  
 
\theo{\TH(pr.1)} 
{\it 
  If $\b < e^{-2} (p!)^{\frac{1}{2}} \equiv \b'_p$, then the annealed 
  free energy   
corresponding to $H$ satisfies 
$$ 
F^{\hbox{\sevenrm an}}_{N, \b}  = \frac{\a \b^2}{2}(1 + {\cal 
O}(N^{-1})).  
\Eq(PR.8) 
$$ 
} 
\specskip 
 
Note that for larger values of $\b$, the annealed free energy 
diverges. Our analysis will be limited to the case when $\b<\b'_p$ where a comparison  
to the SK model is still possible. It is nice to see that this value tends to infinity 
with $p$ very rapidly. Moreover, we shall see that this value becomes much  
larger than the  critical temperature, as $\a$ gets large.  
 
Jensen's inequality implies that the quenched free energy is  
less then or equal to the  annealed free energy, 
$$ 
\E\,F_{N,\b} = \frac{1}{N} \E\, \ln Z_{N,\b}  
\leq \frac{1}{N} \ln \E\,Z_{N,\b} = F^{\hbox{\sevenrm an}}_{N, \b}. 
\Eq(PR.9) 
$$ 
We define the critical temperature to be the infimum of values for 
which equality holds in \eqv(PR.9), i.e.~in terms of $\b$, 
$$ 
\b_p \equiv \sup\big\{\b \geq 0: \limsup_{N \uparrow \infty}\E\, F_{N,\b}  
        = \limsup_{N \uparrow \infty}F^{\hbox{\sevenrm an}}_{N,\b} \big\}. 
\Eq(PR.10) 
$$ 
 Observe 
that in general   
$\lim_N \E\,F_{N,\b}$ need not exist. 
 
By \eqv(PR.6), as a random process on $\SS_N$, $H_N(\s)$ has  
(up to an overall factor) essentially the same  
covariance structure as the  $p$-spin SK Hamiltonian.  
This suggest that as in that case, for $p$ large the model should be similar  
to Derrida's 
 {\it random energy model} (REM) 
[D1,D2]s Recall that in this model, $H_N(\s)\equiv \sqrt NX_\s$, where  
 $\{X_{\s}\}_{\s \in \SS_N}$ are 
i.i.d.\ standard   
normal random variables). Defining the corresponding  partition function 
$ 
Z^{\hbox{\sevenrm REM}}_{N,\b} = \E_{\s} e^{\b \sqrt{N} X_{\s}} 
$, one easily sees that the free energy satisfies  
 [D2] 
$$ 
f_{\b}^{REM} = \lim_{N \rightarrow \infty} \frac{1}{N}\E\,\ln 
Z^{\hbox{\sevenrm REM}}_{N,\b}   
= \cases \b^2/2,&                       
                \hbox{if\ $\b \leq \sqrt{2 \ln 2}$} \cr 
         \b\sqrt{2 \ln 2} - \ln 2,& \hbox{if\ $\b \geq \sqrt{2 \ln 2}$} \cr 
  \endcases 
\Eq(PR.12) 
$$ 
We will show  that as $p$ tends to infinity, $\sqrt{\a} \b_p$ tends to the  
critical value $\sqrt{2 \ln 2}$ of the REM. Moreover, pointwise in $\a,\b$,  
$$ 
\frac{1}{\b} \lim_{p \rightarrow \infty} \lim_{N \rightarrow \infty}  
        \frac{1}{N} \E\,\ln Z_{N,\sqrt\a\b} = \frac{1}{\b} f_{\b}^{REM}. 
\Eq(PR.14) 
$$ 
in analogy to the situation in the $p$-spin SK model [T6]. 
While this may  not be  very surprising, it is also not totally obvious and 
will require some non-trivial computations.  
 
Our next two theorems make these relations precise. 
We will denote by $I(t)$ the                                                      {\it Cram\'er entropy} function, 
$$ 
I(t) =   \frac{1}{2}(1 - t)\ln(1 - t) + \frac{1}{2}(1 + t)\ln(1 + t),  
\Eq(PR.15) 
$$ 
 
\theo{\TH(pr.2)}  
{\it 
  The critical value $\b_p = \b_p(\a)$ satisfies  
$$ 
\b_p(\a)^2 \geq \min\left( \frac{\b'_p{}^2}{4}, \inf_{t \in [0,1]} I(t) \frac{1 + t^p}{\a t^p}\right) 
        \equiv \check{\b}_p(\a)^2. 
\Eq(PR.16) 
$$ 
Furthermore, if $\a \geq \frac{e^4 2\ln 2}{p!} \equiv  \a_p$ then 
$$ 
\b_p(\a)^2 \leq \frac{2 \ln 2}{\a} \equiv \hat{\b}(\a)^2. 
\Eq(PR.17) 
$$ 
} 
\specskip 
\noindent{\bf Remarks:} (i) One can show that the inequality \eqv(PR.17)  
actually strict. In [B2] it is shown that for the SK case, 
$\b_p\geq \sqrt {2\ln 2}(1-c_p)$ with $c_p=2^{-p(4+O(1/p))}$. This follows  
from a corresponding upper bound on the supremum of $H_N(\s)$ which can be  
obtained using standard techniques. These estimates can without doubt be  
carried over to our case.

\indent (ii) The bounds on the critical temperature are essentially  
(up to a factor $\sqrt \a$) the same as for the $p$-spin SK-model ([T4], 
Theorem~1.1). 
\note{Observe that in [T4], the normalization of the Hamiltonian contains an extra 
factor $2^{-1/2}$.} 
\specskip 
\noindent 
By elementary analysis  one finds  that, as $p$ tends to infinity, 
$$ 
\inf_{0 \leq t \leq 1} ( (1 + t^{-1}p) I(t))^{1/2}  
=  \sqrt{2\ln 2} \left(1 - \frac{2^{-p-1}}{\ln 2} \right) +  
{\cal O}(p^3 2^{-2p}). 
\Eq(PR.18) 
$$ 
This, together with the convexity of the free energy in $\b$, 
 will allow us to  prove  
the following statement. 
 
\theo{\TH(pr.3)} 
{\it 
   As $p \rightarrow \infty$, the lower bound 
 $\check{\b}_p \uparrow \hat{\b}$. Moreover, for all $\b\geq 0$ and $\a>0$,  
$$ 
\lim_{p \uparrow \infty} \lim_{N \uparrow \infty} \frac{1}{N} \E\,F_{N,\b}  
        = f^{REM}_{\b \a^{-1/2}}. 
\Eq(PR.19) 
$$ 
} 
The basic strategy used to prove these results are rather general.  
In Chapter~2, we will explain  
them by means of the analogous calculations in the REM.  
For now, we just mention that the hard part is to prove the lower bound  
\eqv(PR.16), whereas 
the upper bound \eqv(PR.17) is comparatively easy and will follow from an 
 estimate on the ground  
state energy.

\specskip 
\noindent

An important point in the study of disordered models is the question of  
self-averaging of  
the free energy. While in many cases this follows from general principles 
[MS,T1] of mass concentration, due to the failure of certain  
convexity properties, it  turns out to be surprisingly difficult to prove the  
following result\note{A sharper estimate can be proven with much less effort  
for the interaction $\bar H_N$, see [Ni1]. } 
 
\theo{\TH(pr.4)}  
{\it For all $\b, n, \t, \varepsilon >0$ there exists $C_n<\infty$ 
(depending only on  
$n$ and $\b$), and $\bar{N}<\infty$ such that 
the free energy satisfies 
$$ 
\P\left[|F_{N,\b} - \E\,F_{N,\b}| \geq  \t \b N^{- \frac{1}{2} + \varepsilon} \right] \leq C N^{- n}
\Eq(PR.19bis) 
$$ 
for all $N \geq \bar{N}$. In particular, 
$$ 
\lim_{N \uparrow \infty} |F_{N,\b} - \E\,F_{N,\b}| = 0, \quad \P-a.s. 
\EQ(PR.19ter) 
$$ 
} 
 
{\bf Remark:} From recent results in the $p$-spin SK-model and the REM [BKL], one actually 
expects that the fluctuations in the small $\b$ region are of much lower order. 

While the critical temperature is defined in terms of the behavior of the  
free energy, it turns out that this phase transition goes along with a change  
in the behavior of the replica overlap parameter, $R_N(\s,\s')$.  
This will eventually lead to rather detailed insight into the properties of  
the Gibbs measures at low temperatures. 
 
The crucial link between the two will be provided by the next theorem.  
 
\theo{\TH(pr.5)} 
{\it Assume that $\b<\frac 12 \b_p'$. Then 
  the replica overlap $R_N(\s,\s')$ satisfies 
$$ 
\E\,\frac{\partial F_{N,\b}}{\partial \b}  
        =  \a \b \left(1 - \E\,\GG_{N,\b}\otimes\GG_{N,\b}[R_N(\s,\s')^p]\right)(1 + {\cal O}(N^{-1})), 
\Eq(PR.20) 
$$ 
} 
 
Note that in the case of the Gaussian SK models, this relation is a trivial  
consequence of the  
 {\it integration by parts formula}  
$$ 
\E\,[g f(g)] = \E\,[g^2] \E\,[f'(g)], 
\Eq(19.threequarter) 
$$ 
which holds for any centered Gaussian random variable $g$ and any function  
$f$ not  
growing faster than some polynomial at infinity.    
To establish this result without the help of this formula turns 
 out to require a considerable effort. Similar tools are also 
instrumental in the proof of Theorem \thv(pr.4). 
 
We then have the following consequence to Theorem~\thv(pr.2) and  
Theorem~\thv(pr.5). 
 
\theo{\TH(pr.6)} 
{\it Assume that  $\a\geq \a_p$.  
  If $\b < \b_p$, then 
$$ 
\limsup_{N \uparrow \infty} \E\,\GG_{N,\b}\otimes\GG_{N,\b}
  [R_N(\s,\s')^p] = 0.  
\Eq(PR.21) 
$$ 
Conversely, if 
$\limsup_N \E\,\frac{\partial F_{N,\b}}{\partial \b} < \a \b$, then 
$$ 
\liminf_{N \uparrow \infty} \E\,
\GG_{N,\b}\otimes\GG_{N,\b}[R_N(\s,\s')^p] > 0.  
\Eq(PR.22) 
$$ 
In particular, \eqv(PR.22) holds for all $\b \in[ \hat{\b},\frac
12\b_p')$. 
} 
\specskip 
\remark 
It seems reasonable that \eqv(PR.22) should hold for all $\b$ above the  
critical 
$\b_p$, but there seems to be no general principal that would  
 prohibit a {\it reentrant 
phase transition}. 
 
Inequality \eqv(PR.22) expresses in a weak  way that below the critical  
temperature, the Gibbs measure gives some mass to a  a small subset of the  
configuration space. This result can be strengthened. As in [T4], we show 
that the overlap between replicas is either very close to one, or to zero: 
 
\theo{\TH(pr.9)}{\it For any $\e>0$ there exists  
$p_0<\infty$ such that for all $p\geq p_0$,  $\a>\a_p$, 
 and 
 for all $0\leq\b<\b_p'$ 
$$ 
\lim_{N\uparrow\infty} 
\E\GG^{\otimes 2}_{N,\b} (|R_N(\s,\s')|\in [\e,1-\e])=0  
\Eq(9.0) 
$$ 
If, moreover, $\b<\check \b_p$, then for any $\e>0$ there exists  
$p_0<\infty$ such that for all $p\geq p_0$, such that for some $\d>0$, 
for all large enough $N$, 
$$ 
\E\GG^{\otimes 2}_N(|R_N(\s,\s')|\in [\e,1])\leq e^{-\d N} 
\Eq(9.01) 
$$ 
} 
\specskip 
 
\remark Note that we prove this result without any restriction on the  
temperature,             
while Talagrand requires some upper bound on $\b$ both in [T4] and in the  
announcement [T5] even though the bound in [T5] is greatly improved.  
We stress that the our  result is also valid   for  the  
$p$-spin SK-model. The same applies for all subsequent results.  
 
The information provided by Theorem's \thv(pr.6) and \thv(pr.9) allow 
gain considerable insight into the nature of the Gibbs measures in the 
low temperature phase. This observation is due to Talagrand.  
 
In [T4] he showed that whenever \eqv(PR.22) and \eqv(9.0) 
hold, it is possible to decompose the state space $\SS_N$ into a 
collection of disjoint subsets $\CC_k$ such that  
\item{(i)} 
$$ 
\lim_{N\uparrow\infty}\E\GG_{N}^{\otimes 
2}\left(\bigl\{(\s,\s')|\,|R_N(\s,\s')|>\e\bigr\}\ba  
\cup_k\CC_k\times \CC_k\right)=0 
\Eq(11.1) 
$$ 
(where the $\CC_k$ depend both on $N$ and on the random parameter!), and 
\item{(ii)} If $\s,\s'\in \CC_k$, then $R_N(\s,\s')\geq 1-\e$.  
 
Note that because of the global spin flip symmetry of our models with 
$p$ even, these lumps necessarily appear in symmetric pairs.  
 
In [T4] Talgrand analyzed the properties of these lumps further using 
the cavity method. He showed that, under a certain hypothesis that we 
shall discuss shortly, for $\b$ not too large this lumps correspond to 
what is known as ``pure states''. While it is very likely that this 
analysis can also be carried over to our models, we will leave this 
question open to further investigation. We find it however interesting 
to discuss the situation of the general hypothesis. Talgrand's 
hypothesis in [T4] concern the distribution of mass on the 
lumps. Roughly, they can be states as  
 
\theo{\TH(L.2)} {\it Assume that $\frac 12\b_p'>\b>\b_p$.  
Let $\CC_k$ be 
ordered such that for all $k$, $\GG_{N,\b}(\CC_k)\geq \GG_{N,\b}(\CC_{k+1})$.  
Then for 
all $k\in \N$, there exists $p_k<\infty$ such that for all $p\geq p_k$,   
$$ 
\lim_{N\uparrow \infty} \E\GG_{N,\b}\left(\cup_{l=1}^k \CC_l\right)<1 
\Eq(11.2) 
$$ 
 except possibly for an 
exceptional set of $\b$'s of zero Lebesgue measure. 
Moreover, for $k$ large, $p_k\sim \frac 23\frac {\ln k}{\ln 2}$. 
} 
 
In [T5] Talagrand has announced a proof of an even stronger  theorem 
in the $p$-spin SK model that makes use of general identities between 
replica overlaps proven by Ghirlanda and Guerra [GG].  We show that at 
least Theorem  
\thv(L.2) also holds in our model.

A final result is particular to the Hopfield model and concerns the  
storage properties of the model. 
Newman has proven in [N1] that for small $\a$, the Hamiltonian has 
deep local minima in the vicinity of each pattern. Here we show a 
somewhat converse result, stating that if $\a$ is not too small, then 
small neighborhoods of the patterns have asymptotically mass zero. In 
other words, none of the patterns falls into one of the 'lumps'. This 
gives the final justification to call the phase transition we have 
observed a transition to a genuine {\it spin glass phase}.  
 
\theo{\TH(pr.7)}  
{\it 
  Suppose that $\a$ satisfies $\a \b_p(\a) > (p!)^{-1/2}$. Then there   
exists a  $\d \in (0,\frac{1}{p})$ and $\bar{N} \in \N$ such that for  
all $N \geq \bar{N}$, 
$$ 
\P[\arg \sup |H_N(\s)| \in \bigcup_{\mu = 1}^{M(N)} B_{\d} (\xi^{\mu})]  
\leq  N^{- m},  
\Eq(PR.23) 
$$ 
where $B_\d(\xi^{\mu})$ is the $N\d$-ball around $\xi^{\mu}$ in the space $\R^N$ with 
respect to the Hamming metric. 
In particular, there exists an $\a_{sp} = \a_{sp}(p)$ such that \eqv(PR.23) holds 
for all $\a > \a_{sp}$. 
Furthermore,  
$$ 
\arg \sup |H_N(\s)| \notin \bigcup_{\mu = 1}^{M(N)} B_{\d} (\xi^{\mu})]  
        \hbox{\tenrm \ eventually} \quad \P-a.s. 
\Eq(PR.23bis) 
$$ 
} 
\specskip 
\noindent 
The proof of this result is based on the comparison between the ground state energy 
of the system and an estimate on the values of the Hamiltonian in the balls around 
the patterns. While the former increases as $N\sqrt{\a}$, the latter is almost constant 
and with high probability close to $N (p!)^{-1/2}$.

The remainder of this paper is organized as follows. In Chapter~2, we explain the ideas behind 
the proof of the bounds on the critical temperature by calculating the corresponding quantities 
in the REM. In Chapter~3, Theorem~\thv(pr.1) is proved. Chapter~4 is devoted to the lower and 
the upper bound on the critical $\b$ (as well as the proof of Corollary~\thv(pr.3)).  
In Chapter~5  we prove Theorem \eqv(pr.4)
In Chapter~6 
we prove the results on the distribution of the  
replica overlap, Theorems \thv(pr.5) to \thv(L.2).  
In Chapter~8 we prove Theorem \thv(pr.7).  
\bigskip 

\chap{2.~Second Moment Method: The REM}2

\bigskip

\noindent This section is meant to give a pedagogical exposition of
Talagrand's truncated second moment method [T3,T4] in the context of
the simplest possible setting, 
the random energy model. A more detailed exposition can
also be found in [B2] and [T6]. 
Since the application of this method in our case will become rapidly
somewhat technical in our case, we still find it useful to give the
reader an outline in a non-technical context\note{Note that of
course much sharper results than those presented here can be obtained
in the REM when making use of its special features. See e.g. [BKL] for
a full analysis of the fluctuations of the free energy.}.   Moreover,
the REM 
provides important bounds for the real model.

We will now show how this method works by using it to compute the free energy 
of the REM. 
Note first that in general,
$$ 
\frac{\partial F_{N,\b}}{\partial \b} = -
\frac{1}{N}\GG_{N,\b}[H_N] \leq \frac{1}{N} \E\,[\sup_{\s}
|H_N(\s)|].  
\Eq(REM.1)
$$
Moreover, since 
$$
\P\,[\sup_{\s}|H_N(\s)| > t N]
\leq 2^N \P\,[|H_N(\s)| > t N] 
\leq 2^{N+1}  e^{- \frac{t^2 N}{2}}.
\Eq(REM.4)
$$
from this it follows easily  that
$$
\eqalign{
\frac{1}{N}\E\,[\sup_{\s} H_N(\s)] 
&\leq \sqrt{2 \ln 2} + 2 \intl_{\sqrt{2 \ln 2}}^\infty e^{- N
(\frac{t^2}{2} - \ln 2)} dt \cr 
&\leq \sqrt{2 \ln 2} + N^{-1} \sqrt{\frac {2}{\ln 2}}.
}
\Eq(REM.5)
$$
This is the upper bound on the derivative of the expectation of the
free energy. Suppose 
now that $\b > \sqrt{2 \ln 2} = \b'$. Convexity of the free energy
then implies that 
$$
\E\,F_{N,\b} \leq \E\,F_{N,\b'} + (\b - \b') \b'
\EQ(REM.6)
$$
and in the limit
$$
\limsup_{N \uparrow \infty} \E\,F_{N,\b} 
\leq - \frac{\b^{\prime 2}}{2} + \b \b'
= \frac{\b^2}{2} - (\b - \b')^2  < \frac{\b^2}{2},
\Eq(REM.7)
$$
which by definition means that $\b' \geq \b_{\hbox{\fiverm REM}}$. In
the case of the  
$p$-spin Hopfield model, the corresponding
 calculations will be identical to those above, except
for the bounds on the  
extrema of the Hamiltonian, where the non Gaussian character induces
somewhat more  involved calculations.

The basic idea behind Talagrand's approach to prove the lower bound
(which he did for the $p$-spin 
SK-model in [T4]), is to obtain a variance estimate on the partition
function. This will imply 
that the expectation of the logarithm behaves like the logarithm of
the expectation of this 
quantity. In the REM, one would naively compute
$$
\eqalign{
\E\,[Z^{\hbox{\fiverm REM}}_{N,\b}{}^2] 
&= \E_{\s,\s'} \E\, e^{\b \sqrt{N} (X_{\s} + X_{\s'})} \cr
&= 2^{-2N} \left( \sum_{\s \neq \s'} e^{N\b^2} + \sum_{\s} e^{2 N
\b^2} \right) \cr 
&= e^{N \b^2} \left[ (1 - 2^{-N}) + 2^{-N}e^{N\b^2}\right].
}
\Eq(REM.10)
$$
The second term in the brackets is exponentially small if and only if
$\b^2 < \ln 2$,  
and this cannot be the critical value since it violates the upper
bound $\b'$ above. 
\note{This is already contained in [D2]}
The point is that while in the 
computation of $\E\,e^{2 \b \sqrt{N} X_{\s}}$, the dominant
contribution comes from the part 
of the distribution of $X_{\s}$ around $X_{\s} = 2 \b \sqrt{N}$,
whereas in  
$\E\,Z^{\hbox{\fiverm REM}}_{N,\b}$ 
the main part is contributed by $X_{\s}$ around $\b \sqrt{N}$. One is
thus led to consider  
  the second moment of a suitably truncated version of
$Z^{\hbox{\fiverm REM}}_{N,\b}$.              
Namely, for $c>0$,
$$
\tilde{Z}^{\hbox{\fiverm REM}}_{N,\b}(c) 
= \E_{\s}e^{\b \sqrt{N} X_{\s}} \1_{\{X_{\s} < c \sqrt{N}\}}.
\Eq(REM.11)
$$
One then finds that (modulo irrelevant prefactors)
$$
\E\,\tilde{Z}^{\hbox{\fiverm REM}}_{N,\b}(c) = 
\cases 
        \rlap{$e^{\frac{\b^2 N}2},$} 
        \hphantom{\frac{1}{\sqrt{N} (\b - c)} e^{N \b c - \frac{N
        c^2}{2}},} 
        \hbox{\ if\ } \b < c, \cr
        \frac{1}{\sqrt{N} (\b - c)} e^{N \b c - \frac{N c^2}{2}},
        \hbox{if\ $\b > c$}. \cr 
\endcases      
\Eq(REM.12)
$$
Moreover, for $\b < c$, 
$$
\E\,\tilde{Z}_{N,\b}(c) =\E\, Z_{N,\b} \left(1 -\frac{
e^{-\frac{1}{2}(c-\b)^2N}}{\sqrt N(c-\b)}\right) 
\Eq(REM.13)
$$
On the other hand,
$$
\E\,\tilde{Z}_{N,\b}(c)^2 = (1 - 2^{-N})
\left(\E\,\tilde{Z}_{N,\b}(c)\right)^2  
                + 2^{-N} \E\,e^{2 \b \sqrt{N} X_{\s}} \1_{\{X_{\s} < c
                \sqrt{N}\}}, 
\Eq(REM.14)
$$
where the second term satisfies
$$
2^{-N} \E\,e^{2 \b \sqrt{N} X_{\s}} \leq
\cases  \rlap{$2^{-N} e^{2 \b^2 N},$}
        \hphantom{2^{-N} e^{2 c \b N - \frac{c^2 N}{2}},}
        \hbox{\ if\ $2\b < c$} \cr
        2^{-N} \frac {(2\b -c)\sqrt N}{e^{2 c \b N - \frac{c^2
        N}{2}}}, \hbox{\ otherwise}, \cr 
\endcases
\Eq(REM.15)
$$
and thus
$$
\eqalign{
2^{-N} &\E\,e^{2 \b \sqrt{N} X_{\s}} \1_{\{X_{\s} < (1 + \varepsilon)\b
\sqrt{N}\}}  \cr 
& \leq (\E\,\tilde{Z}_{N,\b})^2 \times
\cases
        e^{-N(\ln2 - \b^2)},&\,\hbox{if}\,
        \b< \frac{c}{2}, \cr
        \frac{e^{-N(c - \b)^2 - N(\ln 2 - \frac{c^2}{2})}}
{(2\b-c)\sqrt N},&\,\hbox{if}\, \frac{c}{2} < \b
        < c,
 \cr e^{(c^2/2-\ln 2)N}\sqrt N\frac {(\b-c)^2}{2\b -c},& \,\hbox{if}\, \b>c
\endcases
}
\Eq(REM.16)
$$
Hence, for all $c < \sqrt {2\ln 2}$, and all $\b \neq c$
$$
\E\,\frac{(\tilde{Z}_{N,\b}(c) -
\E\,\tilde{Z}_{N,\b}(c))^2}{\E\,[\tilde{Z}_{N,\b}(c)^2]} 
\leq e^{-N g(c,\b)},
\EQ(REM.17)
$$
where $g(c,\b) > 0$. Thus, by Chebyshev's inequality, it is immediate that
$$
\lim_{N \uparrow \infty} \frac{1}{N} \E\,\ln \tilde{Z}_{N,\b}(c) 
= \lim_{N \uparrow \infty} \frac{1}{N} \ln \E\,\tilde{Z}_{N,\b}(c) ,
\quad \forall c < \sqrt{2\ln2}. 
\Eq(REM.18)
$$
Since this gives a lower bound of the free energy that is as close to
the upper bound  
as desired, we see that the upper bound gives in fact the true value.

This is a remarkable feature of the REM: the expectation of the
logarithm of the partition 
function coincides with the log of the expectation of a suitably
truncated partition function. 
While this  is  rather special to the REM, the  method
is general enough to  
provide lower bounds in the far more complicated situations, as we will see.

\bigskip

\chap{3.~The Annealed Free Energy.}3
\bigskip

\noindent
In this Section we compute the anneled free energy. Apart from the intrinsic 
interest this can be seen as the computation of the log-moment generating 
function of the Hamiltonian and this will be a basic input in the sequel. 
While in the SK models this is a two line computation, here even this will
require a considerable effort.
The idea is to use Taylor expansions and to exploit the fact that the 
Hamiltonian is a sum of a very large number of independent random variables.
Namely
$$
\eqalign{
 \E\,Z_{N,\b} &=  \E\, e^{- \b H_N[\o](\s)} 
=  \E\, \exp\left( \b \left(\frac{p!}{N^{2p-2}}\right)^{\frac{1}{2}} 
                \sum_{\mu=1}^{M(N)} \sum_{\II \in \PP_N} \xi_{\II}^{\mu}
                \right) \cr
&= \prod_{\mu=1}^{M(N)}
\left[ \E\,\exp\left(\b \left(\frac{p!}{N^{2p-2}}\right)^{\frac{1}{2}} 
                \sum_{\II \in \PP_N} \xi_{\II}^{\mu}
                \right)\right] \cr
&=\left[ \E\,\exp\left(\b \left(\frac{p!}{N^{p-2}}\right)^{\frac{1}{2}} 
                Y \right)\right]^{M(N)},
}
\Eq(A.1)
$$
where we introduced the abbreviation 
$Y \equiv N^{-\frac{p}{2}}\sum_{\II \in \PP_N} \xi_{\II}^{1}$.
We now expand the exponential function according to the bound
$\left|e^x - 1 - x - \frac{x^2}{2}\right| < |x|^3 e^{|x|}$.
Thus,
$$
\eqalign{
\Bigg| \E\, \left[ \exp\left(\b \left(\frac{p!}{N^{p-2}}\right)^{\frac{1}{2}} Y \right)
        \right] 
 - 1 - \frac{\b^2 N^{2-p}}{2} \Bigg|  & \cr
&\kern-3.5cm\leq \E\,\left[ \b^3 \left(\frac{p!}{N^{p-2}}\right)^{\frac{3}{2}} |Y|^3
        \exp\left(\b \left(\frac{p!}{N^{p-2}}\right)^{\frac{1}{2}} |Y| \right) \right]
        + {\cal O}(N^{1-p}).
}
\Eq(A.2)
$$
Observe that the quadratic term is in fact just $N^{p-1}$ times 
the variance of $H_N$.
We will show in a moment that the expectation on the right-hand side of
 \eqv(A.2) is 
bounded by a constant times $N^{3 - \frac{3p}{2}}$. Assuming this and 
recalling that
$p \geq 4$, it is evident that 
$$
\eqalign{
\ln \E\, Z_N &= M(N) \ln\left(1 + \frac{\b^2 N^{2-p}}{2}(1+O(N^{-1})\right)  
        \cr
&= \frac{\a \b^2 N}{2}(1 +O(N^{-1})).
}
\Eq(A.3)
$$
which is what we want to prove. We now turn to the non-trivial part of the 
proof, the estimate of
 the remainder on the right-hand side of \eqv(A.2).
 To to this, we decompose the exponent
into two factors, and use on one the obvious bound 
$|Y| \leq (p!)^{-1} N^{p/2}$. This yields
$$
\eqalign{
\E\,\Big[|Y|^3
        \exp\Big(\b (p!)^{\frac{1}{2}} N^{\frac{2-p}{2}} 
|Y| \Big) \Big] 
&=  \E\,\left[|Y|^3 \exp\left(\b (p!)^{\frac{1}{2}} N^{\frac{2-p}{2}} |Y|^{\frac{2}{p}} |Y|^{\frac{p-2}{p}}
                \right) \right]\cr
&\leq \E\,\left[|Y|^3 \exp\left(\b (p!)^{\frac{2}{p} - \frac{1}{2}} |Y|^{\frac{2}{p}}\right)\right].
}
\Eq(A.5)
$$
The point is that the term $|Y|^{2/p}$ should behave almost like 
the square of a Gaussian. More precisely, we have the following 
bound.

%
%

\lemma{\TH(a.1)}
{\it
        Let $\{X_i\}_{i = 1,\ldots, N}$ be a sequence of i.i.d.~Bernoulli
 variables, taking 
values $+1$, $-1$ with equal probability. Then $\forall C \in (0,e^{-1})$,
 there exists an 
$\varepsilon'_{C}<\infty$ (depending also on $p$) and an $\bar{N} \in \N$ such that
 for all 
$\varepsilon > \varepsilon'_C$
$$
\P\,\left[ \left|N^{-p/2} \sum_{\II \in \PP_N}
        \prod_{l \in {\cal I}} X_l \right| > \varepsilon \right]
\leq 2 \exp\left(- C^{2}
 \frac{(p!)^{\frac{2}{p}} \varepsilon^{\frac{2}{p}}}{2}\right).
\Eq(A.6)
$$
}
\specskip
\proof
The proof is surprisingly more involved than what one might at first suspect (at least, if
optimal constants are desired). We shall show that $\sum_{\II \in \PP_N} X_{\II}$
is a function of $\sum_{i = 1}^{N} X_i$ only. 
Since the distribution of this latter random variable 
is well known, all we have to do is to find an accurate upper bound for the function relating the 
two quantities. And since we are only interested in the tail behavior, we can restrict our attention
to large values of the sum (large meaning at least of the order of $\sqrt{N}$).

Suppose that $\sum_{i = 1}^N X_i = N - 2l$. Then the quantity $\sum_{\II \in \PP_N} X_{\II}$ is given
by
$$
\sum_{\II \in \PP_N} X_{\II}
= \sum_{k = 0}^p (-1)^k {l \choose k}{{N-l} \choose {p - k}} 
=\lfloor z^p \rfloor \left[(1 + z)^{N-l} (1 - z)^l\right],
\Eq(A.7)
$$
where 
$\lfloor z^p \rfloor(\cdot) \equiv \frac{1}{p!}\frac{\partial^p}{\partial z^p}\cdot \Big|_{z = 0}$ 
is the operator which extracts the coefficient of the term
$z^p$ from a formal power series.
Note that it will be important to take into account that the sum in \eqv(A.7) 
is oscillating to get a useful estimate. To do this, we consider the polynomial
on the right-hand side of \eqv(A.7) as an analytic  function 
 $\C \rightarrow \C$ and use  Cauchy's integral formula
to write
$$
\lfloor z^p \rfloor \left[(1 + z)^{N - l}(1 - z)^{l}\right] = \frac{1}{2 \pi i} \oint_{\CC}
z^{-p-1} (1 + z)^{N - l}(1 - z)^l \, dz,
\Eq(A.8)
$$
for any closed path $\CC$ surrounding the origin counterclockwise. To evaluate this
integral, we  apply the well known saddle point method (see for instance [CH]). We choose
$\CC$ to be a circle around the origin with radius
$$
r = \frac{N - 2l}{2 (N - p)}\left(1 - \sqrt{1 - \frac{4 p (N - p)}{(N - 2l)^2}} \, \right).
\Eq(A.9)
$$
Suppose that $\frac{4 p (N - p)}{(N - 2l)^2}< \k < 1$. Then  the argument of the 
square root is positive. Moreover, the following bounds for $r$ hold,
$$
\frac{p}{N - 2l} \leq r \leq \frac{p}{N - 2l}(1 + C_1(\k)),
\Eq(A.10)
$$
where $C_1$ increases from zero to some finite constant as $\k$ varies from zero to 1. 

Indeed, $\sqrt{1 - x}$ is $C^{\infty}$ for all $|x| < 1$. Therefore, for all $\k < 1$, we can
find a $C>0$ such that $\sqrt{1 - x} \geq 1 - \frac{x}{2} - C x^2$, for all $|x| < \k$. Obviously,
$C$ tends to $\frac{1}{8}$ as $\k$ tends to zero. This implies the upper bound. On the other
hand, $\sqrt{1-x} \leq 1 - \frac{x}{2}$, for all $x \geq -1$, which yields the lower bound.

The contour integral in \eqv(A.8) then becomes
$$
\eqalign{
I \equiv 
&\frac{1}{2 \pi i} \oint_{C} z^{-p-1} (1 + z)^{N - l}(1 - z)^l \, dz \cr
=& \frac{1}{2 \pi} \intl_{- \pi}^{\pi} \exp\left(- i p \vartheta \ln r 
        + (N - l) \ln (1 + r e^{i \vartheta}) + l \ln (1 - r e^{i \vartheta})\right) \,d\vartheta \equiv \frac{1}{2 \pi} \intl_{- \pi}^{\pi} e^{g(\vartheta)} \,d\vartheta.
}
\Eq(A.11)
$$
As usual, we expand the function $g$ around its maximum (which happens to lie at $\vartheta = 0$)
and try to control the error. This yields 
$$
\eqalign{
I
&= \exp\left(g(0)+ \frac{(2\pi)^3}{3!} \sup_{\zeta \in [-\pi,\pi)} g^{(3)}(\zeta)\right) 
        \intl_{- \pi}^{\pi}  e^{\frac{\vartheta^2}{2} g^{(2)}(0)} \,d\vartheta \cr
&= r^{-p}(1 + r)^{N -l} (1 - r)^{l}
        \exp\left(\frac{(2\pi)^3}{3!} \sup_{\zeta \in [-\pi,\pi)} g^{(3)}(\zeta)\right)
        \intl_{- \pi}^{\pi}  e^{\frac{\vartheta^2}{2} g^{(2)}(0)} \,d\vartheta 
}
\Eq(A.12)
$$
The main contribution comes from the term $r^{-p} (1 + r)^{N - l}(1 - r)^l$. Using \eqv(A.10),
this is bounded by 
$$
\eqalign{
r^{-p} (1 + r)^{N - l}(1 - r)^l
&= \exp\left(- p \ln r + (N - l)\ln(1 + r) + l \ln (1 - r) \right) \cr
&\leq \exp\left( - p \ln p + p \ln (N - 2l) + (N - l)r - lr\right) \cr
&\leq \exp \left( - p \ln p + p \ln (N - 2l)  + (N - 2l) r \right) \cr
&\leq \frac{(N - 2l)^p}{p!} \sqrt{p} e^{C_1(\k) p}.
}
\Eq(A.13)
$$
The integral in \eqv(A.12) is explicitly
$$
\eqalign{
\intl_{- \pi}^{\pi}  e^{\frac{\vartheta^2}{2} g^{(2)}(0)} \,d\vartheta 
&\leq \intl_{\R} \exp\left(\frac{\vartheta^2}{2} \left( \frac{lr}{(1 - r)^2} -
        \frac{(N - l) r}{(1 + r)^2} \right) \right) \, d\vartheta 
=\left(\frac{\pi}{ \frac{(N - l) r}{(1 + r)^2} - \frac{lr}{(1 - r)^2} }\right)^{1/2},
}
\Eq(A.13bis)
$$
and can be bounded by (for all $N$ large enough)
$$
\left((N - l) \frac{r}{(1 + r)^2} - l \frac{r}{(1 - r)^2} \right)^{-\frac{1}{2}}
\leq  p^{-\frac{1}{2}} \left( 1 - \frac{p^2}{(N - 2l)^2} \right)
\left(1 - \frac{2\k}{3}\right).   
\Eq(A.14)
$$
Finally, we estimate the error due to the remainder in the Taylor expansion in \eqv(A.12). 
One shows by a straightforward computation that for all $\k, \d > 0$ there exists an 
$\bar{N}_{\k,\d} \in \N$ such that
$$
|g^{(3)}(\vartheta)| \leq  p (1 + C_1(\k)) \left(1 + \k(1 + C_1(\k)) + \d \right)
= p C_3(\k,\d),
\Eq(A.15)
$$
where $C_3 = 1$ for $\k = \d = 0$.
Hence, the error committed can be bounded as (if $N > \bar{N}_{\k,\d}$)
$$
\eqalign{
\exp\left(\frac{(2\pi)^3}{3!} \sup_{\zeta \in [-\pi,\pi)}g^{(3)}(\zeta)\right) & 
\leq \exp\left( \frac{2\pi}{3!} p ( 1 + C_1(\k)) 
        \left(1 + \k(1 + C_1(\k)) + \frac{C_2}{N - 2l}\right)
        \right). 
}
\Eq(A.16)
$$
This follows from the exact expression for $g^{(3)}$,
$$
g^{(3)}(\vartheta) = i r e^{i \vartheta} \left( (N - l) \frac{r e^{ i \vartheta} - 1}
        {(1 + r e^{i \vartheta})^3}
- l \frac{1 + r e^{i \vartheta}}
        {(r e^{i \vartheta} - 1)^3} \right),
\Eq(A.17)
$$
which one gets through straightforward derivation.

Inserting the bounds \eqv(A.13), \eqv(A.14), and \eqv(A.15) into the 
estimate \eqv(A.12) then gives 
$$
I \leq  \frac{(N - 2l)^p}{p!} e^{(C_1(\k) + C_3(\k,\d)) p},
\Eq(A.17bis)
$$ 
and thus 
$$
f\left(\sum_{i \in {\cal N}} X_i\right) \leq  \frac{1}{p!} e^{p(C_1(\k) + C_3(\k,\d)) }
\left(\sum_{i \in {\cal I}} X_i\right)^p, \quad N \geq \bar{N}_{\k,\d}
\Eq(A.18)
$$
Let $\rho(\k,\d) =  e^{(C_1(\k) + C_3(\k,\d))p}$, for $\k \in (0,1)$ and $\d>0$. Then 
$\rho$ is increasing in $\k$ and bounded below by $e^p$.
Thus, for all
$C \in (0,e^{-p})$, we can find $\tilde{\k} \in (0, 1)$ and $\tilde{\d} > 0$
such that $C \leq \rho(\tilde{\k},\tilde{\d})^{-1}$. Let now
$$
\varepsilon_{\k,\d} \equiv \left(\frac{4p}{\k}\right)^{p/2} \frac{\rho(\k,\d)}{p!}.
\Eq(A.29)
$$
Suppose that $\varepsilon > \varepsilon_{\tilde{\k},\tilde{\d}}$ and 
$N \geq \bar{N}_{\tilde{\k},\tilde{\d}}$. Then, we have that
$$
\P\left[N^{-1/2} \sum_{i \in {\cal N}} X_i > \left(\varepsilon p!\, 
                \rho(\tilde{\k},\tilde{\d})^{-1}\right)^{1/p}\right]
        \leq  \exp\left(-\frac{1}{2}\left(\varepsilon p!\, \rho(\tilde{\k},\tilde{\d})^{-1}\right)^{2/p}\right),
\Eq(A.20)
$$
by the standard bound on sums of Bernoulli variables. On the other hand, since 
$$
N^{-1/2} \sum_{i \in {\cal N}} X_i 
> \left(\varepsilon p!\,\rho(\tilde{\k},\tilde{\d})^{-1}\right)^{1/p}  
> \left( \varepsilon_{\tilde{\k},\tilde{\d}} p!\, \rho(\tilde{\k},\tilde{\d})^{-1} \right)^{1/p} 
= \left(\frac{4p}{\tilde{\k}}\right)^{1/2}
\Eq(A.21)
$$
implies that
$$
\frac{4pN}{(N-2l)^2} < \tilde{\k} < 1,
\Eq(A.21bis)
$$
the condition following \eqv(A.9) is satisfied and hence the above bound on 
$f(\sum_{i \in {\cal N}} X_i)$ is valid. Thus
$$
\eqalign{
\P\bigg[N^{-1/2} \sum_{i \in {\cal N}} X_i > \left(\varepsilon p!\, 
                \rho(\tilde{\k},\tilde{\d})^{-1}\right)^{1/p}\bigg] 
&= \P\left[  N^{-p/2}\frac{\rho(\tilde{\k},\tilde{\d})}{p!} \left(\sum_{i \in \NN} X_i \right)^p
        > \varepsilon \right] \cr
&\geq \P\left[ N^{-p/2} f\left(\sum_{i \in \NN} X_i\right) > \varepsilon\right].
}
\Eq(A.22)
$$
Hence, by \eqv(A.20) and \eqv(A.22),
$$
\P\left[ N^{-p/2} f(\sum_{i \in \II} X_i) > \varepsilon\right]
\leq \exp\left(-\frac{1}{2}\left(\varepsilon p!\, 
                \rho(\tilde{\k},\tilde{\d})^{-1}\right)^{2/p}\right)
\leq \exp\left( - \frac{C^{2/p}}{2} (\varepsilon p!\,)^{2/p}\right).
\Eq(A.23)
$$
Thus, we have shown that for all $C \in (0, e^{-p})$, there exists 
$\tilde{\varepsilon}_{C} = \varepsilon_{\tilde{\k},\tilde{\d}}$ such that \eqv(A.23) holds for
all $\varepsilon > \tilde{\varepsilon}_{C}$ and all $N$ large enough. Together with the 
analogue bound for the negative tails, this proves the lemma. 
\endproof
\specskip
\noindent
To finish the proof of the theorem, let us go back to \eqv(A.5). To get the claimed
bound, it is enough to show that the integral on the right-hand side is bounded uniformly in
$N$. Indeed, since the variable $Y$ satisfies the bound \eqv(A.6) of the lemma, we get for 
any $C' < e^{-p}$ 
$$
\eqalign{
\E\,\big[|Y|^3 \exp\big(\b (p!)^{\frac{2}{p} - \frac{1}{2}} 
|Y|^{\frac{2}{p}} \big)\big] 
&\leq \sum_{l \geq 1} \E\,\big[|Y|^3 \1_{\{|Y| \in [l,l+1)\}}\exp\big(\b (p!)^{\frac{2}{p} - 
        \frac{1}{2}} |Y|^{\frac{2}{p}}\big)\big] \cr
&\leq (l+1)^3 \P[|Y| \geq l] \exp\big(\b (p!)^{\frac{2}{p} - \frac{1}{2}}(l+1)^{\frac{2}{p}}\big) \cr
&\leq \intl_{0}^{\infty} (x + 1)^3
        \exp\big(\b (p!)^{\frac{2}{p} - \frac{1}{2}} (x + 1)^\frac{2}{p}
                - C'{}^{2/p}  (p!)^{\frac{2}{p}}x^{\frac{2}{p}} \big) dx \cr
& \quad + (\tilde{\varepsilon}_{p,C'} + 1)^3. 
}
\Eq(A.24)
$$
By the preceding lemma, for any $\b \leq e^{-2} (p!)^{\frac{1}{2}}$, we can find $C' < e^{-p}$ and 
a corresponding $\varepsilon'_{C'}$  such that the above integral is finite. 
Setting $C^p=C'$, this   proves the theorem.
\endproof
\specskip
\noindent
We observe that we could have equally well replaced $H_N$ by in $-H_N$ in the proof of 
Theorem~\thv(pr.1), without changing the result (since only the square of the Hamiltonian 
does enter). We therefore have readily the following result, which we state for further use.

\corollary{\TH(a.2)}
{\it
        If $|\b| <\b'_p$, then
$$
        \E\,\E_{\s} e^{\b H_N} = e^{\frac{\a \b^2 N}{2}(1 + {\cal O}(N^{-1}))}.
\Eq(A.25)
$$
}
\proof
Completely analogous to the proof of Theorem~\thv(pr.1). \endproof
\specskip
We also put a result here, that will be used in the next chapter, but whose proof is
very similar to the above.

\lemma{\TH(c.1)}
{\it
  If $|\b| < \frac{1}{2}\b'_p$, then there exists a constant $C > 0$ such that
$$
\E\,\left[e^{- \b H_N(\s) - \b H_N(\s')}\right] 
  \leq e^{\a N \b^2(1 + R(\s,\s')^p + C)},
\Eq(C.1)
$$
for all $N$ large enough.
}
\specskip
\proof
The proof is actually almost identical to the proof of Theorem~\thv(pr.1). We start by 
expanding the exponential up to order two, with the same error as in the proof of 
Theorem~\thv(pr.1) (inequality \eqv(A.2)). This error term is then treated similarly, 
by first decoupling the terms in $\s$ and $\s'$ with Cauchy-Schwarz. This already shows 
why $\b$ has to be less than half the bound of Theorem~\thv(pr.1). The linear term in 
the expansion vanishes, whereas the quadratic term gives us the covariance term $R(\s,\s')^p$. 
Indeed, if we set 
$Y^{\mu}(\s) = N^{-p/2}\sum_{{\cal I} \subset \NN} \xi_{\II}^{\mu} \s_{\II}^{\vphantom{\mu}}$, 
we get
$$
\eqalign{
\ln \E\,&\Big[ \exp(
-\b H_N(\s) - \b H_N(\s'))\big]  \cr
\leq &\sum_{\mu = 1}^{M(N)} \ln \bigg(1 + \frac{\b^2 p!}{2}N^{2 - p} 
        \E\,\big[\big( Y^{\mu}(\s) + Y^{\mu}(\s')\big)^2\big] \cr&
+ \frac{\b^3(p!)^{\frac{3}{2}}}{3} N^{3 - \frac{3p}{2}} 
        \E\,\Big[|Y^{\mu}(\s) + Y^{\mu}(\s')|^3 
\exp\left(\b (p!)^{\frac{1}{2}} N^{1 
                - \frac{p}{2}}|Y^{\mu}(\s) + Y^{\mu}(\s')|\right)\Big]\bigg).
}
\Eq(C.2)
$$
We now apply the triangle inequality and Cauchy-Schwarz to the error term, which yields
$$
\eqalign{
&N^{3 - \frac{3p}{2}}\E\,\Big[|Y^{\mu}(\s) + Y^{\mu}(\s')|^3 
e^{\b(p!)^{1/2} N^{1 - p/2}|Y^{\mu}(\s) + Y^{\mu}(\s')|}\Big] \cr
&\leq N^{3 - \frac{3p}{2}} \sum_{i=1}^{3} 
        \left(\E\,\left[|Y^{\mu}(\s)|^{2j} 
                \exp\left(2 \b (p!)^{\frac{1}{2}} 
N^{1 - \frac{p}{2}}|Y^{\mu}(\s)|\right)
                 \right] \right)^{\frac{1}{2}}
\cr&\quad\times   \left(\E\,\left[|Y^{\mu}(\s')|^{6 - 2j} 
                \exp\left(2 \b (p!)^{\frac{1}{2}} N^{1 - \frac{p}{2}} |Y^{\mu}(\s')|\right)
                                        \right] \right)^{\frac{1}{2}} \cr
&\leq C_1 N^{3 - \frac{3p}{2}},
}
\Eq(C.3)
$$
if $\b < \frac{1}{2}\b'_p$ and $N$ large enough, by the result in the proof of Theorem~\thv(pr.1) 
(cf.~the remark after \eqv(A.2)). 

The quadratic term in \eqv(C.2) is evaluated easily. One obtains (observing that the covariance
of $H_N$ appears)
$$
\eqalign{
\E\,\Big[ (- Y^{\mu}(\s) - Y^{\mu}(\s'))^2\big] 
&= 2 \E\,[Y^{\mu}(\s)^2] + 2 \E\,[Y^{\mu}(\s)Y^{\mu}(\s')] \cr
&= 2 N^{-p} {N \choose p} + 2 N^{-p} \sum_{{\cal I} \subset \NN}  \s^{\vphantom{'}}_{\II} \s'_{\II} \cr
&=\frac{2}{p!} (1 + R(\s,\s')^p) + {\cal O}(N^{-1}).
}
\Eq(C.4)
$$
Hence,
$$
\eqalign{
\ln \E\,e^{- \b( H_N(\s) + H_N(\s'))}        
&\leq \sum_{\mu = 1}^{M(N)} \ln \Big(1 + \frac{\b^2}{N^{p - 2}}(1 + R(\s,\s')^p) 
                + \frac{C_2}{N^{p-1}} + \frac{C_1}{N^{\frac{3p}{2} - 3}}\Big) \cr
&\leq M(N) (\b^2 N^{2 - p}(1 + R(\s,\s')^p) + C_3N^{1-p}),
}
\Eq(C.5)
$$
that is,
$$
\E\,e^{- \b H_N(\s) - \b H_N(\s')}
\leq e^{\a \b^2 N (1 + R(\s,\s')^p) + C_4 }.
\Eq(C.6)
$$
This proves the lemma.
\endproof

Finally, we have as an application of Corollary~\thv(a.2).

\lemma{\TH(c.9)} 
{\it
        The Hamiltonian satisfies
$$
\P\left[\sup_{\s} |H_N(\s)| > t N\right]
\leq C \cases   \rlap{$\exp\left(- N(\frac{t^2}{2\a} - \ln 2)\right),$}
                \hphantom{\exp \left( - N (\frac{(p!)^{\frac{1}{2}}}{2} t 
                                - \frac{\a p!}{8} - \ln 2)\right),}
                        \quad \hbox{\tenrm if\ } 
t \leq \frac{\a (p!)^{\frac{1}{2}}}{e^2}, \cr
                \exp \left( - N (\frac{(p!)^{\frac{1}{2}}}{e^2} t - 
\frac{\a p!}{2e^4} - \ln 2)\right),
                \quad\hbox{\tenrm otherwise.} \cr 
        \endcases
\Eq(C.66)
$$
}
\proof
We start with a crude bound to extract the supremum. Standard arguments and Chebyshev's inequality
in its exponential form yield
$$
\P[\sup_{\s} |H_N(\s)| > t N] 
\leq  2^{N} \inf_{q > 0} e^{-qtN} \E\,e^{q H_N(\s)}+ 2^N \inf_{q > 0} e^{-qtN} \E\,e^{- q H_N(\s)}.
\Eq(C.67)
$$
We now use Theorem~\thv(pr.1), respectively Corollary~\thv(a.2) to bound the two integrals and 
obtain
$$
\eqalign{
\P\left[\sup_{\s} |H_N(\s)| > tN\right] 
&\leq C_1 2^{N+1} \inf_{q \in (0, \b'_p)} e^{-qtN} e^{\frac{\a q^2 N}{2}} \cr
&= C_2 \cases   \rlap{$\exp\left(- N (\frac{t^2}{2\a} - \ln 2)\right),$}
                \hphantom{\exp \left( - N (\frac{(p!)^{\frac{1}{2}}}{2} t 
                                - \frac{\a p!}{8} - \ln 2)\right),} 
                        \quad \hbox{\tenrm if\ }t \leq 
\frac{\a (p!)^{\frac{1}{2}}}{e^2}, \cr
        \exp \left( - N (\frac{(p!)^{\frac{1}{2}}}{2e^2} t - 
\frac{\a p!}{2e^4} - \ln 2)\right),
                \quad\hbox{\tenrm otherwise.} \cr 
        \endcases
}
\Eq(C.68)
$$
This proves the lemma.
\endproof

\bigskip

\chap{4.~Critical $\boldsymbol\beta$ and Convergence to the REM}4

\bigskip

\subchap{4.1.~Estimates on the Truncated Partition Function.}


\noindent
To get the lower bound for the critical temperature, we would like to compare $\E\,Z_{N,\b}^2$
and $(\E\,Z_{N,\b})^2$. However, as mentioned in the introduction and explained in Chapter~2
it is essential to do this comparison for a truncated partition function. Define therefore
$$
\widetilde{Z}_{N,\b}(c) \equiv \E_{\s} \left[e^{-\b H_N[\o](\s)} 
\1_{\{- H_N(\s) \leq c \a \b N\}}\right],
\Eq(C.7)
$$
for $c>1$.
The key observation is that the truncation has no influence on the expectation of the 
partition function if $c$ is chosen appropriately. This is the content of the following lemma.

\lemma{\TH(c.2)}
{\it
        For all $\b > 0$, $c > 1$ such that $\b c < \b'_p$ there exist $K,K' > 0$ such that
$$
\E\,\widetilde{Z}_{N,\b}(c) 
 \left(1 - K e^{- K' (c-1)^2 N}\right) \E\,Z_{N,\b} .
\Eq(C.8)
$$
}
\proof
Let us set $q = q(N) \equiv \a \b^2 N$. Note that 
$
\E\,Z_{N,\b} - \E\,\widetilde{Z}_{N,\b} = \E\, \left[e^{-\b H_N(\s)}
        \1_{\{- \b H_N(\s) > c q\}}\right]
$ and thus by the exponential Chebyshev inequality
$$
\E\,Z_{N,\b} - \E\,\widetilde{Z}_{N,\b} \leq \E_{\s} \inf_{t > 0} e^{-t c q}
        \E\,\left[e^{- \b (1 + t) H_N(\s)}\right].
\Eq(C.10)
$$
We now use Theorem~\thv(pr.1) with $\b$ replaced by $(1+t)\b$ to estimate the
expectation to get
$$
\inf_{t > 0} e^{-t c q} \E\,\left[e^{-\b (1+t) H_N(\s)}\right]
\leq \inf_{0<t \leq \b/\b'_p-1} e^{- t c q + \frac{(1+t)^2 q}{2} + q C N^{-1}}.
\Eq(C.11)
$$
The exponent is minimized for $t = c - 1$. By assumption, $\b c < \b'_p$, so
that this value falls into the interval over which the inf on the right
is taken. Thus, 
$$
\inf_{t > 0} e^{- t c q} \E\,\left[e^{-\b (1+t) H_N(\s)}\right]
\leq e^{- \frac{q}{2} (c-1)^2 + C q N^{-1}} e^{\frac{q}{2}} 
\leq e^{- \frac{q}{2} (c-1)^2 + C q N^{-1}} \E\,Z_{N,\b},
\Eq(C.12)
$$
This  implies the statement of the lemma.
\endproof
\specskip
\noindent
We now turn to the square of the truncated partition function. We bound 
$$
\E\,\widetilde Z_{N,\b}^2\E\,e^{-\b H(\s) - \b H(\s')}\1_{\{- H_N(\s) \leq c \a \b N\}}\1_{\{- H_N(\s')
 \leq c \a \b N\}}
\EQ(C.12bis)
$$  
by two different functions.
When calculating the expectation with respect to $\s$ and $\s'$, we use one bound for small
values of the replica overlap $R(\s,\s')$, and the other for the rest. Define therefore
$$
S(b) \equiv \E_{\s,\s'} \left[e^{-\b( H_N(\s) + H_N(\s'))} \1_{\{|R(\s,\s')| < b\}}\right]
\Eq(C.13)
$$
and
$$
T(c,b,b') \equiv \E_{\s,\s'} \left[e^{-\b(H_N(\s) + H_N(\s'))}\1_{\{|R(\s,\s')| \in [b,b']\}}
        \1_{\{- \b(H_N(\s) + H_N(\s')) < 2 c\a \b^2 N\}}\right].
\Eq(C.14)
$$
Then 
$$
\widetilde{Z}_{N,\b}(c)^2 \leq S(b) + T(c,b,1),
\Eq(C.15)
$$
for all $b \in (0,1)$. We now control each of the terms separately. We start with $S(b)$.

\lemma{\TH(c.3)}
{\it 
Suppose $\b < \frac{\b'_p}{2}$, and $b$ is such that 
$$
\g \equiv \a \b^2 b^{p-2} < \frac{1}{2}.
\Eq(C.16))
$$
Then for all $\varepsilon \in (0, \frac{1}{2} - \g)$ there exists $N_{\varepsilon} \in \N$ 
such that for all $N > N_{\varepsilon}$,
$$
\E\, S(b) \leq \frac{1}{\sqrt{1 - 2 (\g + \varepsilon)}} e^{\a \b^2 N}.
\Eq(C.17)
$$
}
\proof 
If $\b$ satisfies the above condition, we can apply Lemma~\thv(c.1) to the integrand of 
the right-hand side of \eqv(C.13). One obtains
$$
\E\,\bigg[e^{-\b (H_N(\s) + H_N(\s'))} \1_{\{|R(\s,\s')| < b\}} \bigg]
\leq  \1_{\{|R(\s,\s')| < b\}} e^{\a \b^2  N(1+(R(\s,\s')^p) + C  N^{-1})}.
\Eq(C.18)
$$
Thus,
$$
\eqalign{
\E\,S(b) &\leq \E_{\s,\s'} \bigg[e^{\a \b^2  N(1+R(\s,\s')^p + C N^{-1})} 
                \1_{\{|R(\s,\s')| < b\}} \bigg] \cr
&\leq \E_{\s,\s'} \bigg[e^{\a \b^2 N(1+R(\s,\s'){}^2 b^{p-2} + C N^{-1})} 
                \1_{\{|R(\s,\s')| < b\}} \bigg] \cr
&= e^{\a \b^2 N} \E_{\s,\s'} \bigg[e^{\a \b^2 N(R(\s,\s'){}^2 b^{p-2} + C N^{-1})}
                \bigg]\cr
&\leq\sum_{k=N/2-[bN]}^{N/2+[bN]} 2^{-N}{N\choose k} e^{(\g+\e)N k^2}.
}
\Eq(C.19)
$$
by assumption \eqv(C.16), for any $\e>0$, if $N$ is large enough.
Standard estimates then yield \eqv(C.17).\endproof
\specskip
\noindent
The next result concerns the term $T(c,b,1)$ in \eqv(C.15).

\lemma{\TH(c.4)}
{\it
       Let $I(t)$ be the Cram\`er Entropy as defined in \eqv(PR.15).
Suppose that there exist $c > 1$, $d > 0$, such that
$$
\forall t \in [b,b'],\quad
        2\a \b^2 c \Big(1 - \frac{c}{2(1+t^p)}\Big)
        \leq \a \b^2 + I(t) - d.
\Eq(C.24)
$$
Then, if 
$$
c < \min\big(\frac{1}{2 \b }\b'_p, 1 + b^p\big),
\Eq(C.24bis)
$$ 
there exists $\bar{N} \in \N$ such that
for all $N \geq \bar{N}$,
$$
\E\,T(c,b,1) \leq e^{\a \b^2N}e^{-\frac{Nd}{2}}.
\Eq(C.25)
$$ 
}
\proof
By definition,
$$
\textstyle
\E\,T(c,b,b') = \E_{\s,\s'} \E\,\bigg[e^{-\b (H(\s) + H(\s'))}
        \1_{\{|R(\s,\s')| \in [b,b']\}}
        \1_{\{- \b (H_N(\s) + H_N(\s')) \leq 2 c \a \b^2 N\}} \bigg].
\Eq(C.26)
$$
In a first step, we bound the expectation over the disorder for fixed $\s,\s'$.
Similar to the  proof of Lemma \thv(c.2) we get 
(again $q \equiv \a \b^2 N$), yields
$$
\eqalign{
\E\,\bigg[\1_{\{- \b (H_N(\s) + H_N(\s')) \leq 2 c q\}} e^{-\b (H_N(\s) + H_N(\s'))}\bigg] 
& \cr
&\kern-1cm\leq \inf_{t>0} e^{2 t c q} \E\,\bigg[e^{-\b (1-t) (H_N(\s) + H_N(\s'))} \bigg]
}
\Eq(C.27)
$$
We now use  Lemma~\thv(c.1), with
$\b$ replaced by $\b (1-t)$ to obtain
$$
\inf_{t>0} e^{2 t c q} \E\,\bigg[ e^{-\b (1-t) (H_N(\s)+H_N(\s'))} \bigg]
\leq \inf_{t>1-\b'_p/(2\b)} e^{2 t c q} e^{(1-t)^2 q (1+ R(\s,\s')^p)} e^{C_2 N^{-1} q}.
\Eq(C.28)
$$
The infimum is attained for $t=1-\frac {c}{(1 + R^p)}>1-\frac {\b'_p}{2\b}$
(by assumption \eqv(C.24bis).  Thus
$$
\eqalign{
\E\,\bigg[e^{-\b H_N(\s) -\b H_N(\s')}
&\1_{\{-\b H_N(\s)-\b H_N(\s') \leq 2 c \a \b N\}} \bigg]  \cr
&\leq C_3 \exp \bigg(2 c \a \b^2 N \Big(1 - \frac{c}{2(1+ R(\s,\s')^p)}\Big)\bigg).
}
\Eq(C.29)
$$
Finally, we integrate over all configurations $\s$, $\s'$ satisfying $|R(\s,\s')| \in [b,b']$.
We observe that $R(\s,\s')$ has the same distribution as $S(\s) =
N^{-1} \sum_{i=1}^N \s_i$.  
Hence, 
$$
\eqalign{
\E\,\Big[T(c,b,b')\Big] 
&\leq C_3 \E_{\s,\s'}\bigg[
        \exp \bigg(2 c \a \b^2 \Big(1 - \frac{c}{2(1+
        R(\s,\s')^p)}\Big) 
                \bigg)
        \1_{\{|R(\s,\s')| \in [b,b']\}}\bigg] \cr
&= C_3 \E_{\s} \bigg[\exp \bigg(2 c \a \b^2 N 
        \Big(1 - \frac{c}{2(1+ S(\s)^p)}\Big)\bigg)
        \1_{\{|S(\s)| \in [b,b']\}}\bigg] \cr
&\leq 2 C_3 N \exp \bigg( N \sup_{t \in [b,b']} \bigg[ 2 \a \b^2 c
                \Big(1 - \frac{c}{2(1+ t^p)}\Big) - I(t)  
                \bigg] \bigg) \cr
&\leq 2C_3 e^{N (\a \b^2  + \frac{\ln N}{N} - d)} \leq  e^{N (\a \b^2
        - \frac{d}{2})}. 
}
\Eq(C.30)
$$
The second to last inequality follows from the hypothesis of the
lemma, and the observation 
that we sum over at most $2N$ values of $S(\s)$. The last inequality  holds 
for all $N$ larger 
than a certain $\bar{N} \in \N$. Since this estimate is uniform in
$b'$, we may choose $b'=1$. 
\endproof
\specskip
\noindent
From the preceding results, we now get a variance estimate for the
truncated partition function. 

\proposition{\TH(c.5)}
{\it
  Suppose that $\b < \check{\b}_p$. Then there exist constants $C > 0$
  and $c>1$ such that 
$$
\E\,[\wt {Z}_{N,\b}(c)^2] \leq C (\E\,\wt {Z}_{N,\b}(c))^2.
\Eq(C.31)
$$
and,
$$
\P[\wt {Z}_{N,\b}(c) > \frac{1}{2} \E\,\wt {Z}_{N,\b}(c)] \geq
\frac{3}{4 C}. 
\Eq(C.32)
$$
}
\proof
We first prove that the hypothesis implies that the assumptions of 
Lemmas~\thv(c.2)--\thv(c.4) are satisfied.
Consider therefore $\b < \frac{1}{2} \b'_p$ such that 
$$
\b^2 < \inf_{0 \leq t \leq 1} I(t) \frac{1 + t^p}{\a t^p}.
\Eq(C.33)
$$
Then it is immediate that 
$$
2 \a \b^2 \bigg(1 - \frac{1}{2(1+t^p)}\bigg)
=2\a\b^2\frac {1+2t^p}{2(1+t^p)} < \a \b^2 + I(t),
\Eq(C.34)
$$
for all $t \in [0,1]$. By continuity, there exist $c^* >1$ and $d^* > 0$ such that 
$\forall c \in (1,c^*)$ and $d \in (0,d^*)$
$$
2 c \a \b^2 \bigg(1 - \frac{c}{2(1 + t^p)} \bigg) < \a \b^2 + I(t) - d,\quad \forall t \in [0,1].
 \Eq(C.35)
$$
This implies the hypothesis of Lemma~\thv(c.4).

We now show that $(\E\,[\wt {Z}_N])^2$ is of the order of $\E\,[\wt {Z}_N{}^2]$. We start by
fixing the free parameters $b$, $b'$, and $c$. Choose first $b$ such that $\g(b) = \frac{1}{4}$
(or any other constant less than one half). Then choose $c$ such that
$$
c < \min\left(c^*, \frac{\b'_p}{2 \b}, 1 + b^p \right).
\Eq(C.36)
$$
Then the hypotheses of all preceding lemmas are fulfilled. Finally, choose $b' = 1$.
By Lemmas~\thv(c.3) and \thv(c.4), we then have
$$
\eqalign{
\E\,\big[\wt {Z}_N{}^2\big] \leq \E\,[S(b) + T(c,b,1)] \leq (C_1 + e^{-N d/2}) 
        e^{\a \b^2  N}.
}
\Eq(C.37)
$$
The right-hand side is by Theorem~\thv(pr.1) bounded by 
$$
(C_1 + e^{-Nd/2}) e^{\a \b^2 N} \leq 2 C_2 \Big(\E\,[Z_N]\Big)^2,
\Eq(C.38)
$$
which in turn is of the order of $(\E\,[\wt {Z}_N])^2$ by Lemma~\thv(c.2), so that
$$
(C_1 + e^{-Nd/2}) e^{\a \b^2 N} \leq C_3 \Big(\E\,[\wt {Z}_N]\Big)^2.
\Eq(C.39)
$$
This implies \eqv(C.31).
The second assertion of the proposition follows from  the Paley-Zygmund inequality, 
which states that for a positive random variable $Y$ and any positive constant $g$,
$$
\P\,\Big[Y \geq g \E\,Y \Big] \geq (1-g)^2 \frac{(\E\,Y)^2}{\E\,[Y^2]}.
\Eq(C.41)
$$
This relation gives us a lower bound on the probability that
$\wt {Z}_N \geq g \E\,[\wt {Z}_N]$, which is strictly greater than zero and uniform in $N$. 
Indeed, if we set $g = \frac{1}{2}$ in \eqv(C.41), then, by \eqv(C.31), we get
$$
\P\big[ \wt {Z}_N \geq \frac{1}{2} \E\,\wt {Z}_N\big] \geq \frac{1}{2 C_3}.
\Eq(C.42)
$$
This concludes the proof of the proposition. 
\endproof

\bigskip

\subchap{4.2.~Proof of the Lower Bound.}

\noindent We will now proof the lower bound assuming that Theorem 
\thv(pr.4) holds. This is by now quite standard [T1,T2,T3], but we repeat the
argument for the reader's convenience. Note that by Lemma \thv(c.2) for $N$ 
large enough, for any $\d>0$,
$$
\P\big[\wt {Z}_N \geq \frac{1}{2} \E\,\wt {Z}_N\big] 
        \leq \P\,\big[Z_N \geq \frac{1}{2} (1-\d) \E\,Z_N\big].
\Eq(C.43)
$$
But 
$$
 \P\,\left[Z_N \geq \frac{1}{2} (1-\d) \E\,Z_N\big]
= \P\,\big[F_N-\E F_N \geq   N^{-1} (\ln \E Z_N-\E \ln Z_N    - \ln\left(  \frac{1}{2} (1-\d)\right) \right].
\Eq(C.43.1)
$$
But Theorem \thv(pr.4) implies that this quantity is smaller than
$B^{-n}$, if 
$$
 N^{-1} (\ln \E Z_N-\E \ln Z_N)\geq N^{-1/2+\e}
\Eq(C.44)
$$ 
in contradiction to the lower bound \eqv(C.42). This proves that for 
$\b<\check\b$, 
$$
\lim_{N\uparrow\infty}  N^{-1} (\ln \E Z_N-\E \ln Z_N) =0
\Eq(C.45)
$$
proving the lower bound on $\b_p$. 
\endproof
\specskip 
\remark 
It should be noted that the above argument requires only an upper deviation 
inequality for the free energy. Such an inequality 
 can be obtained in a much
 simpler way than Theorem \thv(pr.4) (in that it does not require the results
of Section 5) on the basis of a result of Ledoux [Le]. The reason is  that 
while the free energy is not a convex function of all the disorder variables,
it is separately convex in each $\xi_i^\mu$. This suffices to apply 
Ledoux's theorem. A proof of the corresponding one-sided inequality 
can be found in [Ni].

%
%
\bigskip
\subchap{\ver.3.~Upper Bound on the Critical $\b$.}

\noindent The proof of the upper bound in Theorem~\thv(pr.2) is considerably
 simpler 
than the lower bound. By \eqv(REM.1),  $\E\,\frac{\partial F_N}{\partial \b} 
\leq  N^{-1}\E\, \sup_{\s} | H_N(\s) |$,
while 
 Lemma \eqv(c.9) yields immediately (see
the argument leading to \eqv(REM.5)) that:

\lemma{\TH(ub.1)}
{\it There exists  $C<\infty$, such that:
   If $\a \geq \frac{8 \ln 2}{p!}$, then
$$
\E\,\sup_{\s}|H_N(\s)| \leq N B_\a +C
\Eq(UB.3)
$$
where 
$$
B_\a=\cases  \sqrt{ 2 \a \ln 2},&\,\hbox{if}\, \a \geq \frac{e^4 2\ln 2}{p!}
\cr\frac {\a\sqrt{ p!}}{2e^2}+
\frac {e^2\ln 2}{\sqrt{p!}},&\,\hbox{if}\, 0\leq \a \leq \frac{e^4 2\ln 2}{p!}
\endcases
\Eq(UB.3bis)
$$
}
 \specskip

%

 Let 
$
\b_{\infty}\equiv  B_\a/\a
$ and assume that $\a\geq \a_p$. 
Now assume that $\b_p>\b_\infty$. Then for $\b_\infty<\b<\b_p$, 
 we have that
$$
\eqalign{
\limsup_{N\uparrow\infty}\E\,F_N(\b) &\leq\limsup_{N\uparrow\infty} \E\,F_N(\b_{\infty}) +
 + (\b - \b_{\infty}) B_\a\cr
=\frac {\a\b^2_\infty}2-(\b-\b_\infty) \a\b_\infty
=\frac {\a\b^2}2-\frac {(\b-\b_\infty)^2}2<\frac {\a\b^2}2
}
\Eq(UB.9)
$$
in contradiction to the assumption that $\b<\b_p$. Thus $\b_p\leq \b_\infty$
which proves the upper bound \eqv(PR.17). \endproof

\bigskip

\subchap{4.4.~Convergence to the REM: Proof of Theorem~\thv(pr.3).}

\noindent The convergence of the free energy as $p\uparrow\infty$ follows 
now from a simple convexity argument. Note that for all $\b<\check\b_p$,
$\lim_{N\uparrow\infty} \E F_{N,\b}=f^{REM}_b$, while for all $\b>\hat
\b_p$, by convexity of $F_{N,\b}$,
$$
\eqalign{
\liminf_{N \uparrow \infty} \E\,F_{N,\b}
&\geq  \liminf_{N \uparrow \infty} \E\,F_{N,\hat\b} 
        + \a \hat\b_p (\b - \hat\b_p) \cr
&= \frac{\a {\hat\b}^2}{2} + \a \hat{\b} (\b - \hat{\b}) \cr
}
\Eq(UB.9ter)
$$
while on the other hand
$$
\eqalign{
\limsup_{N \uparrow \infty} \E\,F_{N,\b}
&\leq  \liminf_{N \uparrow \infty} \E\,F_{N,\hat\b} 
        + \a \check\b_p (\b - \hat\b_p) \cr
= \frac{\a {\hat\b}^2}{2}  + \a \check\b_p (\b - \hat\b_p) \cr
}
\Eq(UB.9quater)
$$
provided $p$ is large enough such that $\a>\a_p$. 
But since $\lim_{p\uparrow\infty}\check\b_p= \lim_{p\uparrow\infty}
\hat\b_p$, the two bounds above both converge to $f^{REM}_\b$, as $p\uparrow
\infty$, for any $a>0$. This proves Theorem \eqv(pr.2).\endproof

\bigskip
... include definition file
%
%
%
%
%
%

\chap{5.\ Fluctuations: Proof of Theorem~\thv(pr.4)}5
\bigskip
\noindent
The main line of reasoning of the proof of the fluctuation theorem is 
as follows. First, for each $N$ we define a set whose complement has 
a very small probability (of the order of $N^{-n}$). On this set, we 
prove the estimates on the deviation with the so-called Yurinskii 
martingale method [Yu]. On the complement, we simply use that the 
free energy is bounded by a polynomial function.
This approach was first used  in the context of the mean field  model in
[PS,ST] for variance estimates and in   [BGP2,B1] for exponential
inequalities, but has later been made obsolete by new concentration of   
measure inequalities provided by Talagrand in [T1]. Unfortunately,
these require convexity of the level sets of the random functions
considered which in the current situation do not appear to
hold. Although, as remarked at the end of Section 4, the 
hypotheses of Ledoux's inequalities from [Le] do hold, these
provide only one-sided deviation estimates which will not be 
sufficient for our later purposes. In this situation the return 
to Yurinskii's method appears to be the only way out. 

Define the decreasing sequence of $\s$-algebras $\FF_k =
\s(\{\xi^{\mu}_{i}\}^{\mu \in \N}_{i \geq k}$. 
Furthermore, for $c, \g > 0$ and $k \in \NN$, let
$$
\AA_k = \AA_{k,c,\g, N} 
\equiv \bigg\{\o \in \O: \left|\GG_{N,\b} \left[H^\mu_k(\s)\right]\right| < c N^{-1 + \g}\bigg\}
\Eq(F.3)
$$
where 
$$
H^\mu_k(\s) \equiv - \frac{(p!)^{\frac{1}{2}}}{N^{p-1}} 
	\sum_{\mu = 1}^{M(N)} \sum_{{\II \ni k} \atop {|\II| = p}}
        \xi^{\mu}_{\II} \s^{\vphantom{\mu}}_{\II}.
\Eq(F.101)
$$
We put 
and $\AA \equiv \AA_{c, \g , N} \equiv \bigcap_{k = 1}^N \AA_k$.
The set $\AA$ will be our `good' set. We first show that its 
measure is large.

\lemma{\TH(f.1)}
{\it
  For all $\g, c, m > 0$, there exists $C > 0$, such that  
$$
\P[\AA_{c,\g,N}] \geq 1 - C N^{-m}.
\Eq(F.4)
$$
}

\proof
Since
$\P[\AA^c] \leq \sum_{k=1}^N \P [\AA_{k}{}^c]$ we only need to show
that   for each $k$,  $\P[\AA_k{}^c]\leq CN^{-m}$, for any $m$. 
By the definition of the sets $\AA_k$, Chebyshev's inequality  and
Jensen's inequality, we have, for any $l\in\N$, 
$$
\eqalign{
\P[\AA_k{}^c] 
= \P\bigg[\Big|\GG\Big[ 
		\sum_{\mu = 1}^{M(N)}H^{\mu}_k\Big]\Big| \geq c N^{ \g}
	\bigg] 
&\leq (c N^{\g})^{-2l}\E\,\bigg( \GG\Big[
	\sum_{\mu=1}^{M(N)} H^\mu_k\Big]\bigg)^{2l} \cr
&\leq (c N^{\g})^{-2l}\E\,\GG\bigg[
	\Big(\sum_{\mu=1}^{M(N)} H^\mu_k\Big)^{2l}\bigg].
}
\Eq(F.5)
$$
If we can show that the expectation on the right-hand side is bounded
by some $N$-independent constant, \eqv(F.5) will prove the lemma.

Expanding the power in the integrand yields, with the usual
multi-index notation, 
$$
\E\,\GG\bigg[\Big(\sum_{\mu=1}^{M(N)} H^\mu_k\Big)^{2l}\bigg]
= \sum_{r: |r| = 2l} c_{2l, r} 
	\E\,\GG\bigg[\prod_{\mu = 1}^{M(N)} (H^\mu_k)^{r_\mu}\bigg],
\Eq(F.102)
$$
where $r$ is a multi-index and the numbers $c_{2l,r}$ are the multinomial coefficients.
The main point in what follows is the realisation that the difficult terms are those
which have at least one $\mu$ with $r_{\mu} = 1$. This is due to the following
observation, which is a simple consequence of a result proven in [Ni2].

\lemma{\TH(f.101)}
{\it
  There exist constants $c, K > 0$ such that for all $N$ large enough,
$$
\sup_{\s\in\SS_N}\sum_{\mu = 1}^{M(N)} (H^\mu_k(\s))^2 \leq c
\Eq(F.103)
$$
with probability at least $1 - e^{-K N^{1/4}}$.
}
\specskip
\proof
We write the left-hand side of \eqv(F.103) as 
$$
\sum_{\mu = 1}^{M(N)} (H^\mu_k(\s))^2 
= \frac{p!}{N^{2p-2}} \sum_{\mu = 1}^{M(N)} \sum_{\II, \JJ \ni k} 
	\s^{\vphantom{\mu}}_{\II} \xi^\mu_{\II} 
	\xi^\mu_{\JJ} \s^{\vphantom{\mu}}_{\JJ} 
= \frac{\a p!}{N^{p-1}} \sum_{\II,\JJ \ni k} 
	\s^{\vphantom{\mu}}_{\II} \sum_{\mu} \frac{\xi^\mu_\II \xi^\mu_\JJ}
	{\a N^{p-1}} \s^{\vphantom{\mu}}_\JJ.
\Eq(F.104)
$$
Consider $\s$ as a vector in an ${{N - 1} \choose {p-1}}$ dimensional 
space, and $ \a^{-1}N^{1-p}\sum_\mu \xi^\mu_\II \xi^\mu_\JJ$ as the 
coefficients of a matrix $P$ representing a map from this space onto itself.
Then, denoting by $\l_{max}$ the operator norm of $P$, uniformly in $\s$,
$$
\sum_{\mu = 1}^{M(N)} (H^\mu_k(\s))^2 
= \frac{\a p!}{N^{1-p}} (\s, P \s) \leq \frac{\a p!}{N^{1-p}} \|\s\|_2^2 \l_{max}
= \frac{\a p!}{N^{1-p}} {{N-1} \choose {p-1}} \l_{max}
\leq \a p \l_{max}.
\Eq(F.105)
$$
In [Ni2, Theorem~2] it is shown that  $\l_{max}$ is bounded by a constant 
with probability at least $1 - e^{-K N^l}$ with $l \in (0,\frac{1}{3})$. 
This proves the lemma.
\endproof
\specskip
\noindent
Returning to \eqv(F.102), we will try to get only terms of the form bounded
by the lemma above, the idea being that we do not really want to integrate,
but rather use a uniform bound for the integrands. We therefore single out
those $\mu$'s for which $r_\mu = 1$. We obtain
$$
\E\,\GG\bigg[\Big(\sum_{\mu = 1}^{M(N)} H^{\mu}_k\Big)^{2l}\bigg]
= \sum_{{\JJ \subset \MM:} \atop {|\JJ| \leq 2l}}
	\sum_{{r: r \triangleleft \JJ} \atop {|r| = 2l}} c_{2l,r} 
	\E\,\GG\left[ \prod_{\mu \in \JJ} H^{\mu}_k 
	\prod_{\mu \in \MM \setminus \JJ} (H^\mu_k)^{r_\mu}\right],
\Eq(F.106)
$$
where the compatibility relation $r \triangleleft \JJ$ means that for 
all $\mu \in \JJ$, $r_\mu = 1$. Since the $\mu \in \MM \setminus \JJ$
will not enter in any of the calculations that follow, we write (the
relation $r \prec \JJ$ now denotes the condition that $\forall \mu \in \JJ$,
$r_\mu = 0$)
$$
\eqalign{
I = 
\E\,\GG\bigg[\Big(\sum_{\mu = 1}^{M(N)} H^{\mu}_k\Big)^{2l}\bigg]
&= \sum_{{\JJ \subset \MM:} \atop { |\JJ| \leq 2l}} 
	\sum_{{r: r \prec \JJ} \atop {|r| = 2l - |\JJ|}}
	c_{2l, q, |\JJ|} \E\,\GG\left[
	\prod_{\mu \in \JJ} H^{\mu}_k 
	\prod_{\mu \in \MM \setminus \JJ} (H^\mu_k)^{r_\mu}\right] \cr
&= \sum_{{\JJ \subset \MM:} \atop { |\JJ| \leq 2l}} 
	\E\,\GG\bigg[ \prod_{\mu \in \JJ} H^{\mu}_k 
	\underbrace{\sum_{{r: r \prec \JJ} \atop {|r| = 2l - |\JJ|}}
			c_{2l, q, |\JJ|}
	\prod_{\mu \in \MM \setminus \JJ} (H^\mu_k)^{r_\mu}}_{\LL_{\JJ}(\s)}\bigg].
}
\Eq(F.107)
$$
At this point, we expand recursively the Boltzmann weights with
respect to the terms $H^\mu_k$, $\mu \in \JJ$. This will generate new
terms which are slightly more complicated than the term we started with.
The procedure stops when no $H^\mu_k$ is left to expand in. In particular,
since $|\JJ|$ does not depend on $N$, this will ensure that none of the  
appearing factors will depend on $N$.
\note{One may ask why we do not expand jointly in all the patterns 
$\mu \in \JJ$ at once. It turns out that one needs a similar recursive scheme
since there will always be error terms which cannot be treated by
Lemma~\thv(f.101).}

We use the following notation. We order the set $\JJ$ in the canonical way, 
i.e.\ $\JJ= \{\mu_1,\ldots,\mu_{n}\}$, with $i<j \Rightarrow \mu_i < \mu_j$. 
Then, we define interpolating Hamiltonians (they will reappear later)
$$
H^{\mu_1,\ldots,\mu_n}_{u_1,\ldots,u_n}(\s) 
= H(\s) - \sum_{i=1}^n (1 - u_i) H^{\mu_i}_k(\s).
\Eq(F.108)
$$
In particular, $H = H^{\mu_1,\ldots,\mu_n}_{1,\ldots,1}$, and 
if $u_j = 0$, then $H^{\mu_1,\ldots,\mu_n}_{u_1,\ldots,u_n}$ is independent
of $\xi^{\mu_j}_k$. The associated Gibbs measures and partition functions
will be denoted by $\GG^{\mu_1,\ldots,\mu_n}_{u_1,\ldots,u_n}$, respectively
$Z^{\mu_1,\ldots,\mu_n}_{u_1,\ldots,u_n}$.

The terms that will appear are of the form
$$
\E\,\GG^{\mu_1,\ldots,\mu_{n'}}_{u_1,\ldots,u_{n'}}{}^{\otimes q}
\left[	\prod_{i =n'+1}^n H^{\mu_i}_k(\s^1)
	\prod_{i=1}^{n'} H^{\mu_i}_k(\s^1) H^{\mu_i}_k(\s^{\pi_{n'}(i)})
	\LL_{\JJ}(\s^1) \right],
\EQ(F.109)
$$
where $q \leq n'$, and the $\pi_j$, $j=1,\ldots,n$ are functions from 
$\{1,\ldots,n'\}$ to $\{1, \ldots, n'\}$. They appear because the expansion 
of the denominator (the partition function) will introduce new copies of 
the measure (hence the power $q$).

The first product in the integrand above contains the $H^{\mu}_k$ with respect
to which the expansion has not yet been done. The second corresponds
to those which have been used. 

The initial expressions on the right of \eqv(F.107) correspond to the case 
$q=1$, $n'=0$, $u_{i}=1, \forall i$, that is,
$$
\E\,\GG\left[ \prod_{\mu \in \JJ} H^{\mu}_k \LL_{\JJ}(\s)\right]
= \E\,\GG^{\mu_1,\ldots,\mu_n}_{1,\ldots,1}\left[
	\prod_{i=1}^n H^{\mu}_k\LL_{\JJ}(\s^1)\right].
\Eq(F.110)
$$
The following provides the basic recursion relation.

\lemma{\TH(f.102)}
{\it
  For all numbers $n' \in \{0,\ldots,n-1\}$, $q \in \N$, and 
$u_1,\ldots,u_{n'}$, and functions $\pi_{n'}$, there exist functions
$(\pi^j_{n'+1})_{j = 1,\ldots, q+1}$, and a number $u_{n'+1} \in [0,1]$
such that 
$$
\eqalign{
\E\,\GG^{\mu_1,\ldots,\mu_{n'}}_{u_1,\ldots,u_{n'}}{}^{\otimes q}
	&\left[	\prod_{i = n' + 1}^n H^{\mu_i}_k(\s^1) 
		\prod_{i = 1}^{n'}H^{\mu_i}_k(\s^1) 
			H^{\mu_i}_k(\s^{\pi_{n'}(i)}) \LL_{\JJ}(\s^1)
	\right] \cr
&\kern-1cm = - \b \sum_{j=1}^q 
	\E\,\GG^{\mu_1,\ldots,\mu_{n'+1}}_{u_1,\ldots,u_{n'+1}}{}^{\otimes q}
	\left[\prod_{i = n' + 2}^n H^{\mu_i}_k(\s^1) 
		\prod_{i = 1}^{n'+1}H^{\mu_i}_k(\s^1) 
			H^{\mu_i}_k(\s^{\pi^j_{n'+1}(i)}) \LL_{\JJ}(\s^1)
	\right] \cr
&\kern-1cm \quad + \frac{\b}{q} 
	\E\,\GG^{\mu_1,\ldots,\mu_{n'+1}}_{u_1,\ldots,u_{n'+1}}{}^{\otimes q+1}
	\left[\prod_{i = n' + 2}^n H^{\mu_i}_k(\s^1) 
		\prod_{i = 1}^{n'+1}H^{\mu_i}_k(\s^1) 
			H^{\mu_i}_k(\s^{\pi^{q+1}_{n'+1}(i)}) \LL_{\JJ}(\s^1)
	\right] 
}
\Eq(F.112)
$$
The functions $(\pi^j_{n'+1})_{j = 1,\ldots, q+1}$ satisfy
$$
\pi^{j}_{n'+1}(i) = \cases \pi_{n'}(i), \quad\hbox{\tenrm if\ }i \leq n'; \cr
			   \rlap{$j,$}
			   \hphantom{\pi_{n'}(i),} 
				\quad\hbox{\tenrm if\ } i = n'+1. \cr
		    \endcases
\Eq(F.113)
$$
}
\proof
We expand the Boltzmann weight of the Gibbs measure on the left-hand
side of \eqv(F.112) in the pattern $\mu_{n'+1}$. Since
$H^{\mu_1,\ldots,\mu_{n'}}_{u_1,\ldots,u_{n'}}
= H^{\mu_1,\ldots,\mu_{n'},\mu_{n'+1}}_{u_1,\ldots,u_{n'},u_{n'+1}}|_{u_{n'+1} = 1}$,
expanding in the variable $u_{n'+1}$ about 0 to zero order with 
remainder of order 1 yields
$$
\eqalign{
\frac{\exp\bigg(-\b \sum_{j'=1}^q H^{\mu_1,\ldots,\mu_{n'}}_{u_1,\ldots,u_{n'}}(\s^{j'})
	\bigg)}
	{(Z^{\mu_1,\ldots,\mu_{n'}}_{u_1,\ldots,u_{n'}})^q}
&= \frac{\exp\bigg(-\b \sum_{j'=1}^q 
		H^{\mu_1,\ldots,\mu_{n'},\mu_{n'+1}}_{u_1,\ldots,u_{n'},0}(\s^{j'})
	\bigg)}
	{(Z^{\mu_1,\ldots,\mu_{n'+1}}_{u_1,\ldots,u_{n'},0})^q}
	\cr
&\kern-1cm- \b\sum_{j = 1}^q 
	\frac{\exp\bigg(-\b \sum_{j'=1}^q 
		H^{\mu_1,\ldots,\mu_{n'},\mu_{n'+1}}_{u_1,\ldots,u_{n'},u_{n'+1}}(\s^{j'})
	\bigg)}
	{(Z^{\mu_1,\ldots,\mu_{n'},\mu_{n'+1}}_{u_1,\ldots,u_{n'},u_{n'+1}})^q}
		H^{\mu_{n'+1}}_k(\s^{j}) \cr
&\kern-1cm+ \frac{\b\exp\bigg(-\b \sum_{j'=1}^q 
		H^{\mu_1,\ldots,\mu_{n'},\mu_{n'+1}}_{u_1,\ldots,u_{n'},u_{n'+1}}(\s^{j'})
	\bigg)}
	{q(Z^{\mu_1,\ldots,\mu_{n'},\mu_{n'+1}}_{u_1,\ldots,u_{n'},u_{n'+1}})^{q+1}}
	\cr
&\kern-1cm\quad\times
	\E_{\s^{q+1}}\exp\bigg( - \b
	 	H^{\mu_1,\ldots,\mu_{n'},\mu_{n'+1}}_{u_1,\ldots,u_{n'},u_{n'+1}}(\s^{q+1})
	\bigg)
	H^{\mu_{n'+1}}_k(\s^{q+1}),
}
\EQ(F.114)
$$ 
for some $u_{n'+1} \in [0,1]$.

The first term on the right does not depend on $\xi^{\mu_{n'+1}}_k$ 
(see the remark after \eqv(F.108)). Hence, when multiplied by the 
products of the $H^\mu_k$, this disorder variable appears exactly once, so
that integration with respect to it yields zero.

The second and third term above give the new terms on the right in \eqv(F.112).
The relations for the functions $\pi^j_{n'+1}$ are easily verified.
\endproof
\specskip
\noindent
Applying this recursion relation $n$ times yields the following decomposition.

\lemma{\TH(f.103)}
{\it
  Let $\JJ = \{\mu_1,\ldots, \mu_n\}$, $n \leq 2l$. Then there exist
numbers $u_1,\ldots,u_n \in [0,1]$ such that 
$$
\E\,\GG\Big[\prod_{i=1}^n H^{\mu_i}_k(\s^1) \LL_{\JJ}(\s)\Big]
= \sum_{q = 1}^n \sum_{\pi\sim q} c_{\pi,q} \b^n
	\GG^{\mu_1,\ldots,\mu_n}_{u_1,\ldots,u_n}{}^{\otimes q}
	\Big[\prod_{i=1}^n H^{\mu_i}_k(\s^1) H^{\mu_i}_k(\s^{\pi(i)}) 
		\LL_{\JJ}(\s^1)
	\Big],
\Eq(F.115)
$$
where the functions $\pi$ permute the indices $i \in \{1,\ldots,n\}$, and
the relation $\pi \sim q$ describes the condition that 
$|\{i\in \{1,\ldots,n\}: \pi(i) \neq 1\}| = q$. The number of such functions 
$\pi$ is thus independent of $N$.
}
\specskip
\proof
The proof follows by applying the recursion relation from Lemma~\thv(f.102)
$n$ times. Observing that each step adds at most one other replica implies
that $q \leq n$.
\endproof
\specskip
\noindent

We finally sum over the sets $\JJ \subset \MM$ on the right of \eqv(F.107).
We obtain
$$
|I| \leq \sum_{{\JJ \subset \MM:} \atop{|\JJ| = 2l}} 
	\sum_{q=1}^{|\JJ|} \sum_{\pi \sim q} c_{\pi, q} \b^n
	\E\,\GG^{\mu_1,\ldots,\mu_n}_{u_1,\ldots,u_n}{}^{\otimes q}
	\Big[\prod_{i=1}^n \Big|H^{\mu_i}_k(\s^1) H^{\mu_i}_k(\s^{\pi(i)}) \Big|
		\Big|\LL_{\JJ}(\s^1)\Big|
	\Big].
\Eq(F.116)
$$
First, we observe that since $|H^{\mu}_k| < 1$, 
$$
|\LL_{\JJ}| 
\leq \sum_{{r: r \prec \JJ} \atop {|r| = 2l - |\JJ|}}
	c_{2l,r,|\JJ|} \prod_{\mu \in \MM\setminus\JJ} |H^{\mu}_k|^{r_\mu} 
\leq \sum_{{r: r \prec \JJ} \atop {|r| = 2l - |\JJ|}}
	c_{2l,r,|\JJ|} \prod_{\mu \in \MM\setminus\JJ} 
		|H^{\mu}_k|^{2 \d_{r_{\mu},2}},
\Eq(F.117)
$$
where $\d_{a,b} = 1$, if $a\geq b$ and zero otherwise. 

For any multi-index $r$, denote by $\#r$ the number of $r_{\mu}$ which
are not zero. Hence, the products on the right-hand side of the above inequality 
are just the completely off-diagonal terms of the form
$\Big(\sum_{\mu \in \MM\setminus\JJ} (H^{\mu}_k)^2\Big)^{\#r}$.
Then, adding the terms which have at least two indices equal (and which 
are obviously positive), yields uniformly in $\s$
$$
|\LL_{\JJ}| 
\leq  \sum_{j=1}^{2l-|\JJ|}c_{j,2l,|\JJ|}
	\Big(\sum_{\mu \in \MM\setminus\JJ} (H^\mu_k)^2\Big)^j 
\leq  \sum_{j=1}^{2l-|\JJ|} c_{j,2l,|\JJ|}
	\Big(\sum_{\mu \in \MM} (H^\mu_k)^2\Big)^j 
\leq C,
\Eq(F.119)
$$
on a set $\BB$ of measure at least $1 - e^{-K N^{1/4}}$ by Lemma~\thv(f.101). 
Using this in \eqv(F.116), we bound 
$I' = \E \1_\BB \GG[(\sum_{\mu} H^\mu_k)^{2l}]$ by
$$
|I'| 
\leq \sum_{{\JJ \subset \MM:} \atop {|\JJ| \leq 2l}}
	\sum_{q = 1}^{|\JJ|} \sum_{\pi \sim q} c_{\pi,q,|\JJ|,\b} 
	\E \1_\BB \GG^{\mu_1,\ldots,\mu_n}_{u_1,\ldots,u_n}{}^{\otimes q}
	\Big[
		\prod_{i=1}^n \Big|H^{\mu_i}_k(\s^1) H^{\mu_i}_k(\s^{\pi(i)}) \Big|
	\Big].
\Eq(F.120)
$$
Since the integrand is non-negative and $|H^{\mu}_k| < 1$, we can change the 
Boltzmann weights back to the original ones (that is, setting all $u_i = 1$), 
and committing at most an error of $e^{\b n}$. Furthermore, the functions
$\pi$ depend only on the size of $\JJ$. Hence, adding again positive terms
in the third step below (and observing that $|\JJ|$ is even), 
$$
\eqalign{
|I'| &\leq C \sum_{{n=0}\atop {\hbox{\sevenrm even}}}^{2l} 
	\sum_{{\JJ \subset \MM:}\atop{|\JJ| = n}}
	\sum_{\pi \sim q}  \sum_{q = 1}^n c_{\pi,q, \b} \E \1_\BB \GG^{\otimes q}\Big[
		\prod_{i=1}^n \Big|H^{\mu_i}_k(\s^1) H^{\mu_i}_k(\s^{\pi(i)}) \Big|
	\Big] \cr
&\leq C \sum_{{n=0} \atop {\hbox{\sevenrm even}}}^{2l} \sum_{\pi \sim q} 
	\sum_{{\JJ \subset \MM:}\atop{|\JJ| = n}}
	\sum_{q = 1}^n c_{\pi,q,\b} \E\1_\BB\GG^{\otimes q}\Big[
		\prod_{i=1}^n \Big|H^{\mu_i}_k(\s^1) H^{\mu_i}_k(\s^{\pi(i)}) \Big|
	\Big] \cr
&\leq C \sum_{{n=0} \atop {\hbox{\sevenrm even}}}^{2l} \sum_{\pi \sim q} \frac{1}{n!} 
	\sum_{\mu_1,\ldots,\mu_n = 1}^{M}
	\sum_{q = 1}^n c_{\pi,q,\b} \E\1_\BB \GG^{\otimes q}\Big[
		\prod_{i=1}^n \Big|H^{\mu_i}_k(\s^1) H^{\mu_i}_k(\s^{\pi(i)}) \Big|
	\Big] \cr
&\leq C \sum_{{n=0}\atop{\hbox{\sevenrm even}}}^{2l} \sum_{\pi \sim q} \frac{1}{n!} 
	\sum_{q = 1}^n c_{\pi,q,\b} \E\1_\BB\GG^{\otimes q}\Big[
		\prod_{i=1}^n \Big(
			\sum_{\mu_i = 1}^M 
			|H^{\mu_i}_k(\s^1)H^{\mu_i}_k(\s^{\pi(i)})| 
		\Big)
	\Big].
}
\Eq(F.121)
$$
Finally, we apply Cauchy-Schwarz to get rid of the absolute value in the sum
over $\mu_i$,
$$
\sum_{\mu=1}^M | H^{\mu_i}_k(\s^1) H^{\mu_i}_k(\s^{\pi(i)})|
\leq \Big(\sum_{\mu_i = 1}^M (H^{\mu_i}_k(\s^1))^2\Big)^{\frac{1}{2}}
	\Big(\sum_{\mu_i = 1}^M (H^{\mu_i}_k(\s^{\pi(i)}))^2\Big)^{\frac{1}{2}}
\leq C
\EQ(F.122)
$$
on $\BB$ by Lemma~\thv(f.101). Inserting the above in \eqv(F.121) shows that
$\E\1_\BB \GG[\sum_{\mu} H^\mu_k]$ is bounded by a number independent
of $N$, since all the remaining sums are over finite sets whose sizes 
do not depend on $N$.

Since  $(\sum_\mu H^\mu_k)^{2l}$ is polynomially bounded in $N$,
uniformly in $\o$, the remaining part $I-I'$, (that is, the integral
on the set $\BB^c$), is  
obviously bounded by an exponentially small number in $N^{1/5}$ (e.g.), and
is thus also smaller than a constant.
	
We use this in \eqv(F.5) which shows that
$$
\P[\AA_k{}^c] \leq c_l c^{-2l} N^{- 2\g l}.
\Eq(F.123)
$$		
Thus for all $\g, m > 0$, there exist $l$ and $C_{l,m}$ such that
$$
\P[\AA_k{}^c] \leq C_{l,m} N^{- m - 1}.
\EQ(F.124)
$$
Summing over all $k = 1,\ldots,N$ shows that indeed $\P[\AA^c] \leq C_{l,m} N^{-m}$.
\endproof
\specskip
\noindent
We now bound the fluctuations of the free energy on the set $\AA$.

\proposition{\TH(f.2)}
{\it
  Let $\tilde{F}_N = N^{-1} \ln Z_N \1_{\AA_{c, \g, N}}$. Then, for all
$\b$, all $ \tau>0$ and all $ \varepsilon  > \g$, there exists $\bar N<\infty$
such that for all $N>\bar N$,  
$$
\P\left[ | \tilde{F}_N - \E\,\tilde{F}_N | > \tau\b N^{-\frac{1}{2} + \varepsilon} \right] 
\leq 3 e^{- N^{\varepsilon/2}}.
\Eq(F.17)
$$
}

\proof
In the sequel, $N, \b, \g, c$ will be fixed, and we will therefore frequently drop the corresponding 
indices. The approach to the proof follows the general idea of [BGP2,B1]. Define a 
decreasing sequence of $\s$-algebras $\{\hat\FF_k\}_{k \in \N}$ by 
$$
\hat\FF_k = \s\left(\{\xi^{\mu}_i\}^{\mu \in \N}_{i \geq k}\right) \vee \AA_{c,\g, N}.
\Eq(F.18)
$$
This allows to introduce a martingale difference sequence (see [Yu])
$$
\tilde{F}^k \equiv \E\,[\tilde{F} | \hat\FF_k] - \E\,[\tilde{F} | \hat\FF_{k+1}].
\Eq(F.19)
$$
By the definition of conditional expectations
$$
\tilde{F} - \E\,\tilde{F} = \sum_{k=1}^{N} \tilde{F}^k \P[\AA].
\Eq(F.20)
$$
The factor $\P[\AA]$ tends to one as $N \uparrow \infty$ by Lemma~\thv(f.1) 
(even polynomially as fast as we want). It is therefore enough to control 
the sum $\sum_{k=1}^{N} \tilde{F}^k$. We observe that
$$
\eqalign{
\P[|\sum_{k=1}^{N} \tilde{F}^k| > z (\P[\AA])^{-1}] 
&\leq 2 \inf_{t > 0} e^{-t z (\P[\AA])^{-1}} \E\, e^{t \sum_{k=1}^{N} \tilde{F}^k} \cr
&= 2 \inf_{t > 0} e^{- t z} 
        \E\,[\E\,[\ldots \E\,[\E\,[e^{t \tilde{F}^1}| \hat\FF_2] e^{t \tilde{F}^2}|\hat\FF_3] \ldots ]
        e^{t \tilde{F}^N} | \hat\FF_{N+1}].
}
\Eq(F.21)
$$
To make use of this inequality, we need bounds on the conditional Laplace transforms, that is, we 
want to show that for some $\LL^k(t)$,
$$
\ln \E\,[e^{t \tilde{F}^k}| \hat\FF_{k+1}] \leq \LL^k(t),
\Eq(F.22)
$$
uniformly in $\hat\FF_{k+1}$. Using a standard second order bound for
the exponential function, we get 
$$
\E\,[e^{t \tilde{F}^k}|\hat\FF_{k+1}] \leq 1 + \frac{t^2}{2} \E\,[(\tilde{F}^k)^2 e^{| t \tilde{F}^k|}
        | \hat\FF_{k+1}].
\Eq(F.23)
$$
To make use of this we need to bound $|\tilde F^k|$.  
A conventional strategy  is to introduce a family of Hamiltonians $\tilde{H}^k(\s,u)$,
defined by
$$
\tilde{H}^k(\s,u) = H(\s) + (1 - u) \frac{(p!)^{1/2}}{N^{p-1}} \sum_{\mu = 1}^{M(N)}
        \sum_{{\II \ni k} \atop {|\II| = p}} \xi^{\mu}_{\II} \s^{\vphantom{\mu}}_{\II}.
\Eq(F.24)
$$
This new Hamiltonian is equal to the original one for $u=1$, and independent of 
$\{\xi_{k}^{\mu}\}^{\mu =1, \ldots ,M}$ for $u = 0$. Denote by $\tilde{Z}^k(u)$
and $\GG^k(u)$ the partition function, respectively the Gibbs measure 
associated to this Hamiltonian.
Observe that the condition on being on the set $\AA$ is stable against
the change in parameter $u \in [0,1]$, that is 
$$
\GG^k(u)\Big[N^{-p}\sum_{\mu=1}^{M(N)} \sum_{\II \ni k} 
	\xi^\mu_\II \s^{\vphantom{\mu}}_\II\Big] \in [-c, c], \quad \forall u \in [0,1],
\Eq(F.25)
$$
on the set $\AA$. Indeed, the derivative of the left-hand side with respect
to $u$ is non-negative, since it is the variance of the integrand with respect
to the measure $\GG(u)$. Moreover, for $u=0$, the Boltzmann weight does not
contain $\s_k$, whence the left is zero for $u=0$. The absolute value of the 
left-hand side thus assumes its maximal value for $u=1$.

Define 
$$
g^k(u) = \frac{1}{N} \1_\AA \ln \tilde{Z}^k(u) 
	- \frac{1}{N} \1_\AA \ln \tilde{Z}^k(0).
\Eq(F.26)
$$
Since $\tilde{Z}^k(0)$ is independent of $\s_k$, this quantity 
relates to $\tilde{F}^k$ via
$$
\tilde{F}^k = \E\,[g^k(1)| \hat\FF_k] - \E\,[g^k(1)| \hat\FF_{k+1}]
\Eq(F.27)
$$
Observe that $g^k(u)$ is convex in $u$, since its derivative is equal to
the expectation of the left-hand side of \eqv(F.25), whose derivative is the 
variance of a random variable with respect to the measure $\GG$. Since by 
its definition $g^k(0) = 0$, and therefore 
$|g^k(1)| \leq \max(|(g^k)'(1)|,|(g^k)'(0)|)$, where the prime denotes the 
derivative with respect to $u$. Moreover, since $\tilde{H}^k(\s,u=0)$ does 
not depend on $\s_k$, it follows that $(g^k)'(0) = 0$, and hence we can 
use $|g^k(1)| \leq |(g^k)'(1)|$. Explicitly, this is
$$
|g^k(1)| 
\leq |(g^k)'(1)| = 
\left|\b\frac{(p!)^{1/2}}{N^p} \GG_{N,\b}[\o] \left( \sum_{\mu = 1}^{M(N)}
        \sum_{{\II \ni k} \atop {|\II| = p}} 
		\xi^{\mu}_{\II} \s^{\vphantom{\mu}}_{\II} \right)\right|
	\1_\AA
\leq c N^{-1 + \g}.
\Eq(F.28)
$$
Inserting this bound into the exponent on the right-hand side of \eqv(F.23) gives
$$
\eqalign{
1 + \frac{t^2}{2} \E\,[ (\tilde{F}^k)^2 e^{|t \tilde{F}^k|} | \hat\FF_{k+1}] 
&\leq 1 + \frac{t^2}{2} \E\,[(\tilde{F}^k)^2 e^{|t g^k(1)|}| \hat\FF_{k+1}] \cr
&\leq 1 + \frac{t^2}{2} e^{2 c t N^{-1 + \g}} \E\,[(\tilde{F}^k)^2 | \hat\FF_{k+1}].
}
\Eq(F.29)
$$
To treat the quadratic term, we observe that by \eqv(F.27), the properties of conditional 
expectations, and Jensen's inequality (see also [B] and [BGP]),
$$
\eqalign{
\E\,[(\tilde{F}^k)^2 | \hat\FF_{k+1}]
&=    \E\,\left[ (\E\,[g^k(1) | \hat\FF_{k}] - 
		\E\,[g^k(1)|\hat\FF_{k+1}])^2 \bigg| \hat\FF_{k+1} \right] \cr
&=    \E\,\left[(\E\,[g^k(1) - \E\,[g^k(1)|\hat\FF_{k+1}]| 
		\hat\FF_{k}])^2 \bigg| \hat\FF_{k+1}\right] \cr
&\leq \E\,\left[\E\,[(g^k(1) - \E\,[g^k(1)| \hat\FF_{k+1}])^2 | 
		\hat\FF_k] \bigg| \hat\FF_{k+1}\right] \cr
&=    \E\,\left[(g^k(1) - \E\,[g^k(1) | \hat\FF_{k+1}])^2 \bigg| \hat\FF_{k+1} \right] \cr
&=    \E\,[(g^k(1))^2 | \hat\FF_{k+1}] - \left( \E\,[g^k(1) | \hat\FF_{k+1}]\right)^2 \cr
&\leq \E\,[(g^k(1))^2 | \hat\FF_{k+1}] \leq \E\,[(g^k(1))'{}^2 | \hat\FF_{k+1}].
}
\Eq(F.30)
$$
The last term is bounded since we are in the set $\AA_k$. Indeed,
$$
\E\,\left[(g^k(1))^{\prime 2} | \hat\FF_{k+1}\right]
= \frac{p!}{N^{2p}} \b^2\E\,\left[\left(\1_\AA \GG \Big[ \sum_{\mu = 1}^{M(N)} 
\sum_{\II \ni k} 
\xi^{\mu}_{\II} \s^{\vphantom{\mu}}_{\II}\Big] \right)^2 \bigg| \hat\FF_{k+1} 
\right]
\leq \b^2C N^{2\g - 2},
\Eq(F.31)
$$
Thus, using the bound \eqv(F.31) in \eqv(F.29), 
$$
1 + \frac{t^2}{2} \E\,[ (\tilde{F}^k)^2 e^{|t \tilde{F}^k|} | \hat\FF_{k+1}] 
\leq 1 + \frac{t^2}{2} e^{2c\b tN^{-1 + \g }} C\b^2 N^{2 \g - 2}
 \leq \exp\left(\frac{t^2}{2} e^{2c\b tN^{-1+ \g}} C\b^2 N^{2\g - 2}\right).
\Eq(F.32)
$$
Inserting this in \eqv(F.21) yields
$$
\P[|\sum_{k=1}^{N} \tilde{F}^k| > z (\P[\AA])^{-1}] 
\leq 2 \inf_{t > 0} \exp\left(- tz + \frac{t^2}{2} e^{2c\b tN^{-1+\g}}
		 C\b^2 N^{2 \g - 1}\right).
\Eq(F.33)
$$
We choose $z = \t\b  N^{-1/2 + \varepsilon}$, and
$t = \frac{1}{z} N^{\frac{\varepsilon}{2}} = \frac 1{\t\b}
 N^{\frac{1 - \varepsilon}{2}}$. This implies that 
$$
\P[|\sum_{k=1}^{N} \tilde{F}^k| > \b\t N^{- \frac{1}{2} + \varepsilon} (\P[\AA])^{-1}] 
\leq 2 \exp\left( - N^{\frac{\varepsilon}{2}} + C \t^{-2}
N^{2 \g - \varepsilon} 
                e^{2 c \t^{-1} N^{-1/2 + \g - \varepsilon/2}}\right).
\Eq(F.35)
$$
Choose $\g < \varepsilon/2$. Then for any $\t>0$, and $N$ large enough,
the right hand side of \eqv(F.35) is bounded by $3e^{-N^{\e/2}}$.
Since $\P[\AA]$ tends to 1 as $1 - N^{-m}$, the claimed estimate  follows.
\endproof
\specskip

\proofof{Theorem~4}
The assertion is now an immediate consequence of Lemma~\thv(f.1) and Proposition~\thv(f.2). Indeed,
$$
| F_N - \E\,F_N | \leq | F_N - \tilde{F}_N| + |\tilde{F}_N - \E\,\tilde{F}_N| 
                + |\E\,\tilde{F}_N - \E\,F_N|.
\Eq(F.40)
$$
The first term is non zero only on $\AA^c$. Also, the last summand is bounded by 
$\P[\AA^c] \sup F_N \leq C N^p \P[\AA^c]$. If we choose $m$ in Lemma~\thv(f.1) 
larger than $p + n + 1$, then this term is eventually less than $N^{-2}$, and thus also
less than $z = \t N^{-1/2 + \varepsilon}$. Thus, for all $n, \tau, \varepsilon > 0$, and $N$ large 
enough,
$$
\eqalign{
\P[|F_N  - \E\,F_N| > z] 
&\leq \P[ |F_N - \tilde{F}_N| > \frac{z}{3}] + \P[|\tilde{F}_N - \E\,\tilde{F}_N| > \frac{z}{3}] \cr
&\leq C N^p \P[\AA^c] + \P[|\tilde{F}_N - \E\,\tilde{F}_N| > \frac{z}{3}] \cr
&\leq C N^{- n - 1} + e^{-N^{\varepsilon}} < N^{-n}.
}
\Eq(F.41)
$$
This concludes the proof of the theorem.
\endproof

\bigskip

%


%
%
%

\bigskip

\chap{6.~Results on the Replica Overlap.}6
\bigskip

\noindent
In this section, we prove the results on the replica overlap, Theorems~\thv(pr.5),
\thv(pr.7), and \thv(pr.9).

\medskip
\subchap{ \ver.1.~Proof of Theorem~\thv(pr.5).}
\medskip
\noindent
By the definition of the free energy, 
$$
\E\,\frac{\partial F_N}{\partial \b} = - \frac{\b}{N} \E\,\GG_{N,\b}[H]
	= - \b \sum_{\mu = 1}^{M(N)} \E\,\GG_{N,\b}[H^\mu(\s)],
\Eq(RO.1)
$$
where 
$$
H^\mu(\s) = - \frac{(p!)^{1/2}}{N^{p-1}} 
	\sum_{{\II \subset \NN} \atop {|\II| = p}} \xi^\mu_{\II} \s^{\vphantom{\mu}}_{\II},
\Eq(RO.2)
$$
is the contribution to the Hamiltonian from pattern $\mu$. 
We introduce the following notation. For any $u \in [0,1]$, we let 
$\bar{H}^\mu_u$ be an interpolating Hamiltonian of the form
$$
\bar{H}^\mu_u = H - (1 - u) H^\mu.
\Eq(RO.2bis)
$$
Observe that for $u=0$, this quantity is independent of the pattern $\mu$, and for
$u=1$, is equal to the original Hamiltonian. The notations $\bar{\GG}^\mu_u$ and 
$\bar{Z}^\mu_u$ refer to the corresponding Gibbs measures and partition functions 
(dropping reference to $N$ and $\b$ for sake of clarity). We now write the Gibbs 
ectation on the right of \eqv(RO.1) as 
$$
\GG_{N,\b}[H^{\mu}(\s)] = \E_{\s} \left[ \frac{e^{-\b \bar{H}^\mu_u}}
		{\E_{\s'}[e^{-\b\bar{H}^\mu_u}]} H^\mu \right]
		\Bigg|_{u=1}.
\Eq(RO.3)
$$
Developping the Boltzmann weights in $u$ about 0 with second order remainder, 
we 
obtain for each term in the sum on the right-hand side of \eqv(RO.1) (for some 
$u \in [0,1]$)
$$
\eqalign{
\GG_{N,\b}[H^\mu] 
&= \E_{\s}\left[\frac{e^{-\b \bar{H}^\mu_0(\s)}}{\bar{Z}^\mu_0} H^{\mu}_0(\s)\right]
		- \b \E_\s \left[\frac{e^{-\b \bar{H}^\mu_0(\s)}}{\bar{Z}^\mu_0} 
			H^{\mu}(\s)^2\right] \cr
&\kern-1cm+\b \E_{\s,\s'}\left[ \frac{e^{-\b \bar{H}^{\mu}_0(\s) - \b \bar{H}^\mu_0(\s')}}
		{\bar{Z}^\mu_0{}^2} H^\mu(\s) H^\mu(\s')\right] 
	+ \frac{\b^2}{2} \underbrace{\E_\s \left[\frac{e^{-\b \bar{H}^\mu_u(\s)}}
		{\E_{\s}[e^{-\b \bar{H}^\mu_u(\s)}]} H^\mu(\s)^3
			\right]}_{R_1} \cr
&\kern-1cm- \frac{3\b^2}{2} \underbrace{\E_{\s,\s'}\left[ \frac{e^{-\b \bar{H}^\mu_u(\s)
	-\b \bar{H}^\mu_u(\s')}}
		{(\E_{\s}[e^{-\b \bar{H}^\mu_u(\s)}])^2}
	H^\mu(\s)^2 H^\mu(\s') \right]}_{R_2} \cr
&\kern-1cm+ \frac{\b^2}{2} \underbrace{\E_{\s,\s',\s''} 
			\left[ \frac{e^{-\b \bar{H}^\mu_u(\s)
	-\b \bar{H}^\mu_u(\s') - \b \bar{H}^\mu_u(\s'')}}
	{(\E_{\s}[e^{-\b \bar{H}^\mu_u(\s)}])^3} 
	H^\mu(\s)H^\mu(\s')H^\mu(\s'')\right]}_{R_3}.
}
\Eq(RO.4)
$$
As remarked above, neither $\bar{H}^\mu_0$ nor $\bar{Z}^\mu_0$ contain any of the variables
$\{\xi^\mu_i\}_{i \in \NN}$. Integration with respect to them (denoted by $\E_\mu$)
thus yields for the linear term,
$$
\E\,\E_{\s}\left[\frac{e^{-\b\bar{H}^\mu_0(\s)}}{\bar{Z}^\mu_0} H^{\mu}(\s)\right]
=\frac{(p!)^{1/2}}{N^{p-1}} \sum_{{\II \subset \NN} \atop {|\II| = p}}
		\E'\,\E_\s\left[ \frac{e^{-\b\bar{H}^\mu_0(\s)}}{\bar{Z}^\mu_0} 
		\E_\mu \xi^\mu_\II \s^{\vphantom{\mu}}_\II
	\right] = 0,
\Eq(RO.5)
$$
and for the second order contribution 
$$
\eqalign{
\E\,\E_\s\left[ \frac{e^{-\b\bar{H}^\mu_0(\s)}}{\bar{Z}^\mu_0} H^\mu(\s)^2\right]
&= \frac{p!}{N^{2p-2}} \E'\E_{\s}\left[\frac{e^{-\b\bar{H}^\mu_0(\s)}}{\bar{Z}^\mu_0}
	\sum_{\II,\JJ} \E_\mu \xi^\mu_\II \xi_\JJ^\mu \s_\II \s_\JJ \right] \cr
&= \frac{p!}{N^{2p-2}} \E'\E_{\s}\left[\frac{e^{-\b\bar{H}^\mu_0(\s)}}{\bar{Z}^\mu_0}
		\sum_{\II} 1 \right] 
= N^{2-2p} (1 + {\cal O}(N^{-1})),
}
\Eq(RO.6)
$$
respectively,
$$
\E\,\E_{\s,\s'} \left[\frac{e^{-\b \bar{H}^{\mu}_0(\s) - \b \bar{H}^\mu_0(\s')}}
		{\bar{Z}^\mu_0{}^2} H^\mu(\s) H^\mu(\s')\right]
= \frac{p!}{N^{2p-2}} \E\,\E_{\s,\s'} \left[ 
	\frac{e^{-\b \bar{H}^{\mu}_0(\s) - \b \bar{H}^\mu_0(\s')}}
		{\bar{Z}^\mu_0{}^2} \sum_{\II} \s_{\II} \s'_{\II} \right].
\Eq(RO.7)
$$
The latter sum is 
$$
\eqalign{
\sum_{\II \subset \NN} \s_{\II} \s'_{\II} 
&= \frac{1}{p!} \sum_{{i_1,\ldots,i_p =1}
		\atop {\hbox{\sevenrm all\ different}}}^{N} \prod_{l=1}^p \s_{i_l}\s'_{i_l} 
= \frac{1}{p!}\left(\sum_{i = 1}^N \s_i \s'_i\right)^p (1 + {\cal O}(N^{-1})) \cr
&= \frac{1}{p!} N^p R(\s,\s')^p (1 + {\cal O}(N^{-1})),
}
\Eq(RO.8)
$$
whence, 
$$
\E\,\E_{\s,\s'} \left[\frac{e^{-\b \bar{H}^{\mu}_0(\s) - \b \bar{H}^\mu_0(\s')}}
		{\bar{Z}^\mu_0{}^2} H^\mu(\s) H^\mu(\s')\right]
= N^{2-2p} E\,\bar{\GG}_0^\mu {}^{\otimes 2} \left[R(\s,\s')^p (1 + {\cal O}(N^{-1}))
	\right].
\Eq(RO.9)
$$
We now show that the remainder terms in \eqv(RO.4) are at least one order (in $N$)
less than the two leading contributions above. We start with a result that shows
that the perturbed partition function 
$\bar{Z}^{\mu}_u = \E_\s [ e^{-\b \bar{H}^\mu_u}]$
is bounded from below by a constant times the partition function 
$\bar{Z} = \bar{Z}^\mu_0$ (that is, the one not containing any of the $\{\xi^\mu_i\}_i$).

\lemma{\TH(ro.2)}
{\it
  For all $\b \geq 0$ there exists a constant $c>0$ such that for all $u \in [0,1]$, 
$$
\bar{Z}^\mu_u \geq c \bar{Z}^\mu_0 = c\, \E_\s[ e^{-\b \bar{H}^\mu_0}].
\Eq(RO.11)
$$
}

\proof
The proof is an immediate consequence of the following result.

\lemma{\TH(ro.3)}
{\it
  Let $\{X_i\}_{i = 1,\ldots,N}$ be a familiy of variables taking values $-1$ and $1$.
Let $\G_{p,N} = N^{-p} \sum_{\II: |\II| = p} X_{\II}$, and $m = N^{-1} \sum_i X_i$.
Then for each even $p$ there exist constants $c_{p,q}$ such that 
$$
\G_{p,N} = \sum_{q = 0}^{\frac{p}{2}} c_{p,2q} m^{2q} N^{q - \frac{p}{2}}
		(1 + {\cal O}(N^{-1})).
\Eq(RO.12)
$$
Moreover, $c_{p,p}$ is positive for all $p$.
}
\specskip
\proof
By induction. For $p=2$, we have 
$$
\eqalign{
\G_{2,N} = N^{-2} \frac{1}{2}\sum_{i = 1}^N \sum_{{j = 1} \atop {j \neq i}}^N
		X_i X_j
&= N^{-2} \frac{1}{2} \sum_{i,j=1}^N X_i X_j - N^{-2} \frac{1}{2} \sum_{i=1}^N 1 \cr
&= \frac{1}{2} m^2 - N^{-1},
}
\Eq(RO.13)
$$
which is of the form claimed in \eqv(RO.12).

Suppose the result is true for all even values $q \leq p$. Then,
$$
\eqalign{
\G_{p+2,N} 
&= N^{-p-2} \sum_{\II: |\II| = p+2} X_{\II}
	= \frac{1}{{{p+2} \choose 2}N^{p+2}} 
		\sum_{\II: |\II| = p} X_{\II} 
		\sum_{{\JJ: |\JJ| = 2} \atop {\II \cap \JJ = \emptyset}} X_{\JJ} \cr
&= c_p N^{-p-2} \sum_{\II: |\II| = p} X_{\II} 
	\sum_{\JJ: |\JJ| = 2} X_{\JJ} 
	- c_p N^{-p-2} \sum_{\II: |\II| = p} X_{\II} 
	\sum_{{\JJ: |\JJ| = 2} \atop {\JJ \cap \II \neq \emptyset}} X_{\JJ}.
}
\Eq(RO.14)
$$
By the induction hypothesis, the first term on the right-hand side is
$$
\eqalign{
c_p N^{-p-2} &\sum_{\II:|\II|= p} X_{\II} \sum_{\JJ: |\JJ| = 2} X_{\JJ}
= c_p \G_{p,N} \G_{2,N} \cr
&= c_p \left(\sum_{q=0}^{\frac{p}{2}} c_{p,2q} m^{2q} N^{q - \frac{p}{2}}
			(1 + {\cal O}(N^{-1}))\right)
	\left(\sum_{q=0}^{1} c_{2,2q} m^{2q} N^{q - 1}(1 + {\cal O}(N^{-1}))\right) \cr
&= \sum_{q=0}^{\frac{p}{2} + 1} c_{p,2q} m^{2q} N^{q - \frac{p}{2}-1} (1 + {\cal O}(N^{-1})).
}
\Eq(RO.15)
$$
The remaining term in \eqv(RO.14) is
$$
\eqalign{
\sum_{\II: |\II| = p} 
	\sum_{{\JJ: |\JJ| = 2} \atop {\JJ \cap \II \neq \emptyset}} X_{\II} X_{\JJ} 
&= \sum_{{\II, \JJ} \atop {|\JJ \cap \II| = 1}} X_{\II} X_{\JJ}
	+ \sum_{{\II, \JJ} \atop {|\JJ \cap \II| = 2}} X_{\II} X_{\JJ} \cr
&= \sum_{\II: |\II| = p} \sum_{i \in \NN\setminus\II} \sum_{j \in \II}
	X_{\II} X_i X_j  
	+\sum_{\II: |\II| = p} \sum_{i,j \in \II} X_{\II} X_i X_j \cr
&= \sum_{\II: |\II| = p} X_{\II} \sum_{i \in \NN \setminus \II} X_i^2
	+ \sum_{\II: |\II| = p-2} X_{\II} \sum_{i,j \in \NN \setminus \II} X_i^2 X_j^2 \cr
&= (N-p) N^p \G_{p,N} + {{N - p} \choose 2} N^{p-2} \G_{p-2, N},
}
\Eq(RO.16)
$$
and hence
$$
 N^{-p-2} \sum_{\II: |\II| = p} X_{\II} 
	\sum_{{\JJ: |\JJ| = 2} \atop {\JJ \cap \II \neq \emptyset}} X_{\JJ}
= N^{-1} \G_{p,N}(1 + {\cal O}(N^{-1})) + N^{-2} \G_{p-2,N}(1 + \cal {O}(N^{-1})).
\Eq(RO.17)
$$
Applying the induction hypothesis to \eqv(RO.17) shows the decomposition \eqv(RO.12).
Positivity of $c_{p,p}$ follows from \eqv(RO.14).
\endproof
\specskip
 From this one concludes that uniformly in $\s$, $\xi$, and for all $N$ large 
enough,
$$
- H^\mu \geq -c.
\Eq(RO.17bis)
$$ 
Indeed, by the preceding result (setting $X_i = \xi^\mu_i \s_i$),
$$
\eqalign{
- H^\mu(\s) 
&= \frac{(p!)^{1/2}}{N^p} N \sum_{\II: |\II| = p} \xi^\mu_\II \s_\II \cr
&= (p!)^{1/2} N \sum_{q = 0}^{\frac{p}{2}} 
	\left(\frac{1}{N} \sum_{i=1}^N \xi^\mu_i \s_i\right)^{2q}N^{q - \frac{p}{2}}
		(1 + {\cal O}\sum_{\II: |\II| = p}(N^{-1})) \cr
&= N \sum_{q = 0}^{\frac{p}{2}} c_{p,2q} (m^\mu)^{2q} N^{q-\frac{p}{2}}
		(1 + {\cal O}(N^{-1})).
}
\Eq(RO.18)
$$
We distinguish two cases. If $m^\mu$ is large, we show that $-H^\mu(\s)$ is positive.
Suppose therefore that $|m^\mu(\s)| > N^{-1/2 + \d}$ for some $\d>0$. Then, 
$$
\eqalign{
- N^{-1} H^\mu(\s)
&\geq c_{p,p} (m^\mu)^p -  \sum_{q=0}^{p/2 - 1} |c_{p,q}| m^{\mu}{}^{2q} N^{q - \frac{p}{2}}
			(1 + {\cal O}(N^{-1})) \cr
&\geq c'_{p,p} N^{ - \frac{p}{2} + p \d}  - \sum_{q=0}^{\frac{p}{2}-1}c'_{p,q} 
			N^{-\frac{p}{2} + 2 q \d}  \cr
&\geq N^{-\frac{p}{2} + p \d}(c'_{p,p} - c''\sum_{q=0}^{\frac{p}{2} - 1} N^{\d(2q - p)}),\cr
}
\Eq(RO.18.1)
$$
which is obviously positive for all $N$ large enough and $\d$ less than $\frac{1}{2}$.

On the other hand, if $m^\mu$ is less than $N^{-1/2 + \d}$, then,
$$
| N^{-1} H^\mu(\s) | 
\leq \sum_{q=0}^{\frac{p}{2}} c'_{p,q} N^{2q(\d - \frac{1}{2})} N^{q - \frac{p}{2}} 
= \sum_{q=0}^{\frac{p}{2}} c'_{p,q} N^{-\frac{p}{2}  + p\d}.
\Eq(RO.19)
$$
Thus, if $\d < \frac{1}{2} - \frac{1}{p}$, then $|H^\mu| = o(1)$, so that the bound
\eqv(RO.17bis) is in fact a gross underestimate.

To prove Lemma~\thv(ro.2), we observe that
$$
\eqalign{
\bar{Z}^\mu_u
&= \E_\s[ e^{-\b\bar{H}^\mu_0(\s) - \b u H^\mu(\s)}] 
		\geq \E_{\s}[ e^{-\b \bar{H}^\mu_0(\s) - \b u \sup H^\mu}] 
\geq \E_{\s}[ e^{-\b \bar{H}^\mu_0(\s) - \b u \d}]
		\geq c_\b \bar{Z}^\mu_0.
}
\Eq(RO.20)
$$
This proves the \eqv(RO.11).
\endproof
\specskip
We apply this result to the error terms in the development \eqv(RO.4). We start with 
$R_1$. By Jensen's inequality,
$$
|R_1| 
= \left| \E_\s\left[\frac{e^{-\b\bar{H}^\mu_u}}{\bar{Z}^\mu_u} 
			H^\mu {}^3\right]\right|
= \left| \bar{\GG}^\mu_u [ H^\mu{}^3]\right| \leq \bar{\GG}^\mu_u [ |H^\mu|^3] 
= \E_\s\left[\frac{e^{-\b\bar{H}^\mu_u}}{\bar{Z}^\mu_u} 
			|H^\mu |^3\right].
\Eq(RO.21)
$$
Since the integrand is a positive function, we may bound the expectation using 
Lemma~\thv(ro.3) in the denominator. We obtain, noting that 
$\bar{H}^\mu_u = \bar{H}^\mu_0 + u H^\mu$,
$$
|R_1| \leq c \E_\s\left[ \frac{e^{-\b\bar{H}^\mu_u}}{\bar{Z}^\mu_0} 
			| H^\mu|^3\right]
= c \, \bar{\GG}^\mu_0[e^{-\b u H^\mu} |H^\mu|^3].
\Eq(RO.22)
$$
We observe that the last Gibbs measure does not depend on the pattern $\mu$. We 
may therefore integrate with respect to $\{\xi^\mu_i\}_i$ ``inside''. In complete
analogy with Chapter~3 (the result about the error term), we get
$$
\E_\mu[e^{-\b u H^\mu} |H^\mu|^3]
\leq \E_\mu[e^{\b u |H^\mu|} |H^\mu|^3]
\leq c N^{3 - \frac{3p}{2}},
\Eq(RO.23)
$$
whenever $\b u < \b_p'$. Since $u \in [0,1]$, this condition is satisfied if $\b < \b_p'$.

The remainder $R_3$ gets essentially the same treatment. By Jensen's inequality,
$$
|R_3| = \left| \bar{\GG}^\mu_u{}^{\otimes 3}[H^\mu(\s)H^\mu(\s') H^\mu(\s'')]\right|
= \left| \bar{\GG}^\mu_u [H^\mu(\s)] \right|^3 \leq \bar{\GG}^\mu_u[|H^\mu|^3] = |R_1|.
\Eq(RO.24)
$$
Hence, 
$$
\E\,|R_3| \leq c N^{3 - \frac{3p}{2}}.
$$
Finally, the term $R_2$. By Lemma~\thv(ro.2),
$$
\eqalign{
|R_2| 
&= |\bar{\GG}^\mu_u[H^\mu{}^2]\, \bar{\GG}^\mu_u[H^\mu]|
	\leq \bar{\GG}^\mu_u[H^\mu{}^2] \, \bar{\GG}^\mu_u[|H^\mu|] \cr
&\leq c \, \bar{\GG}^\mu_0[e^{-\b u H^\mu}H^\mu{}^2]\, \bar{\GG}^\mu_0[e^{-\b u H^\mu}|H^\mu|].
}
\Eq(RO.25)
$$
Thus, by Cauchy-Schwarz and Jensen,
$$
\eqalign{
\E\,|R_2|
&\leq \left| \E\,\left[\bar{\GG}^\mu_0[e^{-\b u H^\mu} H^\mu{}^2] \,
		\bar{\GG}^\mu_0[e^{-\b u H^\mu}|H^\mu|]\right]\right| \cr
&\leq \left( \E\,\left[(\bar{\GG}^\mu_0[e^{-\b u H^\mu}H^\mu{}^2])^2\right]\right)^{\frac{1}{2}}
      \left( \E\,\left[(\bar{\GG}^\mu_0
		[e^{- \b u H^\mu}H^\mu])^2\right]\right)^{\frac{1}{2}} \cr
&\leq \left( \E\,\left[\bar{\GG}^\mu_0[e^{-2 \b u H^\mu}H^\mu{}^4]\right]\right)^{\frac{1}{2}}
	\left( \E\,\left[\bar{\GG}^\mu_0
		[e^{- 2 \b u H^\mu}H^\mu{}^2]\right]\right)^{\frac{1}{2}}.
}
\Eq(RO.26)
$$
Both factors are now treated as $R_1$. Since the integrability of $R_1$ did 
not depend on the power of $H^\mu$, but merely on the exponential factor (this is apparent
from the estimate \eqv(A.24)), we get that whenever $2\b u < \b_p'$,
$$
\E\,|R_2| \leq (c N^{4 - 2p})^{\frac{1}{2}}(c N^{2 - p})^{\frac{1}{2}} = c' N^{3-\frac{3p}{2}}.
\Eq(RO.27)
$$
The above condition is always satisfied if $\b < \frac{1}{2}\b_p'$.

The results above almost prove the theorem. What remains to show is that in the leading
terms, we can replace without harm the Gibbs measure $\bar{\GG}^\mu_0$ by $\GG$. 
More precisely, we claim that
$$
\left| \E\,\bar{\GG}^\mu_0{}^{\otimes 2}[R^p] - \E\, \GG^{\otimes 2}[R^p]\right| 
	\leq c N^{1 - \frac{p}{2}},
\Eq(RO.28)
$$
for some constant $c$.

The proof of this claim is done exactly as before, namely by expanding the Boltzmann
factors, this time, however, only to zero order. We get
$$
\GG^{\otimes 2}[R^p] = \bar{\GG}^\mu_0{}^{\otimes 2}[R^p]
	+ \bar{\GG}^\mu_u{}^{\otimes 2}[R(\s,\s')^p (H^\mu(\s) + H^\mu(\s'))]
	+ \bar{\GG}^\mu_u{}^{\otimes 3}[R(\s,\s')^p H^\mu(\s'')].
\Eq(RO.29))
$$
Since $R^p \in [0,1]$, the second term on the right is bounded by
$$
|\bar{\GG}^\mu_u{}^{\otimes 2}[R(\s,\s')^p (H^\mu(\s) + H^\mu(\s'))]|
= 2 | \bar{\GG}^\mu_u{}^{\otimes 2}[R(\s,\s')^p H^\mu(\s)]|
\leq 2 \bar{\GG}^\mu_u[|H(\s)|].
\Eq(RO.30)
$$
Proceding as above we get,
$$
|\E\,\bar{\GG}^\mu_u{}^{\otimes 2}[R(\s,\s')^p (H^\mu(\s) + H^\mu(\s'))]|
\leq 2 \E\,\bar{\GG}^\mu_u[|H(\s)|] 
\leq 2 c \E,\bar{\GG}^\mu_0[ e^{-\b u H^\mu} |H^\mu|]
\leq 2 c' N^{1 - \frac{p}{2}}.
\Eq(RO.31)
$$
The third term on the right of \eqv(RO.29) is bounded by the same order. Indeed,
$$
|\bar{\GG}^\mu_u{}^{\otimes 3}[R(\s,\s')^p H^\mu(\s'')] |
\leq \bar{\GG}^\mu_u[|H^\mu(\s)|],
\Eq(RO.32)
$$
from which the bound follows again by integration. This proves the claim \eqv(RO.28).

To finish the proof of the Theorem, we sum the contributions we have obtained.
Relation \eqv(RO.1) implies that
$$
\eqalign{
\left| \b \E\,\frac{\partial F_N}{\partial \b} 
	- \a \b^2(1 - \E\,{\GG}^{\otimes 2}[R^p]) \right|
&= \b \left| - \frac{1}{N} \E\,\GG[H] - \sum_{\mu = 1}^{M(N)} \b N^{1-p} 
		+ \sum_{\mu = 1}^{M(N)} \b N^{1-p} {\GG}^{\otimes 2}[R^p]) \right| \cr
& \leq \b \left|- \frac{1}{N} \E\,\GG[H] - \sum_{\mu = 1}^{M(N)} \b N^{1-p} 
	+ \sum_{\mu=1}^{M(N)} \b N^{1-p} \E\,\bar{\GG}^{\mu}_0{}^{\otimes 2}
			[R^p] \right| \cr
&\quad + \b^2 \sum_{\mu=1}^{M(N)} N^{1-p} \left| (\E\,\bar{\GG}^{\otimes 2}[R^p]
			- \E\,\bar{\GG}^\mu_0{}^{\otimes 2} [R^p])\right|
}
\Eq(RO.33)
$$
Using the decomposition \eqv(RO.4), and the results \eqv(RO.5), \eqv(RO.6) and \eqv(RO.9)
in the first term, and the bound \eqv(RO.29) in the second, we get
$$
\eqalign{
\left|\b \E\,\frac{\partial F}{\partial \b} 
	- \a \b^2(1 - \E\,{\GG}^{\otimes 2}[R^p]) \right| 
&\leq c \b \left| \sum_{\mu=1}^{M(N)} (\b N^{1-p} {\cal O}(N^{-1}) + R^\mu_1 
		+ R^\mu_2 + R^\mu_3)\right| \cr
&\quad + c' \b^2 \sum_{\mu=1}^{M(N)} N^{1-p}N^{1 - \frac{p}{2}}.
}
\Eq(RO.34)
$$
We finally insert the bounds \eqv(RO.23), \eqv(RO.24) and \eqv(RO.27) on the errors 
$R_i$, which are valid if $\b < \frac{1}{2} \b_p'$. This yields
$$
\left|\b \E\,\frac{\partial F}{\partial \b} 
	- \a \b^2(1 - \E\,{\GG}^{\otimes 2}[R^p]) \right| 
\leq c_\b N^{-1} + c_\b' N^{2 - \frac{p}{2}} \leq C_\b.
\Eq(RO.35)
$$
This proves Theorem~\thv(pr.5).
\endproof

\bigskip

\subchap{\ver.2.~Condensation: Proof of Theorem~\thv(pr.6).}

\noindent
Theorem~\thv(pr.6) follows now just as the analogous result in [T3] 
from the convexity of the free energy. 
Suppose that $\b < \b_p$. Since we always assume that $\a\geq \a_p$, then 
$$
\limsup_{N \uparrow \infty} \E\,F_N = \frac{\a \b^2}{2}
\Eq(RO.57)
$$
by the definition of $\b_p$. As remarked after their definition in
Chapter~2,  
$\E\,F_N$ is convex for all $N$. It then follows from a standard
result in  
convex analysis ([Ro], Theorem~25.7) that
$$
\limsup_{N \uparrow \infty} \E\,\frac{\partial F_N}{\partial \b}
= \frac{\partial}{\partial \b} \limsup_{N \uparrow \infty} \E\,F_N 
= \a \b.
\EQ(RO.58)
$$
Hence, from Theorem~\thv(pr.5), 
$$
\E\,\GG^{\otimes 2} [R^p] + \E\,\frac{\partial F_N}{\partial \b} = \a
\b + 
                \OO(N^{-1}),
\Eq(RO.59)
$$
and thus, passing to the limit,
$$  
\limsup_{N \uparrow \infty} \E\,\GG^{\otimes 2}[R^p] + \a \b = \a \b,
\Eq(RO.60)
$$
which in turn implies that 
$$
\limsup_{N \uparrow \infty}\E\,\GG_N^{\otimes 2}[R^p] = 0.
\Eq(RO.61)
$$ 
Suppose now that 
$$
\limsup_{N \uparrow \infty} \E\,\frac{\partial F_N}{\partial \b}  < \a
\b. 
\Eq(RO.62)
$$
Then it follows immediately from Theorem~\thv(pr.5) that
$$
\liminf_{N \uparrow \infty} \E\,\GG^{\otimes 2}[R^p] 
        = \a \b - \limsup_{N \uparrow \infty} \E\,\frac{\partial
F_N}{\partial \b} 
        > \a \b - \a \b = 0.
\Eq(RO.63)
$$
This proves \eqv(PR.22). To see where the condition \eqv(RO.62)
actually holds, we  
observe first that by Lemma~\thv(ub.1), it is satisfied for all 
$$
\frac 12\b_p'\b > \hat{\b}_p = \sqrt{\frac{2 \ln 2}{\a}}.
\Eq(RO.64)
$$
This concludes the proof of the Theorem. \endproof

\remark Of course one would expect \eqv(RO.62) starts to hold right after the 
critical temperature. In fact, a weak version of this can be proven.
Namely, Theorem~5.5 in [Ro] implies that the function 
$$
f(\b) = \limsup_{N \uparrow \infty} \E\,F_N
\Eq(RO.65)
$$
is a convex, bounded function on $\UU = [0,\b'_p)$. By Theorem~25.3 in
[Ro] it 
is thus differentiable on an open set $\DD \subset \UU$ which contains
all but  
perhaps countably many points of $\UU$, and its derivative $f'$ is
bounded on  
$\DD$. Lebesgue's integrability criterion   
then implies that
$$
f(\b) = f(\b_p) + \intl_{\b_p}^{\b} f'(u) du, \quad \forall \b > \b_p.
\Eq(RO.66)
$$
Now it is immediate that for all $\b > \b_p$ there must exist a set $I
\subset (\b_p, \b)$  
with strictly positive Lebesgue measure, on which $f'$ is strictly
less than $\a \b$.  
Indeed, were this not the case, then $f \geq \frac{\a \b^2}{2}$, 
which contradicts the definition of $\b_p$. 
Since $\b$ was arbitrary, the relevant condition \eqv(RO.62) is
satisfied on sets of positive 
Lebesgue measure arbitrarily close to $\b_p$.

\specskip
\subchap{ \ver.3~Proof of Theorem \thv(pr.9).}

\noindent
We have shown that in the low temperature phase, the replica overlap is
not concentrated on zero. We will now show that its distribution is 
concentrated on a neighborhood of zero and 1.

\proofof {Theorem \thv(pr.9)}
Let $C_N^+, C_N^-$ be such that 
$$
\P\left[\sup_\s (-H_N(\s))\not\in [NC_N^-,NC_N^+]\right]=p_N= o(1)
\Eq(9.1)
$$
Then 
$$
\eqalign{
\E\GG^{\otimes 2}_N(R_N(\s,\s')\in I)
&\leq\E\1_{\sup_{\s} |H_N(\s)|\leq NC_N }
\frac {\E_{\s,\s'}e^{-\b(H_N(\s)+H_N(\s'))}\1_{R_N(\s,s')\in I}}
{ 2^{-2N}e^{N2C_N^-}} +p_N
\cr
&=
\frac {\E\E_{\s,\s'}e^{-\b(H_N(\s)+H_N(\s'))}\1_{-\b(
H_N(\s)+H(\s'))\leq  
NC_N^+\b}\1_{R_N(\s,\s')\in I}}
{ 2^{-2N}e^{\b N2C_N^-}} +p_N
}\Eq(9.2)
$$
The numerator has been estimated in \eqv(C.30). 
Using this, we get 
$$
\eqalign{
\E\GG^{\otimes 2}_N&(R_N(\s,\s')\in I) \leq 
\sum_{t\in I} C_3 \frac{ \exp \bigg( N \bigg[ 2 \b C_N^+
                \Big(1 - \frac{C_N^+}{2\a\b(1+ t^p)}\Big) - I(t)  
                \bigg] \bigg)}
{e^{\b N2C_N^--2\ln 2}}+p_N \cr
&=\sum_{t\in I} C_3 
 \exp \bigg( N 2 \b (C_N^+-C_N^-)+
               N \Big(2\ln 2 - \frac{(C_N^+)^2}{\a(1+ t^p)} -
                I(t)\big) \bigg)+p_N
}
\Eq(9.3)
$$
Let us note first that from \eqv(9.3) it is obvious that if
we can choose $|C_N^+-C_N^-|\leq N^{-\e}$, then the result cannot
depend on  
$\b$. An obvious candidate for these numbers is thus
$N^{-1}\E  \sup_{\s}(-H_N(\s))\pm\e$. Indeed we have

\lemma{\TH(gap.1)}{\it For any $\e>0$, and for all $N$ large enough,    
$$
\P\left[  |\frac 1N\sup_{\s}(-H_N(\s))-\E  \frac
1N\sup_{\s}(-H_N(\s))|> \e\right] 
\leq  N^{-2}
\Eq(9.3.1)
$$
}

\proof Note first that 
$$
2^{-N}\leq Z_N(\b)e^{\b\sup_\s (-H_N(\s))}\leq 1 
\Eq(9.3.2)
$$
and therefore 
$$
|\frac 1\b  F_N(\b) -\frac 1N\sup_\s (-H_N(\s))|\leq \frac {\ln 2}\b
\Eq(9.3.3)
$$
Therefore, for any $\b<\infty$,
$$
\eqalign{
&\left| \frac 1N\sup_{\s}(-H_N(\s))-\E  \frac
1N\sup_{\s}(-H_N(\s))\right|\cr 
&=
\left| \frac 1N\sup_{\s}(-H_N(\s))-\frac 1\b  F_N(\b) +\frac 1\b  F_N(\b)
 -\E  \frac 1N\sup_{\s}(-H_N(\s))+\E \frac 1\b  F_N(\b) -\E\frac 1\b
 F_N(\b)\right| 
\cr&\leq
\left|\frac 1\b  F_N(\b)-\E \frac 1\b  F_N(\b)\right|
+\frac {2\ln 2}\b
}
\Eq(9.3.4)
$$
By Proposition 6.2, 
$$
\P\left[\left|\frac 1\b  F_N(\b)-\E \frac 1\b  F_N(\b)\right|> N^{-1/2+\e}
\right] \leq C N^{-n} 
\Eq(9.3.5)
$$
from which the claimed result follows by choosing e.g. $\b=\e^{-1}4\ln 2$.  
\endproof

Using this result, and setting $C_N\equiv \E  \frac
1N\sup_{\s}(-H_N(\s))$, we get  
that 
$$
\eqalign{
\E\GG^{\otimes 2}_N(R_N(\s,\s')\in I)& \leq 
\sum_{t\in I} C_3 
 \exp \bigg( N 4 \b\e +
               N \Big(2\ln 2 - \frac{(C_N+\e)^2}{\a(1+ t^p)} - I(t)\big)  
                \bigg)
}
\Eq(9.3.6)
$$
Since $\e$ can be chosen as small as we like, e.g. $\d\b^{-1}$,
we already see that our result will be uniform in $\b$.

It remains to estimate $\E  \frac 1N\sup_{\s}(-H_N(\s))$. 
 We will only consider the case
$\a>\frac{ \ln 2}{2 p!}$. In that case it follows from Lemma 3.4 that
 $C_N\leq \sqrt {2\a\ln 2}+C/N$ from a bound completely analogous to 
\eqv(REM.5). For a lower bound,
 note that 
for any $\b$,
$$
\E\frac{\del}{\del\b}F_N(\b)=\frac 1N\E\GG_N(-H_N(\s))\leq \E\frac
1N\sup_{\s}(-H_N(\s) 
\Eq(9.4)
$$
But we know that for all $\b\leq \b'_p$, 
$\lim_{N\uparrow\infty}\E F_N(\b) =\frac{\a\b^2}2$, and therefore by 
standard results
$\lim_{N\uparrow\infty}\E \frac {\del}{\del\b} F_N(\b)
=\a\b$.
Thus chosing $\b$ as large as possible we see that we see that 
$$
C_N\geq \a\b'_p -\d_N
\Eq(9.6)
$$
where  $\d_N\downarrow 0 $, as $N\uparrow\infty$. 
But  Theorem 1.2 and the estimate \eqv(PR.18)
show that 
$$
C_N\geq \sqrt{2\a\ln 2} - \frac{ 2^{-p-1}\sqrt \a }{\sqrt {2\ln 2}} -\d_N
\Eq(9.7)
$$
Therefore we have that for any $\d>0$, and for $p$ large,
$$
\eqalign{
\E\GG^{\otimes 2}_N(R_N(\s,\s')\in I)
&\leq\sum_{t\in I} C_3 
 \exp \bigg( N\big( \d+ \frac{ 2^{-p}+2\d_N/\sqrt \a +O(2^{-2p})}{1+t^p}\big)+
       \cr&        N \Big(2\ln 2 - \frac{2\ln 2}{1+ t^p} - I(t)\Big)  
                \bigg)+p_N
\cr
&\leq\sum_{t\in I} C_3 
 \exp \bigg( N (\d+ 2^{-p})+
               N \Big(\frac{ 2\ln 2 t^p}{1+ t^p} - I(t)\Big)  
                \bigg)+p_N
}
\Eq(9.8)
$$
The function $\frac{2\ln 2 t^p}{(1+ t^p)} - I(t)$ vanishes at 
zero and at one, and is negative everywhere in the interval $(0,1-z_p)$,
where $z_p\sim 2^{-p}$. 
This implies the main conclusion of Theorem \thv(pr.9), \eqv(9.1). 
Note that since $I(t)\sim t^2$ for small $t$, we can chose the interval
$I$ more precisely of the form $I_p=(C 2^{-p/2}, 1-C 2^p)
$, with $C$ a constant of order $1$.

To proof the estimate \eqv(9.01) in the high-temperature case is
considerably simpler. Since we already have the estimate $\E
T(c,b,1)\leq e^{\a\b^2N-dN/2}$ for some positive $d$, it remains to
show that with sufficently large probability, $Z_{N}^2(\b)\geq
e^{\a\b^2N-dN/4}$. To do so, we use the Paley-Zygmund inequality
\eqv(C.41):
$$
\eqalign{
\P\left[Z_N\geq e^{-dN/8}\E \tilde Z_N\right]\geq
\P\left[\tilde Z_N\geq e^{-dN/8}\E \tilde Z_N\right]\geq (1- e^{-dN/4})C_3
}
\Eq(C.41.1)
$$ 
Given that by Lemma \thv(c.2) and Theorem \thv(pr.1) $\E \tilde
Z_N\geq C e^{N\a\b^2/2}$, \eqv(9.01) follows immediately. This
completes the proof of Theorem \thv(pr.9).
\endproof
\specskip

\subchap{ \ver.4.~Ghirlanda-Guerra identities and lump masses.}

\noindent The techniques used to prove Theorem \thv(pr.5) can also be 
used to derive 
the Ghirlanda-Guerra identities [GG] (see also [AC])    that provide 
relations between distributions of overlaps of a larger number of
replicas. This observation is due to Talagrand [T5]. Note 
that he annopunced  more far-reaching results than those we will prove here.

The basic input is the following slight generalization of Theorem
\thv(pr.5). 

\proposition {\TH(GG.1)} {\it Assume that $\b\leq \frac 12 \b_p'$.
Let $f$ denote any bounded function of 
$n$ spins. Then, for any $k\in \{1,\dots,n\}$,
$$
\eqalign{
&\Bigl|\E \GG_{N,\b}^{\otimes n} \left(N^{-1} 
H_N (\s^{k}) f(\s^1,\dots,\s^n)\right)
\cr
&\quad -\a\b\E \GG_{N,\b}^{\otimes n+1} \left(f(\s^1,\dots,\s^n)\sum_{l
=1}^nR_N^p(\s^k,\s^l)-
n R_N^p(\s^k,\s^{n+1})\right)\Bigr|
\leq C N^{-1}
}\Eq(10.1)
$$
}

\proof The proof of this proposition is an exact rerun of the
inequalities  
\eqv(RO.34), except for the computation of the leading
terms which is however straightforward. We will not repeat the details.
\endproof

As in [GG] it then follows from the concentration result Theorem
\thv(pr.4) and standard arguments that for any bounded function $f$,
$$
\lim_{N\uparrow\infty} \int_{\b'}^{b''}d\b \left|\E
\GG_{N,\b}^{\otimes n} \left( N^{-1} H_N (\s^{k}) f(\s^1,\dots,\s^n) 
\right) -\E \GG_{N,\b} \left( N^{-1} H_N (\s)\right)
\E \GG_{N,\b}^{\otimes n} \left( f(\s^1,\dots,\s^n)
\right) \right| =0
\Eq(10.2)
$$
for any $\b'<\b''$. Combining \eqv(10.1) and \eqv(10.2) with the bounds 
\eqv(10.1), we arrive at the 
identity
$$
\eqalign{
&\lim_{N\uparrow\infty} \int_{\b'}^{b''}d\b
\E \GG_{N,\b}^{\otimes n+1} \left[ f(\s^1,\dots,\s^n)
\left(\sum_{l\neq k}^nR_N^p(\s^k,\s^l)-
n R_N^p(\s^k,\s^{n+1})+ \E \GG_{N,\b}^{\otimes
2}\left(R_N^p(\s^1,\s^2)\right) 
\right)\right]\cr
&\quad=0
}
\Eq(10.3)
$$
which is the analogue of (16) of [GG]. Note that this can be written  
as
$$
\eqalign{
&\lim_{N\uparrow\infty} \int_{\b'}^{b''}d\b
\E \GG_{N,\b}^{\otimes n+1} \left[ f(\s^1,\dots,\s^n)
 R_N^p(\s^k,\s^{n+1}) \right]
\cr
&=\frac 1n\lim_{N\uparrow\infty} \int_{\b'}^{b''}d\b\E
\GG_{N,\b}^{\otimes n}  
\left[ f(\s^1,\dots,\s^n)
\left(\sum_{l\neq k}^nR_N^p(\s^k,\s^l)+ \E \GG_{N,\b}^{\otimes
2}\left(R_N^p(\s^1,\s^2)\right) 
\right)\right]
}
\Eq(10.4)
$$ 
and choosing $f$ to be the indicator function 
$$
f(\s^1,\dots,\s^n)=\1_{ \forall_{k\neq l} R_N(\s^k,\s^l)=q_{kl}}
\Eq(10.5)
$$
This implies
that 
$$
\eqalign{
&\lim_{N\uparrow\infty} \int_{\b'}^{b''}d\b
\E \GG_{N,\b}^{\otimes n+1} \left[
 R_N^p(\s^k,\s^{n+1}) \big | \forall_{k\neq l}
 R_N(\s^k,\s_l)=q_{kl}\right] 
\cr
&=\frac 1n\sum_{l\neq k}^n q_{kl}^p+ \frac 1n 
\lim_{N\uparrow\infty} \int_{\b'}^{b''}d\b
\E \GG_{N,\b}^{\otimes 2}\left(R_N^p(\s^1,\s^2)\right)
}
\Eq(10.6)
$$ 
which is the relation (17) of [GG]. 

\remark While [GG] claim to obtain the same relations also for all other
moments of the replica overlaps, it needs to be said that they tacitly assume
the continuity of the Gibbs measures with respect to certain random 
perturbations of the Hamiltonian that is not only not proven but
is certain to be false in the generality they are announced. Otherwise, the 
argument below could be considerably sharpened and simplified.

The main use of the identities \thv(10.6) is that they allow to draw
conclusions about the distribution of the masses of the Gibbs measures on
the so-called   '
`Talagrand-lumps'.

\proofof {Theorem \thv(L.2)} The starting point of the argument is
that Theorem \thv(pr.5) 
together with Theorem \thv(pr.7) in fact imply that the distribution
of the replica  overlaps has positive mass both near zero and near
one. Let us set 
$$
\eqalign{
p_0
&\equiv
\E \GG_N^{\otimes 2}\left(|R_N(\cdot,\cdot)|\leq
\e_0\right)\cr
p_1
 &\equiv
\E \GG_N^{\otimes 2}\left(|R_N(\cdot,\cdot)|\geq
1-\e_1\right)
}
\Eq(11.3)
$$
Since by convexity  (see \eqv(RO.58))  for all $\b\geq \b_p$, except
possibly for a countable number of exceptional points 
$$
\a\b_p\leq \lim\inf_{N}\E\frac{\del}{\del\b}F_N(\b)\leq 
 \lim\inf_{N}\E\frac{\del}{\del\b}F_N(\b)\leq \sqrt{\a 2\ln 2}\Eq(11.4)
$$
we have on the one hand that 
$$
\limsup_N p_0\leq \frac {\sqrt {\a 2\ln 2}}{\a\b(1-\e_0)}
\Eq(11.5)
$$
and 
$$
\limsup_N p_1\leq \frac {\b-\b_p}{\b(1-\e_1)^p}
\Eq(11.6)
$$
Since we know that $\lim_N(p_0+p_1)=1$, and  this implies what we want
for $\b$ somewhat larger than $\b_p$. Recall  that $\e_0\sim 2^{-p/2}$ 
and $\e_1\sim 2^{-p}$. 

This result shows first of all that it is not possible that the mass
of one single (pair of) lump(s) can be almost equal to one, since in
that case $p_0$ would be close to zero (which is impossible by
\eqv(11.6)).

Now assume that the assertion of Theorem \thv(L.2)  
fails. Then there exists a first instance 
$k^*$
such that 
$$
\lim_{N\uparrow \infty} \E\GG_N\left(\cup_{l=1}^{k^*} \CC_l\right) =1
\Eq(10.7)
$$
Now define  events 
$
\QQ_{\e_0}^{(n)}\in\BB_n$ by 
$$
\QQ_\e^{(n)}\equiv \left\{\un R\in [-1,1]^{n(n-1)/2}|\forall_{1\leq l<k\leq n}
|R_{lk}|\leq \e_0\right\}
\Eq(10.7.1)
$$
The important obervation is that if $\{R_N(\s_l,\s_k)\}_{1\leq l<k\leq k^*}\in
\QQ_\e^{(k^*)}$, then there exists some permutation $\pi\in S_{k^*}$ such that
with probability one $\s^k\in C_{\pi(k)}$ for all $k\leq k^*$. In particular
$$
\eqalign{
& \lim_{N\uparrow\infty} \int_{\b'}^{b''}d\b
\E \GG_{N,\b}^{\otimes k^*+1} \left[
 R_N^p(\s^k,\s^{k^*+1}) \1_{
 \{R_N(\s^l,\s_m)\}_{1\leq l<\leq k^*}\in \QQ_{\e_0}^{(k^*)}}\right] 
\cr
&= \lim_{N\uparrow\infty} \int_{\b'}^{b''}d\b
\E \GG_{N,\b}^{\otimes k^*+1} \left[
 R_N^p(\s^k,\s^{k^*+1}) \1_{
 \exists_\pi \forall_{l=1}^{k^*}\s^l\in \CC_{\pi(l)}}   \right] 
}
\Eq(10.7.2)
$$
But 
$$
\eqalign{
\E &\GG_{N,\b}^{\otimes k^*+1} \left[
 R_N^p(\s^k,\s^{k^*+1}) \1_{
 \exists_\pi \forall_{l=1}^{k^*} \s^l\in \CC_{\pi(l)}}\right]
=\sum_{\pi\in S_{k^*}}
\E \GG_{N,\b}^{\otimes k^*+1} \left[
 R_N^p(\s^k,\s^{k^*+1}) \1_{ \forall_{l=1}^{k^*} \s^l\in
 \CC_{\pi(l)}}\right] 
\cr&
=\sum_{\pi\in S_{k^*}}\sum_{j=1}^{k^*}
\E \GG_{N,\b}^{\otimes k^*+1} \left[
 R_N^p(\s^k,\s^{k^*+1}) \1_{\s_{k^*+1}\in \CC_{\pi(j)}}
\1_{  \forall_{l=1}^{k^*}\s^l\in \CC_{\pi(l)}}\right]
\cr
&=\sum_{\pi\in S_{k^*}}\sum_{j\neq k}^{k^*}
\E \GG_{N,\b}^{\otimes k^*+1} \left[
 R_N^p(\s^k,\s^{j}) \1_{\s_{k^*+1}\in \CC_{\pi(j)}}
\1_{  \forall_{l=1}^{k^*}\s^l\in \CC_{\pi(l)}}\right]\cr
&+\sum_{\pi\in S_{k^*}}
\E \GG_{N,\b}^{\otimes k^*+1} \left[
 R_N^p(\s^k,\s^{k^*+1}) \1_{\s_{k^*+1}\in \CC_{\pi(k)}}
\1_{  \forall_{l=1}^{k^*}\s^l\in \CC_{\pi(l)}}\right]
}
\Eq(10.7.3)
$$
where we used the symmetry betwen replicas in the terms $j\neq k$ to 
exchange $\s^{k^*+1}$ with $\s^j$. 
Note that for the first term we have the obvious (though not very good) bound
$$
\eqalign{
0\leq&\sum_{\pi\in S_{k^*}}\sum_{j\neq k}^{k^*}
\E \GG_{N,\b}^{\otimes k^*+1} \left[
 R_N^p(\s^k,\s^{j}) \1_{\s_{k^*+1}\in \CC_{\pi(j)}}
\1_{  \forall_{l=1}^{k^*}\s^l\in \CC_{\pi(l)}}\right]\cr
&\leq \e_0^p \E \GG_{N,\b}^{\otimes k^*} 
\left[\1_{\forall_{l=1}^{k^*}\s^l\in \CC_{\pi(l)}}\right]\cr
&= \e_0^p \E \GG_{N,\b}^{\otimes k^*} \left[\QQ_\e^{k^*}\right]\cr
}
\Eq(10.7.4)
$$
while the second satisfies
$$
\eqalign{\sum_{\pi\in S_{k^*}}
&\E \GG_{N,\b}^{\otimes k^*+1} \left[
 R_N^p(\s^k,\s^{k^*+1}) \1_{\s_{k^*+1}\in \CC_{\pi(k)}}
\1_{  \forall_{l=1}^{k^*}\s^l\in \CC_{\pi(l)}}\right]
\cr&\geq (1-\e)^p
\sum_{\pi\in S_{k^*}}\E \GG_{N,\b}^{\otimes k^*+1} 
\left[ \1_{\s_{k^*+1}\in \CC_{\pi(k)}}
\1_{  \forall_{l=1}^{k^*}\s^l\in \CC_{\pi(l)}}\right]
\cr
&=\frac 1{k^*}(1-\e_1)^p \E \GG_{N,\b}^{\otimes k^*+1} 
\left[\1_{  \forall_{l=1}^{k^*}\s^l\in \CC_{\pi(l)}}\right]
\cr&=\frac 1{k^*}(1-\e_1)^p \E \GG_{N,\b}^{\otimes k^*} 
\left[\QQ_{\e_0}^{k^*}\right]\cr
}
\Eq(10.7.5)
$$
where we used the obvious permutation symmetry among the first $k^*$ replicas.
Let us now use \eqv(10.4) with $f$ the indicator function of the event 
$\QQ^{(k^*)}_{\e_0}$. clearly we get 
$$
\eqalign{
\lim_{N\uparrow\infty}& \int_{\b'}^{b''}d\b
\E \GG_{N,\b}^{\otimes k^*+1} \left[
 R_N^p(\s^k,\s^{k^*+1}) \1_{
 \{R_N(\s^l,\s_m)\}_{1\leq l<\leq k^*}\in \QQ_{\e_0}^{(k^*)}}\right] \cr
&\leq \frac 1{k^*} \lim_{N\uparrow\infty} \int_{\b'}^{b''}d\b
\E \GG_{N,\b}^{\otimes k^*+1} \left[
  \1_{\{R_N(\s^l,\s_m)\}_{1\leq l<\leq k^*}\in \QQ_{\e_0}^{(k^*)}}\right]
\left((k^*-1)\e_0^p+ \E \GG_{N,\b}^{\otimes 2}R^p(\s,\s')\right)
}\Eq(10.7.6)
$$
Comparing \eqv(10.7.4), \eqv(10.7.5) to \eqv(10.7.6) we see that  
$$
(1-\e_1)^p\leq 
(k^*-1)\e_0^p+ \lim_{N\uparrow\infty}
\E \GG_{N,\b}^{\otimes 2}R^p(\s,\s')
\leq (k^*-1+p_0) \e^p +p_1 
\Eq(10.7.7)
$$
This implies the lower bound
$$
k^*\geq \frac {(1-\e_1)^p-p_1}{\e_0^p}
\Eq(10.7.8)
$$
Quantitatively, this estimate can be refined to 
$$
k^*\geq C^{-1} 2^{3p/2} ((1-C2^{-p})^p-p_1)= 2^p p_0 (1-O(2^{-2p})
\Eq(10.7.9)
$$
This proves the theorem. \endproof

\specskip

%
%
%

\bigskip 

\chap{8.~Spin Glass Phase: Proof of Theorem~\thv(pr.7)}8

\noindent Having established the existence of an infinity of lumps that
carry the Gibbs measure in the low temperature phase, one would like to know 
whether these are in any way related to the original patterns. Recall that 
in the standard Hopfield model at small $\a$ the Gibbs measure concentrates 
on small balls around the patterns $\xi^\mu$. Of course the reader will 
expect that this will not be the case here. To prove this fact, we first obtain
 two  estimate the value of the Hamiltonian
in the vicinity of each pattern.

\lemma{\TH(sp.2)}
{\it
  The Hamiltonian evaluated at the patterns satisfies
$$
\P\left[ |H_N(\s = \xi^{\mu}) | \geq \frac{N}{(p!)^{\frac{1}{2}}} + zN \right]
\leq C \cases \rlap{$e^{-\frac{z^2 N}{2\a}}$,}
                \hphantom{e^{- \b'_p(z - \frac{\a \b'_p}{2}) N},}
                \quad \hbox{\tenrm if\ }z \leq \b'_p, \cr
              e^{- \b'_p(z - \frac{\a \b'_p}{2}) N}, \quad{\hbox{\tenrm otherwise}}. \cr
       \endcases
\Eq(SP.10)
$$
}
\proof
The Hamiltonian at the pattern $\xi^{\mu}$ is given by
$$
\eqalign{
H(\s = \xi^{\mu}) 
&= - \frac{(p!)^{\frac{1}{2}}}{N^{p-1}} \sum_{\II} \xi^{\mu}_{\II} \xi^{\mu}_{\II} 
             - \frac{(p!)^{\frac{1}{2}}}{N^{p-1}} \sum_{\nu \neq \mu} 
             \sum_{\II} \xi^{\nu}_{\II} \xi^{\mu}_{\II} \cr
&= - \frac{(p!)^{\frac{1}{2}}}{N^{p-1}} {N \choose p} - \frac{(p!)^{\frac{1}{2}}}{N^{p-1}} 
             \sum_{\nu \neq \mu} \sum_{\II} \xi^{\nu}_{\II} \xi^{\mu}_{\II},
}
\Eq(SP.11)
$$
which implies that
$$
-H_N(\xi^{\mu}) \leq \frac{N}{(p!)^{\frac{1}{2}}} + \frac{(p!)^{\frac{1}{2}}}{N^{p-1}}
               \sum_{\nu \neq \mu} \sum_{\II} \xi^{\nu}_{\II} \xi^{\mu}_{\II}.
\Eq(SP.12)
$$
We estimate the random part in \eqv(SP.12) by 
 the same method used in the proof
of Theorem~\thv(pr.1). By Chebyshev's exponential inequality, conditional 
independence of 
$\sum_{i=1}^N\xi^{\nu}_i \xi^{\mu}_i$ and $\sum_{i = 1}^N \xi^{\nu'}_i\xi^{\mu}_i$ 
(for $\nu \neq \mu$), and expansion of the exponential, we get for $z > 0$
$$
\eqalign{
\P\left[ \left|\frac{(p!)^{\frac{1}{2}}}{N^{p-1}} \sum_{\nu \neq \mu} 
          \sum_{\II} \xi^{\nu}_{\II} \xi^{\mu}_{\II} \right| \geq z N \right]
&\leq \inf_{t > 0} e^{-tzN} \prod_{\nu \neq \mu} \E\,\left[\exp\left(
          t \frac{(p!)^{\frac{1}{2}}}{N^{p-1}} \sum_{\II} \xi^{\nu}_{\II}\xi^{\mu}_{\II}
          \right)\right] \cr
&\leq \inf_{t > 0} e^{-tzN} \prod_{\nu \neq \mu} \bigg( 1 + \frac{t^2 p!}{2 N^{2p-2}}
          {N \choose p} \cr
&\quad\quad\quad\quad+ \frac{t^3 (p!)^{\frac{3}{2}}}{3!\,N^{3p - 3}}\E\,\bigg[ \Big| 
                  \sum_{\II}\xi^{\nu}_{\II}\Big|^3
                  e^{t \frac{(p!)^{\frac{1}{2}}}{N^{p-1}} \big| \sum_{\II} \xi^{\nu}_{\II}\big|}
          \bigg] \bigg).
}
\Eq(SP.13)
$$
The error term can be written as 
$$
\frac{1}{N^{3p - 3}}\E\,\left[ \Big| 
                  \sum_{\II}\xi^{\nu}_{\II}\Big|^3
                  e^{t \frac{(p!)^{\frac{1}{2}}}{N^{p-1}} \big| \sum_{\II} \xi^{\nu}_{\II}\big|}
          \right]
= \frac{1}{N^{\frac{3p}{2} - 3}} \E\,\left[ \left|
                N^{-\frac{p}{2}}\sum_{\II}\xi^{\nu}_{\II}\right|^3
                e^{t \frac{(p!)^{\frac{1}{2}}}{N^{\frac{p}{2}-1}}
                        | N^{-\frac{p}{2}}\sum_{\II} \xi^{\nu}_{\II}|}
                \right]
\Eq(SP.14)
$$
This latter term is exactly the same as in \eqv(A.2) (with $\b$ replaced by $t$). Hence,
we get (compare \eqv(A.3))
$$
\P\left[\frac{(p!)^{\frac{1}{2}}}{N^{p-1}} \sum_{\nu \neq \mu} 
          \sum_{\II} \xi^{\nu}_{\II} \xi^{\mu}_{\II} \geq z N \right]
\leq \inf_{t \in (0,\b'_p)} e^{-tzN + \frac{\a t^2 N}{2} + C_1}.
\Eq(SP.15)
$$
Minimizing the exponent yields
$$
\eqalign{
\P\left[ - H_N(\s = \xi^{\mu})  \geq \frac{N}{(p!)^{\frac{1}{2}}} + zN \right]
&\leq \P\left[\frac{(p!)^{\frac{1}{2}}}{N^{p-1}} \sum_{\nu \neq \mu} 
          \sum_{\II} \xi^{\nu}_{\II} \xi^{\mu}_{\II} > z N \right] \cr
&\leq C_2 \cases \rlap{$e^{-\frac{z^2}{2\a}N}$,}
                 \hphantom{e^{- \b'_p(z - \frac{\a \b'_p}{2})N},}
                 \quad \hbox{\tenrm if\ } 0 < z \leq \a \b'_p, \cr
                 e^{- \b'_p(z - \frac{\a \b'_p}{2})N}, \quad\hbox{\tenrm otherwise}.
          \endcases
}
\EQ(SP.16)
$$
This proves the claim.
\endproof
\specskip
\noindent
The next result shows that the Hamiltonian does not fluctuate much around a pattern.
This result  was already proven by Newman [N1] for the Hamiltonian
$\bar{H}$. In our case this is even simpler. 
Define $B_{\d}(\s)$ to be the $(N\d)$-ball around the configuration $\s$ in the Hamming 
distance. Then we have the following

\lemma{\TH(sp.3)}
{\it
  If $\d < \frac{1}{p}$, then there exists a constant $C > 0$ such that  
$$
\P\Big[ \exists \s \in B_{\d}(\xi^{\mu}): |H_{N}(\s) - H_N(\xi^{\mu})| 
\geq ({2^{p-1}}{(p!)^{-\frac{1}{2}}}\d + z) N\Big]
\leq C e^{ - N (f_{\d}(z) + \d \ln \d + (1 - \d) \ln (1 - \d))},
\Eq(SP.17)
$$
where
$$
f_{\d}(z) =  \cases \rlap{$\frac{z^2}{2^p \a \d}$,}
                        \hphantom{e^{- \b'_p N (z - \frac{\a \b'_p}{2 (p-1)!})},}
                        \quad \hbox{\tenrm if\ } z \leq 2^{p-1} \a \d \b'_p; \cr
                e^{- \b'_p N (z - \frac{\a \b'_p}{2 (p-1)!})}, \quad \hbox{\tenrm otherwise.}   \cr
        \endcases
\Eq(SP.17bis)
$$
}
\proof
By standard arguments (see also [N1], in particular inequality (2.3) and surrounding
comments),
$$
\eqalign{
\P\Big[ \exists \s \in B_{\d}(\xi^{\mu}): |H_{N}(\s) - H_N(\xi^{\mu})| 
                                \geq (\d + z) N\Big] \cr
&\kern-2.5cm\leq \sum_{q = 1}^{\lfloor \d N \rfloor} {N \choose q} 
        \P[|H_{N}(\zeta^q) - H_N(\xi^{\mu})| \geq (\d + z) N],
}
\Eq(SP.18)
$$
where
$$
\zeta^q_i = \cases -\xi^{\mu}_i, \quad \hbox{\tenrm if\ } i \leq q; \cr
                   \hphantom{-}\xi^{\mu}_i, \quad \hbox{\tenrm if\ } i \geq q + 1. \cr
              \endcases
\Eq(SP.19)
$$
We start by calculating the difference $|H(\zeta^q) - H(\xi^{\mu})|$. Let 
$\JJ = \JJ_q =\{1,\ldots, q \}$. One obtains
$$
\eqalign{
H(\zeta^q) - H(\xi^{\mu}) 
&= - \frac{(p!)^{\frac{1}{2}}}{N^{p-1}} \sum_{\nu = 1}^{M(N)} \sum_{\II} 
        \left(\zeta^q_{\II} \xi^{\vphantom{q}\nu}_{\II} 
        - \xi^{\mu}_{\II} \xi^{\vphantom{\mu}\nu}_{\II}\right) \cr
&= - \frac{(p!)^{\frac{1}{2}}}{N^{p-1}} \sum_{\nu = 1}^{M(N)} 
                \sum_{\II: |\II \cap \JJ|\, \hbox{\sevenrm odd}}
        \left( \zeta^q_{\II} \xi^{\vphantom{q}\nu}_{\II}
        - \xi^{\mu}_{\II} \xi^{\vphantom{\mu}\nu}_{\II} \right) \cr
&= 2 \frac{(p!)^{\frac{1}{2}}}{N^{p-1}} \sum_{\nu = 1}^{M(N)}
                \sum_{\II: |\II \cap \JJ|\, \hbox{\sevenrm odd}}
     \xi^{\mu}_{\II} \xi^{\vphantom{\mu}\nu}_{\II} \cr
&= 2 \frac{(p!)^{\frac{1}{2}}}{N^{p-1}} \sum_{\II: |\II \cap \JJ| \hbox{\sevenrm odd}} 1
        + 2 \frac{(p!)^{\frac{1}{2}}}{N^{p-1}} \sum_{\nu \neq \mu} 
                \sum_{\II: | \II \cap \JJ |\, \hbox{\sevenrm odd}}
        \xi^{\mu}_{\II} \xi^{\vphantom{\mu}\nu}_{\II} 
}
\Eq(SP.20)
$$
Explicitly, this is
$$
H(\zeta^q) - H(\xi^{\mu}) 
= 2 \frac{(p!)^{\frac{1}{2}}}{N^{p-1}} \sum_{r=1,\,\inrm{odd}}^{p-1} 
                { {N - q} \choose {p - r}} {q \choose r}
        + 2 \frac{(p!)^{\frac{1}{2}}}{N^{p-1}} \sum_{\nu \neq \mu} \sum_{\II: |\II \cap \JJ| \, \inrm{odd}} 
                \xi^{\mu}_{\II} \xi^{\vphantom{\mu}\nu}_{\II}
\Eq(SP.21)
$$
Let us treat the random term in \eqv(SP.21) first. By the usual procedure, we get
$$
\eqalign{
\P\bigg[\bigg| \frac{(p!)^{\frac{1}{2}}}{N^{p-1}} 
                \sum_{\nu \neq \mu} \sum_{\II: | \II \cap \JJ| \, \inrm{odd}}
                \xi^{\mu}_{\II} \xi^{\vphantom{\mu}\nu}_{\II} \bigg| \geq 
                        z N \bigg]  \cr
&\kern-3cm \leq 2 \inf_{t > 0} e^{-tzN} \prod_{\nu \neq \mu} \bigg\{1 + \frac{t^2 p!}{2 N^{2p-2}} 
                \sum_{\II: | \II \cap \JJ| \, \inrm{odd}} 1 \cr
&\kern-3cm \quad + \frac{t^3 (p!)^{\frac{3}{2}}}{3!\,N^{3p-3}} 
                \E\,\Big[\Big|\sum_{\II: |\II \cap \JJ| \, \inrm{odd}}
                \xi^{\mu}_{\II} \xi^{\vphantom{\mu}\nu}_{\II} \Big|^3   
                \exp\Big( \frac{t (p!)^{\frac{1}{2}}}{N^{p-1}} 
                \Big|\sum_{\II: | \II \cap \JJ| \, \inrm{odd}}
                        \xi^{\mu}_{\II} \xi^{\vphantom{\mu}\nu}_{\II} 
                                \Big| \Big) \Big] \bigg\} \cr
&\kern-3cm \leq 2 \inf_{t \in (0,\b'_p)} e^{-tzN} \prod_{\nu \neq \mu} \bigg\{1 + 
        \frac{t^2 p!}{2 N^{2p - 2}} \sum_{r = 1\,\inrm{odd}}^{p-1} { {N - q} \choose {p - r}}
                {q \choose r} 
+ C_1 N^{3 -\frac{3p}{2}} \bigg\}.
}
\Eq(SP.22)
$$
The last line follows from the usual bound on the error term (see the proof of 
Theorem~\thv(pr.1) in Chapter~3; in fact, $t$ can even be chosen somewhat larger than 
$\b'_p$, since the sum over sets $\II$ contains fewer terms than we had there).

To treat products of binomial coefficients in last expression, observe that 
if $q \leq \lfloor \d N \rfloor < \frac N2$, then the following inequality holds,
$$
\eqalign{
p! \sum_{r = 1, \,\inrm{odd}}^{p-1} {{N - q} \choose {p-r}} {q \choose r}
&\leq \sum_{r = 1,\, \inrm{odd}}^{p-1} {p \choose r} (N - q)^{p - r} q^r \cr
&\leq (N - q)^{p - 1} q \sum_{r = 1,\, \inrm{odd}}^{p-1} {p \choose r} = 2^{p-1} (N - q)^{p - 1} q .
}
\Eq(SP.23)
$$
Using \eqv(SP.23) in \eqv(SP.22) yields
$$
\eqalign{
\P\bigg[\bigg| \frac{(p!)^{\frac{1}{2}}}{N^{p-1}} 
        \sum_{\nu \neq \mu} \sum_{\II: | \II \cap \JJ| \, \inrm{odd}}
                \xi^{\mu}_{\II} \xi^{\vphantom{\mu}\nu}_{\II} \bigg| \geq z N \bigg] 
& \cr
&\kern-1cm\leq 2 \inf_{t \in (0,\b'_p)} e^{-tzN} \exp\bigg( \frac{\a t^2}{2 N^{p-1}} 2^{p-1}
                (N - q)^{p-1} q + C_1 \bigg).
}
\Eq(SP.24)
$$
The deterministic term in \eqv(SP.21) is given by (again using \eqv(SP.23))
$$
\eqalign{
\frac{(p!)^{\frac{1}{2}}}{N^{p-1}} \sum_{r = 1, \, \inrm{odd}}^{p-1} {{N - q} \choose {p - r}}{q \choose r} 
&\leq \frac{1}{(p!)^{\frac{1}{2}}N^{p-1}} \sum_{r = 1,\, \inrm{odd}}^{p-1} {p \choose r}
                (N - q)^{p-r} q^r \cr
&\leq \frac{2^{p-1}}{(p!)^{\frac{1}{2}}N^{p-1}} (N - q)^{p-1} q.
}
\Eq(SP.25)
$$
If $\d < \frac{1}{p}$, then the last line is bounded by the term for the maximum $q$. That is
$$
\frac{(p!)^{\frac{1}{2}}}{N^{p-1}} \sum_{r = 1, \, \inrm{odd}}^{p-1} {{N - q} \choose {p - r}}{q \choose r}
\leq \frac{2^{p-1}}{(p!)^{\frac{1}{2}}N^{p-1}} (N - \lfloor \d N \rfloor )^{p-1} \lfloor \d N \rfloor 
\leq \frac{2^{p-1}}{(p!)^{\frac{1}{2}}} N \d.
\Eq(SP.26)
$$
Collecting \eqv(SP.24) and \eqv(SP.26), we get
$$
\eqalign{
\P\big[ | H(\zeta^q) - H(\xi^{\mu})| \geq \frac{2^{p-1}}{(p!)^{\frac{1}{2}}}\d N + z N] \cr
&\kern-1cm\leq 2 \inf_{t \in (0,\b'_p)} e^{-tzN} \exp\bigg( \frac{\a t^2}{2 N^{p-1}} 2^{p-1}
                (N - q)^{p-1} q^{r} + C_1 \bigg).
}
\Eq(SP.27)
$$
Plugging this into \eqv(SP.18) gives
$$
\eqalign{
\P[ \exists \s \in B_{\d}(\xi^{\mu}): |H_{N}(\s) - H_N(\xi^{\mu})| \geq
&(\frac{2^{p-1}}{(p!)^{\frac{1}{2}}}\d + z) N]  \cr
&\kern-2cm\leq 2 \sum_{q = 1}^{\lfloor \d N \rfloor} {N \choose q} 
       \inf_{t \in (0,\b'_p)} e^{-tzN} \exp\bigg( \frac{\a t^2}{2 N^{p-1}} 2^{p-1}
                (N - q)^{p-1} q + C_1 \bigg).
}
\Eq(SP.28)
$$
It is straightforward to check that under our assumptions on $\d$ and for fixed $t$, 
the ratio between two consecutive terms in the above sum is larger than 2, and 
therefore the whole sum is at most twice the maximum term,
$$
\eqalign{
\P[ \exists \s \in B_{\d}(\xi^{\mu}): |H_{N}(\s) - H_N(\xi^{\mu})| > 
&(\frac{2^{p-1}}{(p!)^{\frac{1}{2}}}\d + z) N]  \cr
&\kern-1cm\leq 4 {N \choose {\lfloor \d N \rfloor}} \inf_{t \in (0,\b'_p)}
        e^{-tzN} \exp\bigg( \frac{2^{p-1}\a t^2}{2} N \d
                + C_1 \bigg).
}
\Eq(SP.29)
$$
Minimizing with respect to $t$ and using Stirling's formula for the binomial factor
concludes the proof of Lemma~\thv(sp.3).
\endproof
\specskip
\proofof{Theorem~\thv(pr.7)}
We observe the following elementary fact. By the definition of the free energy 
$$
F_N(\b) \leq \frac{\b}{N} \sup_{\s} |H_N(\s)|.
\Eq(SP.30)
$$
Hence, by Theorem~\thv(pr.4), for any $\b, m, z > 0$ there exists  $\bar{N} \in \N$ such 
that 
$$
\P[\frac{1}{N} \sup_{\s} |H_N(\s)| < \frac{1}{\b} \E\,F_N(\b) - z]
\leq \P[F_N(\b) < \E\,F_N(\b) - z] \leq C N^{-m},
\Eq(SP.31)
$$
for all $N \geq \bar{N}$.
Suppose that $\a \b_p(\a) > \frac{1}{(p!)^{\frac{1}{2}}}$. Then there exists $\b > 0$ such that
$$
\a > \frac{1}{(p!)^{\frac{1}{2}} ( \b_p - \frac{\b_p^2}{2\b})},
\Eq(SP.32)
$$
which is equivalent to
$$
\frac{1}{(p!)^{\frac{1}{2}}} 
< \frac{1}{\b}(\a \b \b_p - \frac{\a \b_p^2}{2}) 
\leq \frac{1}{\b} \E\,F_N(\b) + C_1 N^{-1}.
\Eq(SP.33)
$$
The second inequality follows from the convexity of $F_N(\b)$  and the
definition of $\b_p$. But then we can find $\d \in (0,\frac{1}{p})$ and $z > 0$ such that 
(for all $N$ sufficiently large)
$$
\frac{2^{p-1}}{(p!)^{\frac{1}{2}}} \d + 3z < \frac{1}{\b} \E\,F_N(\b) - \frac{1}{(p!)^{\frac{1}{2}}},
\Eq(SP.34)
$$
and (with the definition of $f_{\d}$ from Lemma~\thv(sp.3))
$$
f_{\d}(z) + \d \ln \d + (1 - \d)\ln (1 - \d) > 0.
\Eq(SP.35)
$$
By Lemma~\thv(sp.2), resp.\ \thv(sp.3), for any $m > 0$, we can find an $\bar{N} \in \N$
such that for all $N \geq \bar{N}$
$$
\eqalign{
\P[\exists \s \in \bigcup_{\mu = 1}^{M(N)} B_{\d}(\xi^{\mu}): 
        |H_N(\s)| \geq N(\frac{1}{(p!)^{\frac{1}{2}}} + \d  + 2z)] &\cr
&\kern-4cm\leq \P[\exists \s \in \bigcup_{\mu = 1}^{M(N)} B_{\d}(\xi^{\mu}): 
         |H_N(\s) - H_N(\xi^{\mu})| \geq N(\d + z)] \cr
&\kern-4cm\quad+ \P[\sup_{\mu} | H_N(\xi^{\mu})| \geq N(\frac{1}{(p!)^{\frac{1}{2}}} + z)] \cr
&\kern-4cm\leq N^{-m}.
}
\Eq(SP.36)
$$
On the other hand, the inequality \eqv(SP.31) implies that
$$
\P[\sup_{\s} |H_N(\s)| \leq N \frac{\E\,F_N(\b)}{\b} - z N] 
\leq N^{- m}, 
\Eq(SP.37)
$$
for all $N$ large enough, so that finally, by standard arguments,
$$
\eqalign{
\P[\arg \sup |H_N(\s)| \in \bigcup_{\mu = 1}^{M(N)} B_{\d}(\xi^{\mu})]
&\leq \P[\exists \s \in \bigcup_{\mu = 1}^{M(N)} B_{\d}(\xi^{\mu}): 
        |H_N(\s)| \geq N(\frac{1}{(p!)^{\frac{1}{2}}} + \d  + 2z)] \cr
&\quad + \P[\sup_{\s} |H_N(\s)| \leq N \frac{\E\,F_N(\b)}{\b} - z N] \cr
&\leq N^{-m},
}
\Eq(SP.38)
$$
for all $N$ larger than some $\bar{N} \in \N$.

To show the existence of an $\a_{sp}$, we observe that the bounds \eqv(PR.16) and 
\eqv(PR.17) on the critical $\b$ imply that the quantity $\a \b_p(\a) \sim \sqrt{\a}$ 
and is thus eventually larger than any fixed number. This concludes the proof of 
Theorem~\thv(pr.7).
\endproof
%

\chap{References \PG(appendix.references)}{1}
\bigskip
\item{[ALR]} M.~Aizenman, J.~L.~Lebowitz, D.~Ruelle, {\it Some rigorous results
        on the Sherrington-Kirkpatrick model}, Commun.\ Math.\ Phys. {\bf 112},
        3--20 (1987)    
%
%
\item{[AC]}  M.~Aizenman, P. Contucci, {\it On the stability of the quenched
state in mean field spin glass models}, J. Stat. Phys. {\bf 92}, 
765-783 (1998).
%
%
%
%
%
\item{[B1]} A.~Bovier, {\it Self-averaging in a class of generalized Hopfield Models},
        J.~of Physics {\bf A 27}, 7069--7077 (1994).
\item{[B2]} A.~Bovier, {\it Statistical mechanics of disordered systems},
        MaPhySto Lecture Notes 10, (2001).
%
%
%
\item{[BG1]} A.~Bovier, V.~Gayrard,
        {\it Hopfield models as generalized random mean field models}, in
        {\it Mathematical aspects of spin glasses and neural networks},
        A.~Bovier and P.~Picco (eds.), Progress in Probability, Birkh\"auser,
        Boston-Basel-Berlin (1998).
\item{[BG2]} A.~Bovier, V.~Gayrard,
        {\it The retrieval phase of the Hopfield model: a rigorous analysis of
        the overlap distribution}, Prob.\ Theory Rel.\ Fields {\bf 107}, 61--98 (1997).
\item{[BG3]} A.~Bovier, V.~Gayrard, {\it Metastates in the Hopfield model 
        in the replica symmetric regime}, Math.\ Phys.\ Anal.\ Geom.  {\bf 1}, 107-144
        (1998).
\item{[BG4]} A.~Bovier, V.~Gayrard, {\it An almost sure central limit theorem for the Hopfield model},
        Markov Proc.\ Related Fields {\bf 3}, 151--173 (1997).
\item{[BGP1]} A.~Bovier, V.~Gayrard, P.~Picco,
        {\it Gibbs states of the Hopfield model in the regime of perfect memory},
        Prob.\ Theory Rel.\ Fields {\bf 100}, 329--363 (1994).
\item{[BGP2]} A.~Bovier, V.~Gayrard, P.~Picco, {\it Gibbs states of the Hopfield model with 
        extensively many patterns}, J.~Stat.\ Phys.\ {\bf 79}, 395--414 (1995).
%
%
%
\item{[BKL]} A.~Bovier, M.~L\"owe, I.~Kurkova, {\it Fluctuations of the free energy in 
        the REM and the $p$-spin SK models}, to appear in Ann. Probab. (2002).
%
%
%
%
%
\item{[CH]} R.~Courant, D.~Hilbert, {\it Methoden der mathematischen Physik I}, $3^{\hbox{\sevenrm rd}}$
        ed., Springer, Berlin-Heidelberg-New York  (1968).
%
%
\item{[D1]} B.~Derrida, {\it Random energy model: limit of a family of disordered
        models}, Phys.\ Rev.\ Letts.\ {\bf 45}, 79--82 (1981).
\item{[D2]} B.~Derrida, {\it Random energy model: An exactly solvable model of disordered
        systems}, Phys.\ Rev.\ B {\bf 24}, 2613--2626 (1984).
%
%
%
\item{[FP1]} L.~A.~Pastur, A.~L.~Figotin, {\it Exactly soluble model of a spin glass},
        Sov.\ J.~Low Temperature Physics {\bf 3}({\bf 6}), 378--383 (1977).
\item{[FP2]} L.~A.~Pastur, A.~L.~Figotin, {\it On the theory of disordered spin systems}, 
        Theor.\ Math.\ Phys.\ {\bf 35}, 403--414 (1978).
%
\item{[GG]} S.~Ghirlanda and F.~Guerra, {\it General properties of the overlap
           probability distributions in disordered spin systems.Towards Parisi
           ultrametricity}, J. Phys. A {\bf 31}, 9144-9155 (1998).
%
%
\item{[Ho]} J.~J.~Hopfield, {\it Neural networks and physical systems with emergent collective
        computational capabilities}, Proc.\ Natl.\ Acad.\ Sci.\ USA {\bf 79}, 2554--2558 (1982).
\item{[Ko]} H.~Koch, {\it A free energy bound for the Hopfield model}, J.~Phys.\ {\bf A 26},
        L353--L355 (1993).
%
%
\item{[Le]} M.~Ledoux, {\it On Talagrand's deviation inequalities for product measures},
        ESAIM: Probability {\bf 1},  63--87 (1996).
%
%
\item{[Lee]} Y.~C.~Lee, G.~Doolen, H.~H.~Chen, G.~Z.~Sun, T.~Maxwell, H.~Y.~Lee, and C.~L.~Gilles,
        {\it Machine learning using higher order correlation networks}, Physica~D {\bf 22},
        276--306 (1986).
\item{[MPV]} M.~M\'ezard, G.~Parisi, M.~A.~Virasoro,
        {\it Spin glass theory and beyond}, World Scientific, Singapore (1987).
\item{[MS]} V.~D.~Milman, G.~Schechtman, {\it Asymptotic theory of finite dimensional
        normed spaces}, Lecture notes in Mathematics {\bf 1200}, Springer, 
        Berlin-Hei\-del\-berg-New York (1986).
\item{[N1]} C.~M.~Newman, {\it Memory capacity in neural network models: rigorous lower bounds},
        Neural Networks {\bf 1}, 223--238 (1988).
\item{[Ni1]} B.~Niederhauser, {\it Mathematical aspects of Hopfield models}, PhD Thesis,
\item{[Ni2]} B.~Niederhauser, {\it Norms of certain random matrices}, preprint IME-USP (2001).
\item{[PS]} L.~Pastur, M.~Shcherbina, {\it Absence of self-averaging of the order parameter
        in the Sherrington-Kirkpatrick model}, J.~Stat.\ Phys.\ {\bf 74}, 1161--1183 (1994).
\item{[PN]} P.~Peretto, J.~J.~Niez, {\it Long term memory storage capacity of multiconnected
        neural networks}, Biolog.~Cybernetics {\bf 39}, 53--63 (1986).
\item{[Ro]} R.~T.~Rockafellar, {\it Convex Analysis}, Princeton University Press, 
        Princeton (1972).
\item{[ST]} M.~Shcherbina, B.~Tirozzi, {\it The free energy for a class of Hopfield models},
        J.~Stat.\ Phys.\ {\bf 72}, 113--125 (1992).
%
%
\item{[SK]} D.~Sherrington, S.~Kirkpatrick, {\it Solvable model of a spin-glass},
        Phys.\ Rev.\ Lett.\ {\bf 35}, 1792--1796 (1975).
%
%
\item{[T1]} M.~Talagrand, {\it A new look at independence}, Ann.\
Prob.\  {\bf 24}, 1--34 (1996). 
\item{[T2]} M.~Talagrand, {\it The Sherrington-Kirkpatrick model: a
challenge for mathematicians}, 
        Prob.\ Theory Rel.\ Fields {\bf 110}, 109--176 (1998).
\item{[T3]} M.~Talagrand, {\it Rigourous results for the Hopfield
model with many patterns}, 
        Prob.\ Theory Rel.\ Fields {\bf 110}, 177--276 (1998).
\item{[T4]} M.~Talagrand, {\it Rigorous low temperature results for
the mean field $p$-spin 
        interaction model}, Prob.\ Theory Rel.\ Fields {\bf 117},
        303--360 (2000).   
\item{[T5]}  M.~Talagrand, {\it On the $p$-spin interaction model at
low temperature}, C.R.A.S.  
                         {\bf 331}, 939--942  (2000).
\item{[T6]}    M.~Talagrand, {\it A first course on spin glasses.}
                         Lectures at the \'Ecole d'\'et\'e de St.
                         Flour, 2000 (availble at request from the author).
\item{[T7]} M.~Talagrand, {\it Exponential inequalties and convergence of moments 
in the replica-symmetric regime of the Hopfield model}, Ann. Probab.{\bf 28}, 1393--1469 (2000).
%
%
\item{[Yu]} V.~V.~Yurinskii, {\it Exponential inequalities for sums of random vectors},
        J.~Multivariate Analysis {\bf 6}, 473--499 (1976).
%

%
\end